\documentclass[usegraphicx,usenatbib,useAMS]{mn2e}
  \usepackage{graphicx}
  \usepackage{bm,amssymb,amsmath}
  \usepackage{times}
  \usepackage{natbib}
  \usepackage{float}
  \usepackage{color}
  \usepackage{fixltx2e} %to display {figure*}s in right order
  \usepackage[normalem]{ulem} %%required for various underlining styles

  \graphicspath{{./fig/}{./png/}{./ps/}}

  \newcommand{\phm}{{\phantom{0}}} %empty pace, 0-wide, useful to align table entries
  \newcommand{\vect}[1]{{{\mbox{\boldmath $#1$}}}}%also makes bold Greek letters
  \newcommand\sfrac[2]{{\textstyle{\frac{#1}{#2}}}}
  \newcommand\deriv[2]{\displaystyle\frac{\upartial #1}{\upartial #2} }
  \newcommand\average[1]{{\langle #1 \rangle}} %volume average
  \newcommand\mean[1]{\overline{#1}}         %ensemble average
  \newcommand{\cool}{{_{\rm cool}}}  %radiative cooling  
  \newcommand{\mesh}{{_{\Delta}}}
  \newcommand{\tphi}{{\Phi}}  %ensemble-averaged filling factor
  \newcommand{\Mesh}{{\Delta}}
  \newcommand{\rms}{{_0}}
  \newcommand{\rrms}{{_\mathrm{rms}}}
  \newcommand{\sound}{{_{\mathrm{s}}}}  %sound
  \newcommand{\SN}{_{\rm SN}}     %supernova
  \newcommand{\dd}{{\rm d}}       %for derivatives
  \newcommand{\HI}{{\rm H}{\sc i}}
  \newcommand{\HII}{{\rm H}{\sc ii}}

  \newcommand{\Pe}{{\rm Pe}}								%Peclet
  \newcommand{\Rey}{{\rm Re}}								%Reynolds
  \newcommand{\Op}{{WSWa}}
  \newcommand{\OpH}{{WSWah}}
  \newcommand{\WSWa}{{WSWb}}

  \newcommand{\str}{\mathcal{D}}
  \newcommand{\corr}{\mathcal{C}}
  \newcommand{\cm}{\,{\rm cm}}
  
  \newcommand{\cmcube}{\,{\rm cm^{-3}}}
  \newcommand{\dyn}{\,{\rm dyn}}
  \newcommand{\erg}{\,{\rm erg}}
  \newcommand{\g}{\,{\rm g}}

  \newcommand{\km}{\,{\rm km}}
  \newcommand{\kms}{\,{\rm km\,s^{-1}}}

  \newcommand{\K}{\,{\rm K}}
  \newcommand{\kpc}{\,{\rm kpc}}
  
  \newcommand{\Myr}{\,{\rm Myr}}
  \newcommand{\Gyr}{\,{\rm Gyr}}

  \newcommand{\Msol}{{M_{\sun}}}
  \newcommand{\p}{\,{\rm pc}}
  
  \newcommand{\s}{\,{\rm s}}
  \newcommand{\yr}{\,{\rm yr}}

  \definecolor{burntorange}{RGB}{255,97,0}
  \definecolor{violet}{RGB}{105,20,200}
  \definecolor{royalblue}{RGB}{0,137,255}
  \definecolor{mygreen}{RGB}{0,120,0}

  \newenvironment{replyb}
    {\color{black} }

\pagerange{\pageref{firstpage}--\pageref{lastpage}}
\label{firstpage}
\pubyear{2012}

\begin{document}

\title[SN-regulated ISM. I.]{The supernova-regulated ISM. I. The multi-phase structure}
\author[Gent et al.]{F.~A.~Gent,$^{1,4}$\thanks{E-mails: F.Gent@sheffield.ac.uk, Anvar.Shukurov@ncl.ac.uk,
                        And\-rew.Flet\-cher@ncl.ac.uk, G.R.Sarson@ncl.ac.uk and
                        Maarit.Mantere@helsinki.fi}
            A.~Shukurov,$^1$
            A.~Fletcher,$^1$
            G.~R.~Sarson,$^1$
            M.~J.~Mantere$^{2,3}$\\
$^1$School of Mathematics and Statistics, Newcastle University,
        Newcastle upon Tyne NE1~7RU, UK\\
$^2$Physics Department, University of Helsinki, PO BOX 64, Helsinki, FI-00014,
        Finland\\
$^3$Department of Information and Computer Science, Aalto University, 
PO Box 15400, FI-00076 Aalto, Finland\\
$^4$School of Mathematics and Statistics, University of Sheffield,
        Sheffield S3~7RH, UK
}

\maketitle

%--------------------------------------------------------------
\begin{abstract}
We simulate the multi-phase interstellar medium (ISM) randomly heated
and stirred by supernovae (SNe), with gravity, differential rotation and other parameters 
of the solar neighbourhood. 
Here we describe in detail both numerical and physical aspects of
the model, including injection of thermal and kinetic energy by SN explosions, radiative
cooling,  photoelectric heating and various transport processes.
With a three-dimensional domain extending $1\times1\kpc^2$ horizontally and
$2\kpc$ vertically (symmetric about the galactic mid-plane),
the model routinely spans gas number densities  $10^{-5}$--$10^2\cmcube$, 
temperatures  10--$10^8\K$, local velocities up to $10^3\kms$
(with Mach number up to 25). The working numerical resolution of 4\,pc has been selected via 
simulations of a single expanding SN remnant, where we closely reproduce, at this resolution, analytical solutions for
the adiabatic and snowplough regimes.
{\replyb{
The feedback of the the halo on the disc cannot be
captured in our model where the domain only extends to the height of $1\kpc$
above the mid-plane.
We argue that to reliably model the disc-halo connections
 would require extending the 
domain horizontally as well as vertically due to the increasing horizontal 
scale of the gas
flows with height.
}}

The thermal structure of the modelled ISM is classified
{{by inspection of the joint probability density of the gas number 
density and temperature.}}
We confirm that most of the complexity can be captured in terms of just three phases, separated by temperature borderlines at about $10^3\K$ and $5\times10^5\K$. 
The 
{{probability}} distribution of gas density within each phase is approximately lognormal.
We clarify the connection between the fractional volume of a phase and its various proxies, and derive an exact relation between the fractional volume and the filling factors defined in terms of the volume and probabilistic averages. 
These results are discussed in both observational and computational contexts. 
The correlation scale of the random flows is calculated from the velocity autocorrelation function; it is of order 100\,pc and tends to grow with distance from the mid-plane. 
We use two distinct parameterizations of radiative cooling to show that the multi-phase structure of the gas is robust, as it does not depend significantly on this choice.

\end{abstract}

%--------------------------------------------------------------------------

%--------------------------------------------------------------------------
\section{Introduction}
%--------------------------------------------------------------------------

The multi-phase structure of the interstellar medium (ISM) affects almost all
aspects of its dynamics, including its evolution, star formation, galactic winds and
fountains, and the behaviour of magnetic fields and cosmic
rays. In a widely accepted picture \citep{CS74,MO77}, most of the volume is
occupied by the hot ($T\simeq10^6\K$), warm ($T\simeq10^4\K$) and cold
($T\simeq10^2\K$) phases. The concept of the multi-phase ISM in pressure equilibrium 
has endured with modest refinement \citep{Coxreview05},
{{e.g., deviations from
thermal pressure balance 
have been detected}}
\citep[][and references therein]{KKRev09}. 
Dense molecular
clouds, while binding most of the total mass of the interstellar gas 
and being of key importance for star formation,
occupy a negligible fraction of the total volume
\citep[e.g.][]{Kulkarni87, Kulkarni88, Spitzer90, McKee95}. 
The main sources of energy
maintaining this complex structure are supernova (SN) explosions and stellar
winds \citep[][and references therein]{MLK04}. The clustering of SNe in OB
associations facilitates the escape of the hot gas into the halo thus reducing
the volume filling factor of the hot gas in the disc, perhaps down to 10\% at
the mid-plane \citep{NI89}. The energy injected by the SNe not only produces
the hot gas but also drives ubiquitous compressible turbulence in all phases,
as well as driving outflows from the disc, associated with the galactic fountain or wind, as first suggested by \citet{Bregman80}. Thus turbulence, the
multi-phase structure, and the disc-halo connection are intrinsically related
features of the ISM.

A comprehensive description of the complex dynamics of the multi-phase ISM has
been significantly advanced by numerical simulations in the last three decades,
starting with \citet{CP85}, followed by many others including
\citet{Rosen93,RB95,V-SPP95,PV-SP95,RBK96,Korpi99,GP99,WN99,Avillez00,WN01,AB01,
AM-L02,WMN02,AB04,Balsara04,AB05a};\citet{AB05b,Slyz05,M-LBKA05,Joung06,AB07,WN07,Gressel08}.
Numerical simulations of this type are demanding even with the best computers
and numerical methods available. The self-regulation cycle of the ISM includes
physical processes spanning enormous ranges of gas temperature and density,
and of spatial and temporal scales, as it involves star formation in the
cores of molecular clouds, assisted by gravitational and thermal instabilities
at larger scales, which evolve against the global background of transonic
turbulence driven, in turn, by star formation \citep{MLK04}. It is
understandable that none of the existing numerical models covers the whole
range of parameters, scales and physical processes known to be important.

Two major approaches in earlier work focus either on the dynamics of diffuse
gas or on dense molecular clouds. Our model belongs to the former class, where
we are mainly concerned with the ISM dynamics in the range of scales of order
$10\p$--$1\kpc$. Numerical constraints prevent us (like many other authors)
from fully including the gravitational and thermal instabilities which involve
scales of less than 1\p. In order to assess the sensitivity of our results to
the parameterization of radiative cooling, we consider models with thermal
instability, but reduce its efficiency using a sufficiently strong thermal
conductivity to avoid the emergence of structures that are unresolvable at our
numerical resolution.
The results are compared to models with no thermally unstable branch
over the temperature range between the cold and warm phases. To our
knowledge, no direct study addressing the difference between these two
kinds of 
{{the cooling}}
parameterizations has been made. We note, however, that
\citet{VS00} compared their thermally unstable model to
a different model by \citet{Scalo98}, who used a thermally stable cooling
function. Similarly, \citet{AB04} and \citet{Joung06} compared results
obtained with different cooling functions, but again comparing different
models: here we compare models with different cooling functions but which are
otherwise the same.

An unavoidable consequence of the modest numerical resolution available, if we
are to capture the dynamics on $1\kpc$ scales, is that star formation,
manifesting itself only through the ongoing SN activity in our model,
has to be heavily parameterized.  We do, however, ensure that
individual supernova remnants are modelled accurately, since this is
essential to reliably reproduce the injection of thermal and kinetic
energy into the ISM. In particular, our model reproduces with high
accuracy the evolution of supernova remnants from the Sedov--Taylor
stage until the remnant disintegrates and merges into the ISM
(Appendix~\ref{EISNR}).

The dimensionless parameters characteristic of the ISM, such as the kinetic and magnetic
Reynolds numbers (reflecting the relative importance of gas
viscosity and electrical resistivity) and the Prandtl number (quantifying
thermal conductivity), are too large to be simulated with current computers.
Similarly to most numerical simulations of this complexity, our numerical
techniques involve a range of artificial transport coefficients for momentum
and thermal energy (such as shock-capturing viscosity). We
explore and report here the sensitivity of our results to the artificial
elements in our basic equations.

This paper is the first of a planned series, in which we aim to clarify which
components and physical processes control the different properties of the ISM.
Our next step is to add magnetic fields to the model, to study both their origin
and role in shaping the ISM. But in order to identify where the magnetic field
is important and where it is not, we first must understand what the properties
of a purely hydrodynamic ISM would be.

The structure of the paper is as follows. In Section~\ref{NI} we
present our basic equations, numerical methods, initial and boundary
conditions, as well as the physical ingredients of the model, such as our
modelling of SN activity and heating and cooling of the ISM. Our results
are presented in Sections~\ref{REF}--\ref{SMP}, including an overview of the multi-phase
structure of the ISM, the correlation length of random flows, and their sensitivity
to the cooling function and numerical resolution.
Our results are discussed in a broader context in
Section~\ref{disc}, where our conclusions are also summarised.
Detailed discussion of important technical and numerical aspects of the model,
and the effects of the unavoidable unphysical assumptions adopted,
can be found in Appendices: the accuracy of our modelling of individual
supernova remnants in Appendix~\ref{EISNR}, our control of numerical
 dissipation in Appendix~\ref{BCND}, and sensitivity to thermal instability
in Appendix~\ref{TI}.

%--------------------------------------------------------------------------
\section{Basic equations and their numerical implementation}\label{NI}
%--------------------------------------------------------------------------

%----------------------------------------------------------------------------
\subsection{Basic equations}
We solve numerically a system of hydrodynamic equations using the
{\sc Pencil Code} (http://code.google.com/p/pencil-code) which is designed
for fully nonlinear, compressible magnetohydrodynamic (MHD) simulations. We
consider only the hydrodynamic regime for the purposes of this paper; MHD
simulations, which are in progress, will be reported elsewhere. 
Nor do we include cosmic rays, which we subsequently plan to add to the MHD
simulations. 

The basic equations include the mass conservation equation, the Navier--Stokes
equation (written here in the rotating frame), and the heat equation written
in terms of the specific entropy:\footnote{{For the reader's convenience, 
Appendix~\ref{notation} contains a list of variables used in the text with their 
definitions.}}
\begin{align}
\label{eq:mass}
\frac{D\rho}{Dt} &=-{\replyb{\rho \nabla \cdot \vect{u}}}+\dot{\rho}\SN,\\
\label{eq:mom}
\frac{D\vect{u}}{Dt} &=-\rho^{-1}\nabla{\sigma}\SN
             -c_\mathrm{s}^2\nabla\left({s}/{c_{p}}+\ln\rho\right)\nonumber\\
&\mbox{}-\nabla\Phi-Su_x\bm{\hat{y}}-2\vect{\Omega}\times\vect{u}\nonumber\\
&\mbox{}+\nu \left( \nabla^{2} \vect{u}
             + \sfrac{1}{3}\nabla \nabla \cdot \vect{u}
             + 2{\mathbfss W}\cdot\nabla \ln\rho\right)\nonumber\\
&\mbox{}  +\zeta_{\nu}\left(\nabla\nabla \cdot \vect{u} \right),\\
\label{eq:ent}
\rho T\frac{D s}{Dt} &=
             \dot{\sigma}\SN+\rho\Gamma-\rho^2\Lambda+\nabla\cdot\left(c_p\rho\chi\nabla T\right)
             +2 \rho \nu\left|{\mathbfss W}\right|^{2}\nonumber\\
      &\mbox{}+\zeta_\chi\rho\left(\nabla\cdot\vect{u}\right)^2,
\end{align}
where $\rho$, $T$ and $s$ are the gas density, temperature and specific
entropy, respectively, $\vect{u}$ is the deviation of the gas velocity from
the background rotation profile 
(here called the \textit{velocity perturbation\/}), $c\sound$ is the adiabatic speed of sound, $c_p$
is the heat capacity at constant pressure, 
$S$ is the velocity shear rate
associated with the Galactic differential rotation at the angular velocity
$\vect{\Omega}$ assumed to be aligned with the $z$-axis  (see below).
The Navier--Stokes
equation includes 
viscosity $\nu$ and the rate of
strain tensor $\mathbfss W$ whose components are given by
\begin{equation}
\label{eq:str}
 2 W_{ij}= \frac{\upartial u_{i}}{\upartial x_{j}}+
    \frac{\upartial u_{j}}{\upartial x_{i}}
  -\frac{2}{3}\delta_{ij}\nabla \cdot\vect{u},
\end{equation}
as well as the shock-capturing viscosity $\zeta_\nu$.
The system is driven by SN energy injection,
at the rates $\dot{\sigma}\SN$ (per unit volume) in the form of
kinetic energy in Eq.~(\ref{eq:mom}) and thermal energy in Eq.~(\ref{eq:ent}). 
Energy injection
{{is applied in a single time step and}}
is confined to the interiors of newly introduced SN remnants, and
the total energy injected per supernova is denoted $E\SN$. The mass of the SN
ejecta is included in Eq.~(\ref{eq:mass}) via the source $\dot{\rho}\SN$. The
forms of these terms are specified and further details are given in
Section~\ref{MSN}. The heat equation also contains a thermal energy source due
to photoelectric heating $\rho\Gamma$, energy loss due to optically thin radiative
cooling $\rho^2\Lambda$, heat conduction with the thermal diffusivity $\chi$
(with $K=c_p\rho\chi$ the 
thermal conductivity), viscous heating
(with $|{\mathbfss W}|$ the determinant of ${\mathbfss W}$),
and the shock-capturing thermal diffusivity $\zeta_\chi$.

The advective derivative,
\begin{equation}
\frac{D}{Dt}= \frac{\upartial}{\upartial t} + \left( \vect{U} +
    \vect{u} \right) \cdot \nabla,
\label{eq:advection}
\end{equation}
includes transport by an imposed shear flow $\vect{U}=(0,Sx,0)$\label{shear} 
in the local Cartesian coordinates (taken to be linear across the local simulation box), with the
velocity $\vect{u}$ representing a deviation 
from the overall rotational
velocity $\vect{U}$.
{{As will be discussed later, due to anisotropies (e.g. density 
stratification, anisotropic turbulence), large-scale flows will be generated 
in the system; one example is the systematic vertical outflow discussed 
at length in this paper. Therefore, the perturbation velocity 
$\vect{u}$ consists of two parts, 
a mean flow and random velocities.
Here we consider a mean flow
{obtained by Gaussian smoothing} 
 \citep{G92}:
  \begin{align}\label{eq:Bxgauss}
  \average{\vect{u}}_\ell(\vect{x})
	&=\int_{V}\vect{u}(\vect{x}')G_\ell(\vect{x}-\vect{x}')\,\dd^3\vect{x}',\\
    G_\ell(\vect{x})&=\left(2\upi \ell^2\right)^{-{3}/{2}}
	\exp\left[-{\vect{x}^2}/({2\,\ell^2})\right],\nonumber
  \end{align}
where we use a smoothing scale $\ell\simeq50\p$,
necessarily somewhat shorter than the flow correlation length $l_0$
obtained in Section~\ref{CORR} \citep[for details, see][]{FG12a}. 
The random flow is then $\vect{u_0}=\vect{u}-\average{\vect{u}}_\ell$. }}\label{u0}
The differential rotation of the galaxy is modelled with a background shear
flow along the local azimuthal ($y$) direction, $U_y=Sx$. The
shear rate is $S=r\upartial\Omega/\upartial r$ in terms of galactocentric
distance $r$, which translates into the $x$-coordinate of the local Cartesian
frame. In this paper we consider models with rotation and shear
similar to those in the solar neighbourhood, $\Omega=-S=25\kms\kpc^{-1}$.
{
We do not expect the 
{gas velocities and thermal
structure discussed} here to depend strongly 
on the rotation and shear parameters, 
although other aspects of the solution will be more sensitive to these.
Future papers will consider the rotation and shear in more detail;
and will also include magnetic fields, whose generation may depend strongly
on these parameters.} 

We consider an ideal gas, with thermal pressure given by
\[
    \label{eq:eos}
    p = \frac{k_\mathrm{B}}{\mu m_\mathrm{p}}\rho T,
\]
where $k_\mathrm{B}$  is the Boltzmann constant, $m_\mathrm{p}$ is the proton
mass, and $\mu=0.62$ is the mean molecular weight of a fully ionised gas of
the Solar chemical composition.

In Eq.~(\ref{eq:mom}), $\Phi$ is the gravitational potential produced by
stars and dark matter. For the Solar vicinity of the Milky Way,
\citet{Kuijken89} suggest the following form of the vertical gravitational
acceleration \citep[see also][]{F01}:
\begin{equation}
    \label{eq:grav}
 g_{z}=-\deriv{\Phi}{z}
  =-\frac{a_1}{\sqrt{z_1^{2}+z^{2}}}-a_2\frac{z}{z_2},
  \end{equation}
with $a_1=4.4\times10^{-16}\km\s^{-2}$, $a_2=1.7\times10^{-16} \km\s^{-2}$,
$z_1=200\p$ and $z_2=1\kpc$. We neglect self-gravity of the interstellar gas
because it is subdominant at the scales of interest.

%------------------------------------------------------------------------
\subsection{Modelling supernova activity}\label{MSN}

We include both Type~II and Type~I SNe in our simulations, distinguished only
by their frequency and vertical distribution. The SNe frequencies are those in
the Solar neighbourhood \citep*[e.g.][]{TLS94}. Type~II SNe are introduced at
a rate, per unit surface area, of $\nu_\mathrm{II}=25\kpc^{-2}\Myr^{-1}$
($0.02\yr^{-1}$ in the whole Galaxy), with fluctuations of the order of
$10^{-4}\yr^{-1}$ at a time scale of order $10\Myr$. Such fluctuations
in the Type~II SN rate are natural to introduce; there is some evidence
that they can enhance dynamo action in MHD models
\citep{Hanasz04,Balsara04}. The surface density rate of Type~I SNe is
$\nu_\mathrm{I}=4\kpc^{-2}\Myr^{-1}$ (interval of 290 years between
Type~I SN explosions in the Galaxy).
{{We do not explicitly include any spatial clustering of the SNe.}}

Unlike most other ISM models of this type, the SN energy in the injection site
is split
between thermal and kinetic parts, in order to reduce artificial
temperature and energy losses at early stages of the SN remnant evolution.
Thermal energy density is distributed within the injection site as 
$\exp[-(r/r\SN)^6]$,
with $r$ the local spherical radius and $r\SN$ {{(of order $10\p$ -- see below)}} the nominal location of the
remnant shell 
(i.e.\ the radius of the SN bubble) 
at the time of injection. Kinetic energy is injected by adding
a spherically symmetric velocity field $u_r\propto\exp[-(r/r\SN)^6]$;
subsequently, this rapidly redistributes matter into a shell. 
To avoid a discontinuity in $\vect{u}$ at the
centre of the injection site, the centre is simply placed midway between grid points.
We also inject $4\Msol$ as stellar ejecta, with density profile 
$\exp[-(r/r\SN)^6]$.
Given the turbulent environment, there are significant random motions
and density inhomogeneities
within the
injection regions. Thus, the initial kinetic energy is not the same in each
region, and, injecting part of the SN energy in the kinetic form results 
in the total kinetic energy varying between SN remnants. 
We therefore record the energy added for every
remnant so we can fully account with the rate of energy injection. For example,
in Model~{\Op} we obtain the energy per SN in the range
\[
0.5 < E\SN< 1.5\times10^{51}\erg,
\]
with the average of $0.9\times10^{51}\erg$.

The SN sites are randomly distributed in the horizontal coordinates $(x,y)$.
Their vertical positions are drawn from the Gaussian distributions in $z$ with 
the scale heights of $h_\mathrm{II}=0.09\kpc$ for Type~II and
$h_\mathrm{I}=0.325\kpc$ for Type~I SNe. Thus, Eq.~(\ref{eq:mass}) contains
the mass source of $4\Msol$ per SN,
\[
\dot\rho\SN\simeq4\Msol\left(\frac{\nu_\mathrm{II}}{2 h_\mathrm{II}}
        +\frac{\nu_\mathrm{I}}{2 h_\mathrm{I}}\right), 
\]
whereas Eqs.~(\ref{eq:mom}) and (\ref{eq:ent}) include kinetic and thermal
energy sources of similar strength adding up to $E\SN$ per SN:
\[
{\dot\sigma}\SN \simeq \tfrac12 E\SN\left(\frac{\nu_\mathrm{II}}{2 h_\mathrm{II}}
        +\frac{\nu_\mathrm{I}}{2 h_\mathrm{I}}\right).
\]
The only other constraints applied when choosing SN sites are to
reject a site if an SN explosion would result in a local temperature
above $10^{10}\K$ or if the local gas number density exceeds
$2\cmcube$. The latter requirement ensures that the thermal energy
injected is not lost to radiative cooling before it can be converted
into kinetic energy in the ambient gas. More elaborate prescriptions
can be suggested to select SN sites
\citep{Korpi99a,Avillez00,Joung06,Gressel08}; we found this
unnecessary for our present purposes.

Arguably the most important feature of SN activity, in the present
context, is the efficiency of evolution of the SN energy from thermal to
kinetic energy in the ISM, a transfer that occurs via the
shocked, dense shells of SN remnants. Given the relatively low resolution of
this model (and most, if not all, other models of this kind), it is essential
to verify that the dynamics of expanding SN shells is captured 
correctly{:
inaccuracies in the SN remnant evolution would indicate that
our modelling of the thermal and kinetic energy processes was unreliable.}
Therefore, we present in Appendix~\ref{EISNR} detailed numerical simulations
of the dynamical evolution of an individual SN remnant at spatial grid resolutions
in the range $\Delta=1$--$4\p$. We allow the SN remnant to evolve from the Sedov--Taylor stage
(at which SN remnants are introduced in our simulations) for  $t\approx3.5\Myr$.
The remnant enters the snowplough regime,
with a final shell radius exceeding $100\p$, and we
compare the numerical results with the analytical solution of \citet{Cioffi98}. The
accuracy of the numerical results depends on the ambient gas density $n_0$:
larger $n_0$ requires higher resolution to reproduce the
analytical results. We show that agreement with
\citet{Cioffi98} in terms of the shell radius and expansion speed is {
  excellent} at resolutions $\Delta\leq2\p$ for $n_0\simeq1\cmcube$,
  and also {{very good}} at $\Delta=4\p$ for $n_0\approx0.1$ and
$0.01\cmcube$.
{
Comparisons with models of higher resolution 
\citep{AB04,Joung09}, in Section~\ref{COMP}, 
also indicate that our basic $\Delta=4\p$ resolution is adequate.}

Since shock waves in the immediate vicinity of an SN site are usually
stronger than anywhere else in the ISM, these tests also confirm that our
handling of shock fronts is sufficiently accurate and that the shock-capturing
diffusivities that we employ do not unreasonably affect the shock evolution.

Our standard resolution is $\Delta=4\p$. To be minimally resolved, the initial radius
of an SN remnant must span at least two grid points. Because the origin is set
between grid points, a minimum radius of 7\,pc for
the energy injection site is sufficient.
 The size of the energy
injection region in our model must be such that the gas temperature is above
$10^6\K$ and below $10^8\K$: at both higher and lower temperatures, energy
losses to radiation are excessive and adiabatic expansion cannot be
established. Following \citet{Joung06}, we adjust the radius of the energy
injection region to be such that it contains $60\Msol$ of gas. 
For example, in model~{\Op} this results in a mean $r\SN$ of $35\p$, 
with a standard deviation of $25\p$ and a maximum of $200\p$.
The distribution of radii appears approximately lognormal,
so $r\SN>75\p$ is very infrequent and the modal value is about $10\p$; this corresponds 
to the middle of the Sedov--Taylor phase of the SN expansion.  
Unlike \citet{Joung06}, we found that mass redistribution within the injection site
was not necessary. Therefore we do not impose uniform site density, particularly as
it may lead to unexpected consequences in the presence of magnetic fields 
in our MHD simulations (described elsewhere).

%------------------------------------------------------------------------
\subsection{Radiative cooling and photoelectric heating}
\label{sect:ssvrb}

%-----------------------------------------------------------------------------
\begin{table}
\centering
\caption{The cooling function of \citet{Wolfire95} at $T<10^5\K$, 
joined to that of \citet{Sarazin87} at higher temperatures, 
with $\Lambda=0$ for $T<10\K$.
This cooling function is denoted WSW in the text 
(and in the labels of our numerical models).}  \label{table:coolSS}
\begin{tabular}{ccc}
\hline
$T_k$ [K]&$\Lambda_k\ [\!\erg\g^{-2}\s^{-1}\cm^3\K^{-\beta_{k}}]$ &$\beta_k$\\
\hline
10                    & $3.70\times10^{16}$ &   \phantom{$-$}2.12 \\
141                   & $9.46\times10^{18}$ &   \phantom{$-$}1.00 \\
313                   & $1.18\times10^{20}$ &   \phantom{$-$}0.56\\
6102                  & $1.10\times10^{10}$ &   \phantom{$-$}3.21\\
$10^{5}$              & $1.24\times10^{27}$ &              $-0.20$\\
$2.88\times 10^{5}$   & $2.39\times10^{42}$ &              $-3.00$\\
$ 4.73\times 10^{5}$  & $4.00\times10^{26}$ &              $-0.22$\\
$2.11\times 10^{6}$   & $1.53\times10^{44}$ &              $-3.00$\\
$3.98\times 10^{6}$   & $1.61\times10^{22}$ &   \phantom{$-$}0.33\\
$ 2.00\times 10^{7}$  & $9.23\times10^{20}$ &   \phantom{$-$}0.50\\
\hline
\end{tabular}
\end{table}
%-----------------------------------------------------------------------------

We consider two different parameterizations of the optically thin radiative
cooling appearing in Eq.~(\ref{eq:ent}), both of the piecewise power-law form
$\Lambda=\Lambda_{k}T^{\beta_{k}}$ within a number of temperature ranges 
$T_{k}\le T<T_{k+1}$, with $T_k$ and $\Lambda_k$ given in
Tables~\ref{table:coolSS} and \ref{table:coolRB}. Since this is just a crude
(but convenient) parameterization of numerous processes of recombination and
ionisation of various species in the ISM, there are several approximations
designed to describe the variety of physical conditions in the ISM. Each of the
earlier models of the SN-driven ISM adopts a specific cooling curve, often
without explaining the reason for the particular choice or assessing its
consequences. In this paper, we discuss the sensitivity of the results to the
choice of the cooling function.

One parameterization of radiative cooling,
labelled WSW and shown in Table~\ref{table:coolSS}, consists of two
parts. For $T<10^5\K$, we use the cooling function fitted by \citet{Sanchez02}
to the `standard' equilibrium pressure--density relation of \citet[][cf. Fig.~3b therein]{Wolfire95}. For higher temperatures, we adopt the cooling
function of \citet{Sarazin87}. This part of the cooling function (but
extended differently to lower temperatures) was used by \citet{Slyz05} to
study star formation in the ISM. The WSW cooling function was also used by
\cite{Gressel08}. It has two thermally unstable ranges: at $313\leq T<6102\K$, the
gas is isobarically unstable ($\beta_k<1$); at $T>10^5\K$, gas is
isochorically or isentropically unstable ($\beta_k<0$ and $\beta_k<-1.5$,
respectively).

%-----------------------------------------------------------------------------
\begin{table}
\centering
\caption{The cooling function of \citet{Rosen93}, labelled RBN in the text
(and in the labels of our numerical models),
with $\Lambda=0$ for $T<10\K$.}
\label{table:coolRB}
\begin{tabular}{ccc}
\hline
$T_k$ [K]   &$\Lambda_k\ [\!\erg\g^{-2}\s^{-1}\cm^3\K^{-\beta_k}]$ &$\beta_k$\\
\hline
  10        &$9.88\times10^5\phantom{^5}$             &\phantom{$-$}6.000 \\
  300       &$8.36\times10^{15}$                      &\phantom{$-$}2.000 \\
  2000      &$3.80\times10^{17}$                      &\phantom{$-$}1.500 \\
  8000      &$1.76\times10^{12}$                      &\phantom{$-$}2.867 \\
  $10^{5}$  &$6.76\times10^{29}$                      &           $-0.650$ \\
  $10^{6}$  &$8.51\times10^{22}$                      &\phantom{$-$}0.500 \\
\hline
\end{tabular}
\end{table}
%-----------------------------------------------------------------------------

%-----------------------------------------------------------------------------
\begin{figure}
     \begin{center}
        \includegraphics[width=0.9\columnwidth]{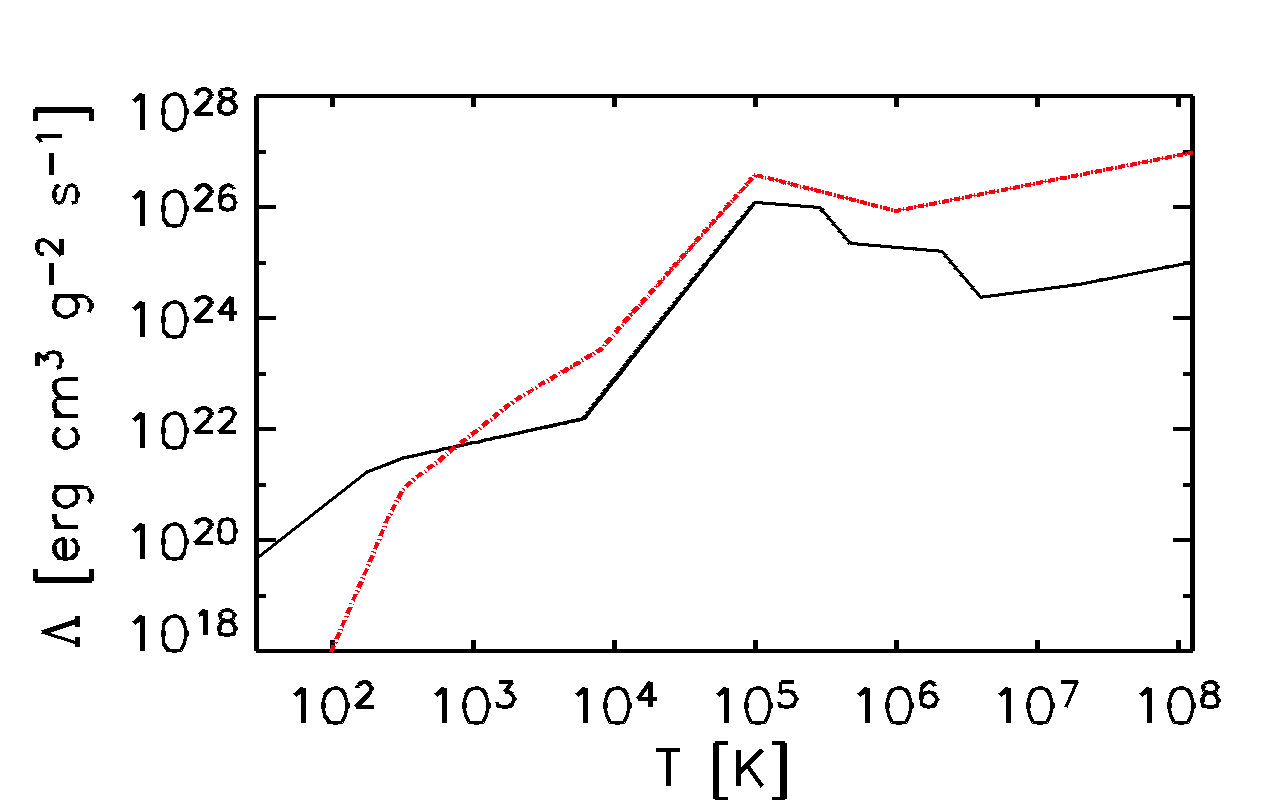}
\caption{The cooling functions WSW (solid, black) and RBN (red, dash-dotted), with
parameters given in Tables~\ref{table:coolSS} and \ref{table:coolRB},
respectively.\label{fig:cool} }
       \end{center}
  \end{figure}
%-----------------------------------------------------------------------------

Results obtained with the WSW cooling function are compared with those
using the cooling function of \citet{Rosen93}, labelled RBN,
whose parameters are shown in
Table~\ref{table:coolRB}. This cooling function has a thermally unstable part
only above $10^5\K$. \citet{Rosen93} truncated their cooling function at
$T=300\K$. Instead of abrupt truncation, we have smoothly extended the
cooling function down to $10\K$. This has no palpable physical consequences
as the radiative cooling time at these low temperatures 
becomes longer {$(10\Myr)$} than other time scales in the model, 
so that adiabatic cooling dominates.
The minimum temperature reported in
the model of \citet{Rosen93} is about $100\K$. Here, with better spatial
resolution, the lowest temperature is {typically below} $50\K$.

We took special care to accurately ensure the continuity of the cooling
functions, as small discontinuities may affect the performance of the
code; hence the values of $\Lambda_k$ in Table~\ref{table:coolSS} differ
slightly from those given by \citet{Sanchez02}. The two cooling functions are
shown in Fig.~\ref{fig:cool}. The cooling function used in each numerical
model is identified with a prefix RBN or WSW in the model label
(see Table~\ref{table:models}). The purpose of
Models {RBN} and {\WSWa} is to assess the impact of the choice of the
cooling function on the results (Section~\ref{COOL}). Other models employ the
WSW cooling function.

We also include photoelectric heating in Eq.~(\ref{eq:ent}) 
via the stellar far-ultraviolet (UV) radiation, $\Gamma$, 
following \citet{Wolfire95} and allowing for its decline
away from the Galactic mid-plane with a length scale comparable to the
scale height of the stellar disc near the Sun \cite[cf.][]{Joung06}:
\[
  \label{eq:zheat}
  \Gamma(z)=\Gamma_0\exp\left(-|z|/300\p\right),
\qquad
\Gamma_0=0.0147\erg\g^{-1}\s^{-1}.
\]
This heating mechanism is smoothly suppressed at $T>2\times10^4\K$, since the
photoelectric effect due to UV photon impact on PAHs (Polycyclic Aromatic
Hydrocarbons) and small dust grains is impeded at high temperatures
\cite[cf.][]{Wolfire95}.
%------------------------------------------------------------------------
\subsection{Numerical methods}\label{NS}
%------------------------------------------------------------------------

%------------------------------------------------------------------------
\subsubsection{{The computational domain}\label{CD}}
%------------------------------------------------------------------------
We model a relatively small region within the galactic disc and lower halo
with parameters typical of the solar neighbourhood. Using a three-dimensional
Cartesian grid, our results have been obtained for a region
$1.024\times1.024\times 2.24\kpc^{3}$ in size, with $1.024\kpc$ in the radial
and azimuthal directions and $1.12 \kpc$ vertically on either side of the
galactic mid-plane. Assuming that the correlation length of the interstellar
turbulence is $l_0\simeq0.1\kpc$ (see Section~\ref{CORR}),
the computational domain encompasses about
2,000 turbulent cells, so the statistical properties of the ISM can be reliably captured.
We are confident that our computational domain is sufficiently broad
to accommodate comfortably even the largest SN remnants at large heights, so as to
exclude any self-interaction of expanding remnants through the periodic boundaries.

Vertically, our reference model accommodates ten scale heights of the cold
\HI\ gas, two scale heights of diffuse \HI\ (the Lockman layer), and one scale
height of ionised hydrogen (the Reynolds layer). The vertical size of the domain
in the reference model is insufficient to include the scale height of the hot
gas,
{{and it would be preferable to 
{consider a computational box of a larger vertical size, $2L_z$}.
Indeed, some similar ISM models use a vertically elongated computational box 
with the horizontal 
size of $1\kpc\times1\kpc$ but the top and bottom boundaries at {$L_z=10\kpc$} 
\citep[e.g.,][and references therein]{AB07}. 
However, the horizontal
size of the domain $L_\perp$ in a taller box {\replyb{may need to}} be 
increased to keep
{its} aspect ratio
of order unity,
so as to avoid 
{introducing other} unphysical behaviour at $|z|\gtrsim 
L_\perp$.

This constraint arises mainly from the periodic (or sliding periodic) boundary conditions
in the horizontal planes as they preclude divergent flows at scales comparable to $L_\perp$.
However, the scale of the gas flow unavoidably increases with $|z|$ because of the density 
stratification.
{The steady-state continuity 
equation for a gas stratified in $z$, $\nabla\cdot\vect{u}=-u_z \upartial\ln\rho/\upartial z$,
leads to the following estimate of the horizontal perturbation velocity arising due
to the stratification:}
\begin{equation}\label{cont_perp}
u_\perp\simeq u_z\frac{l_\perp}{H},
\end{equation}
{where $H$ is the density scale height, $\upartial\ln\rho/\upartial z \simeq -H^{-1}$, and 
$l_\perp$ is  the horizontal scale of the flow, introduced via 
$|\upartial u_x/\upartial x|,|\upartial u_y/\upartial y|\simeq u_\perp/l_\perp$.
Here we have neglected the vertical variation of $u_z$, so that 
$\nabla\cdot\vect{u}\simeq \upartial u_x/\upartial x + \upartial u_y/\upartial y$:
this is justified for the hot and warm gas, since their
vertical velocities vary weakly with $z$ at $|z|\gtrsim0.3\kpc$ (see Fig.~\ref{fig:uzm}). 
Assuming for the
sake of simplicity that $u_\perp$ is a constant, 
in Eq.~\eqref{cont_perp}, where $l_{\perp0}$ is the horizontal correlation length
of $u_\perp$ at $z=0$, we obtain the following estimate of the horizontal correlation
length at $|z|=L_z$, the top of the domain:}
\[
{l_\perp}\left|_{|z|=L_z}\right.\simeq l_{0}+u_\perp t \simeq l_{0}(1+L_z/H),
\] 
{where the time available for the expansion is taken as $t=L_z/u_z$, 
  $l_{0}$ is the horizontal correlation length
 of $u_\perp$ at $z=0$. 
We find $l_{0}\simeq0.1\kpc$ (Table~\ref{table:CORR:l}) and
$H\simeq0.5\kpc$ (Fig.~\ref{fig:zrho_rho}), so that the correlation scale of
the velocity perturbation
at the top and bottom boundaries of our domain, $L_z\approx1\kpc$, follows as}
\[
l_\perp|_{|z|=L_z}\simeq 3l_{0}\simeq0.3\kpc.
\]
{Indeed,} we find the correlation scale of the random flow
increases to {$200$--$300\p$} at $z=0.8\kpc$ (Table~\ref{table:CORR:l}), 
so that the diameter of the correlation
cell, 400--$600\p$ becomes comparable to the horizontal size of the domain, 
{$L_\perp=1\kpc$}.
At larger heights, the periodic boundary conditions would suppress the horizontal flows,
so that that the continuity equation could only be satisfied via an unphysical increase in the 
vertical velocity with $|z|$. In addition, the size of SN remnants also increases with $|z|$
as the ambient pressure decreases. Thus, the gas 
{velocity field} (and other results) obtained in a model with 
periodic boundary conditions in $x$ and $y$ becomes unreliable at heights significantly exceeding 
the horizontal size of the computational domain.

{\replyb{{The 
lack of a feedback of the halo on the gas dynamics in the disc can,
potentially, 
affect our results. However, we believe that this is not a serious problem
and, anyway, it {\replyb{would not necessarily be}} resolved by using a taller box of a horizontal
size of only 1--$2\kpc$. 
The gas flow from the halo is expected to be in the form of relatively
cool, dense clouds, formed at large heights via thermal instability or accreted
from the intergalactic space} 
\citep[e.g.][]{WW97,PPJ12}.
{A strong direct (as opposed to a long-term) effect of this gas on the 
multi-phase gas structure in the disc
is questionable, as it provides just a fraction of the disc's star 
formation rate,
0.1--$0.2M_\odot\yr^{-1}$ versus 0.5--$5M_\odot\yr^{-1}$} 
\citep{PPJ12}.  
{Anyway, a taller computational domain would not help to include the accreted 
intergalactic gas in simulations of this type. 
In a galactic fountain, gas returns to the disc at a galactocentric distance at
 least $3\kpc$ away from where is starts}
 \citep{Bregman80},
{and this could not be accounted for in models with tall computational 
boxes that are only 
1--$2\kpc$ big horizontally.}}}

In light of these concerns, and since it is not yet possible to expand our domain significantly 
in all three dimensions, we prefer to restrict ourselves to a 
box of height $L_z\approx1\kpc$, thus retaining an aspect ratio of order unity.
This choice of a short box requires great care in the choice of vertical
boundary conditions (which might also introduce unphysical behaviour).
We discuss our boundary conditions in detail in appendix~\ref{BCND},
but briefly note here that we use modified open boundary conditions on the
velocity at $z=\pm L_{z}$.
These conditions allow for both inflow and outflow, and so are to some
extent capable of simulating gas 
{\replyb{exchange between the disc and the halo, driven by processes within the disc.
{More specifically, matter and energy are free to flow out of and into the computational domain
across the top and bottom surfaces if the internal dynamics so require.
(An inflow occurs when pressure beneath the surface is lower than at the surface
 or in the ghost zones).}
}}

%------------------------------------------------------------------------
\subsubsection{{Numerical resolution}}\label{NR}
%------------------------------------------------------------------------

For our standard resolution (numerical grid spacing) $\Delta x = \Delta y =
\Delta z=\Delta= 4\p$, we use a grid of $256\times256\times560$ 
(excluding `ghost' boundary zones). 
We apply a sixth-order finite difference scheme for spatial vector
operations and a third-order Runge--Kutta scheme for time stepping. We also
investigate one model at doubled resolution, $\Delta=2\p$, labelled {\OpH} in
Table~\ref{table:models};
the starting state for this model is obtained by remapping a snapshot 
from the standard-resolution Model {\Op} at $t=600$\,Myr 
(when the system has settled to a statistical steady state) 
onto a grid $512\times512\times1120$ in size.

Given the statistically homogeneous structure of the ISM in the horizontal
directions at the scales of interest (neglecting arm-interarm variations), we
apply periodic boundary conditions in the azimuthal ($y$) direction.
Differential rotation is modelled using the shearing-sheet approximation with
sliding periodic boundary conditions \citep{WT88} in $x$, the local analogue
of cylindrical radius. We apply slightly modified open vertical boundary
conditions, described in some detail in Appendix~\ref{BCND}, to allow for 
the free movement of gas to the halo without preventing inward flows at
the upper and lower boundaries.
{{In the calculations reported here, outflow exceeds inflow on average, 
and there is a net loss of mass from our domain, of order 15\% of the total 
mass per \Gyr.  
We do not believe that this slow loss of mass significantly
affects our 
results}}
%------------------------------------------------------------------------
\subsubsection{{Transport coefficients}}\label{TTC}
%------------------------------------------------------------------------

The spatial and temporal resolutions attainable impose lower limits on the
kinematic viscosity $\nu$ and thermal diffusivity $\chi$, which are,
unavoidably, much higher than any realistic values. These limits result from
the Courant--Friedrichs--Lewy (CFL) condition which requires that the
numerical time step must be shorter than the crossing time over the mesh
length $\Delta$ for each of the transport processes involved. It is desirable 
to avoid unnecessarily high viscosity and thermal diffusivity. The cold and 
warm phases have relatively small perturbation gas speeds (of order $10\kms$), so
we prescribe $\nu$ and $\chi$ to be proportional to the local speed of sound,
$\nu=\nu_1 c_s/c_1$ and $\chi=\chi_1 c_s/c_1$.  
 We ensure {that} the {{maximum}} Reynolds and P\'eclet numbers based on
the mesh separation $\Delta$ are always close to unity throughout the
computational domain (see Appendix~\ref{BCND}):  
$\nu_1\approx4.2\times10^{-3}\kms\kpc$, $\chi_1\approx4.1\times10^{-4}\kms\kpc$ and $c_1=1\kms$.
This gives, for example, $\chi=0.019\kms\kpc$ at $T=10^5\K$ and $0.6\kms\kpc$
at $T=10^8\K$. 
Thus, transport coefficients are larger in the hot gas where typical
temperature and perturbation velocity are of order $10^6\K$ and $100\kms$,
respectively.
In all models $\chi\simeq0.1\nu$, 
{{i.e., the Prandtl number $\Pr\simeq10$.}} 
{{The corresponding fluid Reynolds and P\'eclet numbers, based on the correlation scale of the flow, fall in the range 20--40 in the models presented here.}}

Numerical handling of the strong shocks widespread in the ISM needs special care.
To ensure that they are always resolved, we include shock-capturing diffusion
of heat and momentum, with the diffusivities $\zeta_{\chi}$ and $\zeta_{\nu}$,
respectively, defined as
\begin{equation}
\zeta_\chi=
\begin{cases}
c_{\chi}\Delta x^2\max_5|\nabla\cdot\vect{u}| &\text{if }\nabla\cdot\vect{u}<0,\\
0                                                          &\text{otherwise}
\end{cases}\label{shockdiff}
\end{equation}
(and similarly for $\zeta_{\nu}$, but with a coefficient $c_\nu$),
where ${\rm max_5}$ denotes the maximum value occurring at any of the five nearest mesh points (in each coordinate).
Thus, the shock-capturing diffusivities are proportional
to the maximum divergence of the velocity in the local neighbourhood, and
are confined to the regions of convergent flow. Here, $c_{\chi}=c_\nu$ is a
dimensionless coefficient which we have adjusted empirically to 10. This
prescription spreads a shock front over sufficiently many (usually, four) grid
points. Detailed test simulations of an isolated expanding SN remnant in
Appendix~\ref{EISNR} confirm that this prescription produces quite accurate
results, particularly those which are relevant to our goals: most importantly,
the conversion of thermal to kinetic energy in SN remnants.

%-----------------------------------------------------------------------------
\begin{table*} 
\caption{
Selected parameters of the numerical models explored in this paper, named in Column (1).
 {Columns (2)--(3) give input {parameters}}:
numerical resolution {$\Delta$ and} initial mid-plane gas number 
density $n_0$.
 {{The remaining columns} give output {parameters}:}
(4) time span over which the models have been in steady state 
 {({in the units of $\tau=L_x/u_{0,\mathrm{rms}}$,} the typical horizontal
crossing time 
{based on the root-mean-square (r.m.s.) random speed $u_0$ given in Column (9) and 
$L_x\approx1\kpc$})};
(5) average kinematic viscosity $\langle\nu\rangle$;
(6) average sound speed $\langle c\sound\rangle$; 
(7)--(8) average Reynolds numbers defined at the grid spacing, $\Delta$,
and based on the correlation scale of the random flow, $l_0\simeq100\p$;
(9)--(10) r.m.s.\ perturbation velocity  {$u_\mathrm{rms}$}, and
{r.m.s.\ random} velocity 
 {$u_{0,\mathrm{rms}}$};
(11) thermal energy density $e_\mathrm{th}$;
(12) kinetic energy density $e_\mathrm{kin}$; {and}
 {(13) volume fractions $f_V$ of {cold (C), warm (W) and hot (H)} gas at 
$|z|\le200\p$.
}
\label{table:results}\label{table:models}}
\begin{tabular}[htb]{@{}lccccccccccccccc@{}}
(1)      &(2)      &(3)                &(4)                      &(5)                       &(6)        &(7)                            &(8)                            &(9)         &(10)         &(11) &(12) &(13)\\
\hline
Model    &$\Delta$ &$n_0$         &$\Delta t$   &$\average{\nu}$&$\average{c\sound}$ 
	&$\average{\Rey_\mesh}$  &$\average{\Rey}$ &${u\rrms}$               &{$u_{0,\mathrm{rms}}$}       
		&$e_{\rm th}$&$e_{\rm kin}$ &$f_V$, {C:W:H}\\
&[pc]    &$\left[\displaystyle\frac{1}{\rm cm^3}\right]$ &[$\tau$]  
  &$\displaystyle\left[\rm\frac{km\,kpc}{s}\right]$   &$[\!\kms]$ & & &$[\!\kms]$  
  		&$[\!\kms]$ &$\displaystyle\left[\frac{E\SN}{\rm kpc^{3}}\right]$   
  				 &$\displaystyle\left[\frac{E\SN}{\rm kpc^{3}}\right]$ &[\%] \\
\hline                                          
\Op      &4        &1.8           &3.9       &$0.44$             &$108$               &$0.88$	&$22$
             	&$\phantom{1}76$			&$\phantom{1}26$		&$30$    		&$13$ 		&2~:~60~:~38 \\
\OpH     &2        &1.8           &0.5       &$0.77$             &$186$               &$0.85$	&{$43$} 
							&$103$          			&$\phm34$						&$19$    		&$10$  		&3~:~51~:~46 \\
RBN      &4        &2.1           &2.7       &$0.24$             &$\phantom{2}58$     &$1.18$	&$30$
              &$\phantom{1}37$			&$\phantom{1}18$		&$25$    		&$\phm9$ 	&3~:~82~:~15  \\
\WSWa    &4        &2.1           &4.0        &$0.27$            &$\phantom{2}65$     &$0.97$	&$24$
              &$\phantom{1}45$			&$\phantom{1}20$		&$29$    		&$13$   	&3~:~70~:~27 \\
\hline
\end{tabular}
 \end{table*}
%-----------------------------------------------------------------------------

With a cooling function susceptible to thermal instability, thermal
diffusivity $\chi$ has to be large enough as to allow us to resolve its most
unstable normal modes:
\[
\chi\geq\frac{1-\beta }{\gamma~\tau\cool}\left( \frac{\Delta}{2 \upi}\right)^2,
\]
where $\beta$ is the cooling function exponent in the thermally unstable range,
$\tau\cool$ is the radiative cooling time and $\gamma=5/3$ is the adiabatic index.     
{{{Figure~\ref{fig:pdf2d} makes} it evident that,
{in} our models,
$\tau\cool$ typically {exceeds} 1\,Myr in the thermally 
unstable regime.}}
Further details can be found in Appendix~\ref{TI} where we demonstrate that, 
with the parameters chosen in our models, thermal instability is well resolved
by the numerical grid.

The shock-capturing diffusion broadens the
shocks and increases the spatial 
spread of density around them. An undesirable effect of
this is that the gas inside SN remnants cools faster than it should, thus
reducing the maximum temperature and affecting the abundance of the hot phase.
Having considered various approaches while modelling individual SN remnants in
Appendix~\ref{EISNR}, we adopt a prescription which is
numerically stable, reduces gas cooling within SN remnants, and confines extreme
cooling to the shock fronts. Specifically, we multiply the term
$(\Gamma-\rho\Lambda)T^{-1}$ in Eq.~(\ref{eq:ent}) by
\begin{equation}\label{coolxi} 
\xi=\exp(-C|\nabla\zeta_\chi|^2),
\end{equation} 
where $\zeta_\chi$ is the shock diffusivity
defined in Eq.~(\ref{shockdiff}). Thus, $\xi\approx1$
almost anywhere in the domain
but reduces towards zero in strong shocks, where $|\nabla\zeta_\chi|^2$
is large. The value of the additional empirical parameter,
$C\approx0.01$, was chosen to ensure numerical stability with minimum change 
to the basic physics.  We have verified that, acting
together with other artificial diffusion terms, this does not prevent
accurate modelling of individual SN remnants (see Appendix~\ref{EISNR}).

%-----------------------------------------------------------------------------
\subsubsection{{Initial conditions}}\label{ICons}
%-----------------------------------------------------------------------------

We adopt an initial density distribution corresponding to
\textit{isothermal\/} hydrostatic equilibrium in the gravity field of
Eq.~(\ref{eq:grav}):
\begin{equation}
\label{eq:initrho}
\rho(z)= \rho_0 \exp\left[a_1 \left(z_1- \sqrt{z_1^2+z^2}-
    \frac{a_2}{2a_1}\,\frac{z^2}{z_2}\right)\right].
\end{equation}
Since our present model does not contain magnetic fields or cosmic rays, which
provide roughly half of the total pressure in the ISM (the remainder coming 
from thermal and turbulent pressures), we expect the gas scale height to be
smaller than that observed. Given the limited spatial resolution of our
simulations, the correspondingly weakened thermal instability and
neglected self-gravity, it is not quite clear in advance whether the
gas density used in our model should include molecular hydrogen
or, alternatively, include only diffuse gas.

We used $\rho_0=3.5\times10^{-24}\g\cmcube$ for models~RBN and {\WSWa},
corresponding to gas number density, $n_0=2.1\cmcube$ at the mid-plane.
This is the total interstellar gas density,
including the part confined to molecular clouds.
These models, discussed in Section~\ref{MOL}, exhibit
unrealistically strong cooling. Therefore, the subsequent models \Op\ and \OpH\
have a smaller amount of matter in the computational domain
(a 17\% reduction), with
$\rho_0=3.0\times10^{-24}\g\cmcube$, or $n_0=1.8\cmcube$, accounting only
for the atomic gas \citep[see also][]{Joung06}.

As soon as the simulation starts, 
 density-dependent heating and cooling
 {{affect the gas temperature, so it is no longer isothermal and}}
$\rho(z)$ given in
Eq.~(\ref{eq:initrho}) is not a hydrostatic distribution.
To avoid unnecessarily long initial transients, we impose
a non-uniform initial temperature distribution so as to be near static
equilibrium:
\begin{equation}
\label{eq:initT}
T(z)=\frac{T_{0}}{z_1}\left(\sqrt{z_1^2+z^2}
                +\frac{a_2}{2a_1}\frac{z^2}{z_2}\right),
\end{equation}
where $T_0$ is obtained from
\[
\Gamma(0)=\rho_{0}\Lambda(T_{0})\approx 0.0147\erg\g^{-1}\s^{-1}.
\]
The value of $T_0$ therefore depends on $\rho_0$ and the choice of the cooling
function.

%-----------------------------------------------------------------------------
  \begin{figure}
  \centering
  \includegraphics[width=.82\columnwidth]{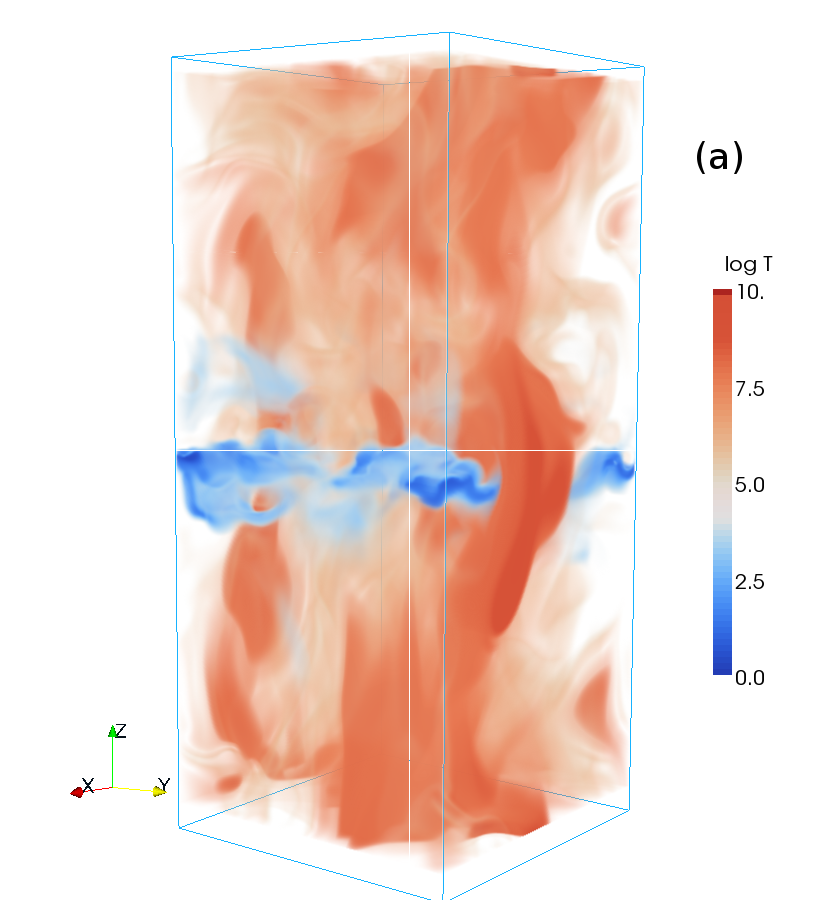}
\hfill
  \includegraphics[width=.82\columnwidth]{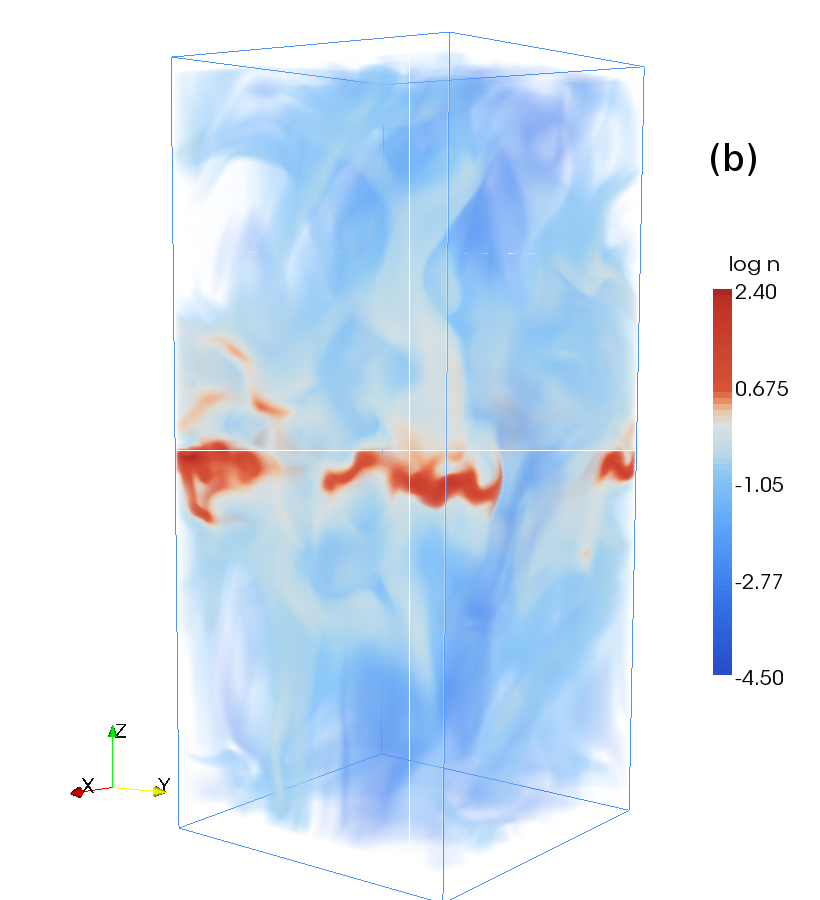}
\caption{A 3D rendering of \textbf{(a)} temperature $T$ and \textbf{(b)} density
  $n$ in Model~{\Op} at $t=551\Myr$. Cold, dense gas is
mostly restricted to near the mid-plane, whereas hot gas extends towards the
boundaries. 
To aid visualisation of 3D structure, warm gas ($10^3<T<10^6\K$) in the 
panel (a) and diffuse ($n<10^{-2}\cm^{-3}$) in panel (b) have high
transparency. Thus the extreme temperatures or dense structures are emphasised.
\label{fig:tsnap}}
  \end{figure}
%------------------------------------------------------------------------

%-----------------------------------------------------------------------------
\begin{figure}
\includegraphics[width=\columnwidth]{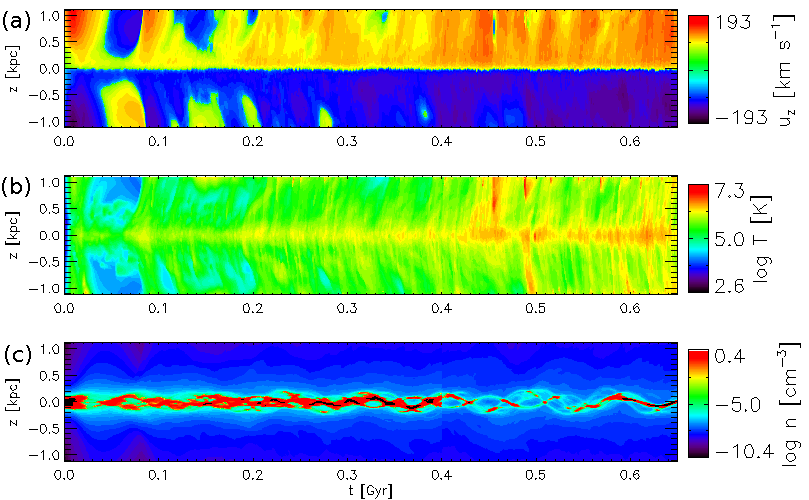}
\caption{Horizontal ($xy$) averages of \textbf{(a)} the vertical velocity, \textbf{(b)}
temperature and \textbf{(c)} gas density as functions of time for
Model~{\Op} (Model {\WSWa} up to 0.4\,Gyr).\label{fig:pav}}
\end{figure}
%---------------------------------------------------------------------------

%-----------------------------------------------------------------------------
\subsection{Models explored}\label{Models}
%-----------------------------------------------------------------------------
We considered four numerical models, with {{relevant}} 
input parameters listed in Table~\ref{table:models}, along with some output 
parameters describing the results.
The models are labelled with prefix RBN or WSW according to the cooling function
used.
Angular brackets in Table~\ref{table:models} denote averages over the whole volume, 
{taken from eleven snapshots (10 for \OpH) within the}
 statistical steady state. 
The time span, $\Delta t$, is given in Column~4, normalised by 
$\tau=L_x/{{{u_{0,\rm rms}}}}$, where 
{{$u_{0,\rm rms}$}} is the root-mean-square
random velocity and $L_x\approx1\kpc$ is the horizontal size of the
computational domain
{{(e.g., $\tau\approx38\Myr$ in Model~WSWa)}}.
As $\nu$ is set proportional to the speed of sound $c\sound$, it is variable 
and the table presents its average value 
$\left<\nu\right>=\nu_1 \left<c\sound\right>/c_1$, where $\nu_1=0.004\kms\kpc$ 
and $c_1=1\kms$ in all models.
The numerical resolution is adequate when the mesh Reynolds number,
$\Rey\mesh = u\,\Mesh/\nu$, does not exceed a certain value (typically between
1 and 10) anywhere in the domain,
{{where $\Mesh$ is the grid spacing (4\,pc for all models, except for Model~{\OpH}, where
$\Mesh=2\p$).}} 
Therefore, we ensure that $u_\mathrm{max}\,\Mesh/\nu<5$, where $u_\mathrm{max}$
is the maximum perturbation velocity at any time and any grid point.
The indicative values 
in Table~\ref{table:models} {{are averages of the mesh Reynolds number,
$\langle\Rey\mesh\rangle=\average{u\rms/c\sound}\Mesh c_1/\nu_1$,
and the Reynolds number, $\langle\Rey\rangle=\average{u\rms/c\sound}l_0 c_1/\nu_1$.}}
{{The Reynolds number based on the correlation scale of the random flow, 
$l_0\simeq100\p$, is thus 25 times larger than $\Rey\mesh$ in all models explored here except for
Model~\OpH, where the difference is a factor of 50.}}

The quantities shown in Table~\ref{table:models} have been calculated as follows. 
{{In Column 9, 
{the r.m.s.\ perturbation velocity $u\rrms$} is derived from the total perturbation
velocity field $\vect{u}$, which excludes only the overall galactic rotation
$\vect{U}$.
In Column~10, 
{the r.m.s.\ random velocity $u_{0,\rm rms}$} is obtained 
with the mean flows $\average{\vect{u}}_\ell$, defined in 
Eq.~\eqref{eq:Bxgauss}, deducted from $\vect{u}$. 
In Columns~11 and 12,
$e_\mathrm{th}=\langle \rho e \rangle$ and 
$e_\mathrm{kin}=\langle\tfrac12\rho u^2\rangle$ are the average
thermal and kinetic energy densities, respectively;
the latter includes the perturbation velocity $\vect{u}$ and both are 
normalised to the SN energy $E\SN$.
The {values of the} volume fractions of the cold, warm and hot phases (defined in Section~\ref{TMPS}) near
the mid-plane are {given} in Column~13.}}

The reference model, \Op, uses the WSW cooling function but with lower gas
density than {\WSWa}, to exclude molecular hydrogen 
(see Section~\ref{REF}).
Model \OpH, which differs from {\Op} only in its
spatial resolution, is designed 
to clarify the effects of resolution on the results. 
We also analyze two models which
differ only in the cooling function, {RBN}  and {\WSWa}, to assess the
sensitivity of the results to this choice.

%-----------------------------------------------------------------------
\section{The reference model}\label{REF}
%-----------------------------------------------------------------------

Model~{\Op} is taken as a reference model; it has rotation corresponding to a
flat rotation curve with the Solar angular velocity, and gas density reduced to
exclude that part which would have entered molecular clouds. 
Results for this model were obtained by the continuation of the Model~{\WSWa}, 
in which the mass from molecular hydrogen had been included: at
$t\approx400\Myr$, the mass of gas in the domain was changed to that of
Model~{\Op} by reducing gas density by $15\%$ at every mesh point.
The effect 
of this change of the total mass is discussed in 
Section~\ref{MOL}.

Figure~\ref{fig:tsnap} shows typical temperature and density
distributions in this model at $t=551\Myr$ (i.e., $151\Myr$ after the restart from Model~{\WSWa} with reduced density). Supernova remnants appear as
irregularly shaped regions of hot, dilute gas. A hot bubble breaking through
the cold gas layer extends from the mid-plane towards the lower boundary,
visible as a vertically stretched region in the temperature snapshot near the
$(x,z)$-face. Another, smaller one can be seen below the mid-plane near the
$(y,z)$-face. Cold, dense structures are restricted to the mid-plane and occupy
a small part of the volume. Very hot and cold regions exist in close proximity.

Horizontally averaged quantities 
{{are shown}}
in Fig.~\ref{fig:pav} {{as functions of $z$ and time}}
for Model \WSWa\ at $t<400\Myr$, and \Op~at later times, 
showing the effect of reducing the total mass of gas at the transition time.
{{Average quantities may have limited physical significance 
because the multi-phase gas
{has an extremely wide range of velocities, temperatures and densities.}
For example,
panel~(b) shows that the average temperature
near the mid-plane, $|z|\la0.35\p$, is, perhaps unexpectedly, generally higher 
than that at the larger heights.
This is due to Type~II SN remnants, which contain very hot gas with $T\ga10^8\K$ and 
are concentrated near the mid-plane; even though their total volume is small,
they significantly affect the average temperature.}}
 
{{Nevertheless these help to illustrate some global properties of the 
{multi-phase structure}.}}
Before the system settles into a quasi-stationary state at about $t=250\Myr$,
it undergoes a few large-scale transient oscillations involving quasi-periodic
vertical motions. 
{{The period of approximately 100\Myr, is consistent with the breathing
modes identified by \citet{WC01} 
  {and attributed to oscillations in the gravity field.}
  Gas falling from high altitude  
  {overshoots the midplane and thus oscillates around it.} 
  Turbulent 
  {and molecular viscosities} dampen these modes.}}
At later times, a systematic outflow develops with an average
speed of about $100\kms$; we note that the vertical velocity increases very
rapidly near the mid-plane and varies much less at larger heights.
The result of the reduction of gas density at $t\approx400\Myr$ is clearly
visible, as it leads to higher mean temperatures and a stronger and more regular
outflow, together with a less pronounced and more disturbed layer of cold gas.

%-----------------------------------------------------------------------------
\section{The multi-phase structure}\label{TMPS}
%-----------------------------------------------------------------------------

%-----------------------------------------------------------------------------
\begin{figure}
\centering
\includegraphics[width=1.0\columnwidth]{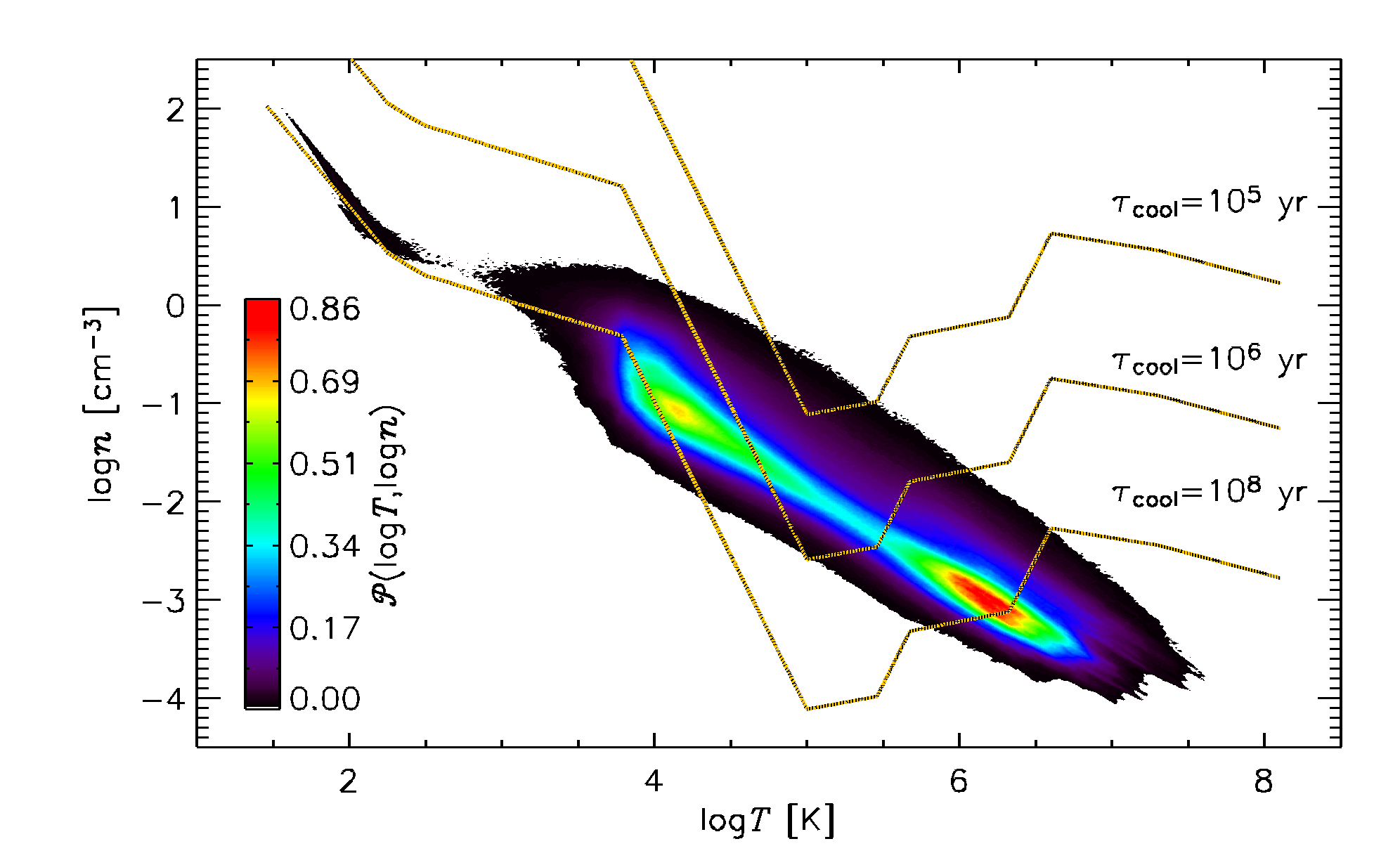}
\caption{
{{{The joint probability density of the gas number density and temperature, shown
for the whole computational domain using}
11 snapshots of Model~{\Op} 
{in a statistically steady state for $634\leq t \leq644\Myr$.}
Contours of constant cooling time ${\tau\cool=}10^5,~10^6$ and $10^8\yr$, are 
 {shown}
to 
{clarify the importance of radiative cooling in the model.}}}
\label{fig:pdf2d}
}
\end{figure}
%-----------------------------------------------------------------------------

All models discussed here have a
well-developed multi-phase structure apparently similar to that observed in
the ISM. Since the ISM phases are not genuine, thermodynamically distinct
phases \citep[e.g.][]{VS12}, their definition is tentative, with the typical temperatures of the
cold, warm and hot phases usually 
{{adopted as}} $T\simeq10^2\K$, $10^4$--$10^5\K$ and
$10^6\K$, respectively. 
{{However, the borderline temperatures
(and even the number of distinct phases) can be model-dependent,
and they are preferably determined by considering the results,
rather than \textit{a priori}.
Inspection of the probability distribution of gas number density and
temperature, displayed in Fig.~\ref{fig:pdf2d}, reveals three 
distinct concentrations at $(T[\K],n[\cmcube])=(10^2,10),~(10^4,10^{-1})$ and 
$(10^6,10^{-3})$.
Thus, we can confirm that the gas structure in this model can
be reasonably well described in terms of three distinct phases. Moreover,
we can identify the boundaries between  
them as the temperatures corresponding to the minima of the joint
probability distribution
at about $500\K$ and $5\times10^5\K$.

The curves of constant cooling time, also shown in Fig.~\ref{fig:pdf2d}, 
suggest that the distinction between the warm and hot gas is due to the
maximum of the cooling rate near $T=10^5\K$ (see also Fig.~\ref{fig:cool}),
whereas the cold, dense gas, mainly formed by compression (see below),
closely follows the curve $\tau\cool\approx10^8\yr$.
}}

%-----------------------------------------------------------------------------
  \begin{figure}
  \centering
  \includegraphics[width=0.7\columnwidth,clip=true,trim=0 0 0 9mm]{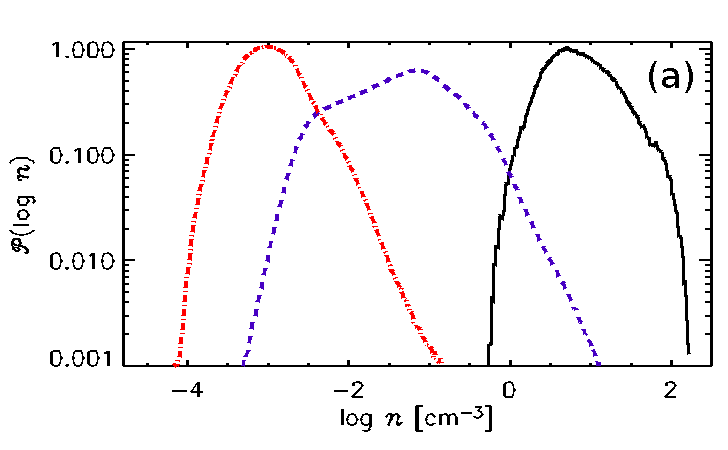}
  \includegraphics[width=0.7\columnwidth,clip=true,trim=0 0 0 9mm]{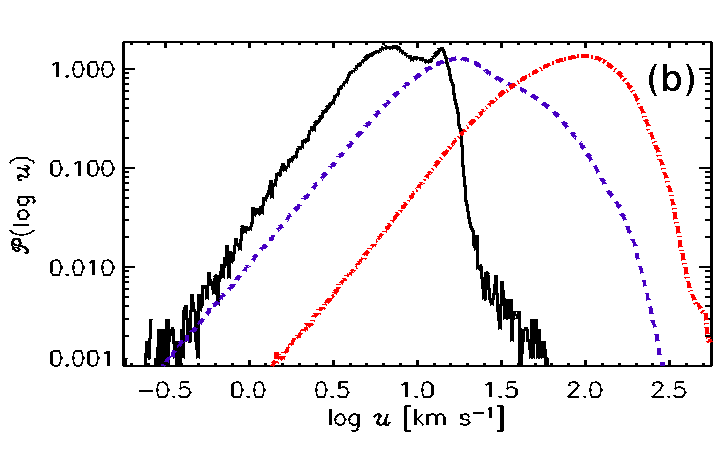}
  \includegraphics[width=0.7\columnwidth,clip=true,trim=0 0 0 9mm]{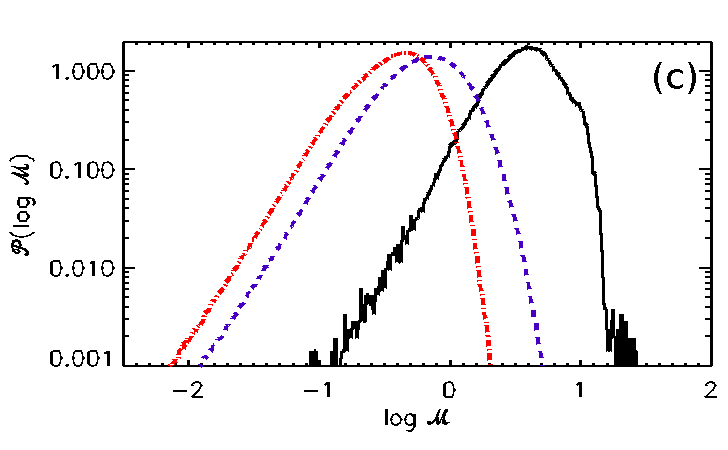}
  \includegraphics[width=0.7\columnwidth,clip=true,trim=0 0 0 9mm]{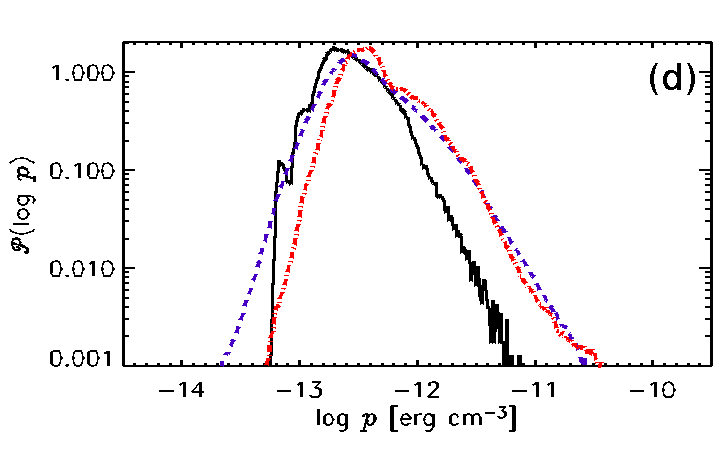}
  \includegraphics[width=0.7\columnwidth,clip=true,trim=0 0 0 9mm]{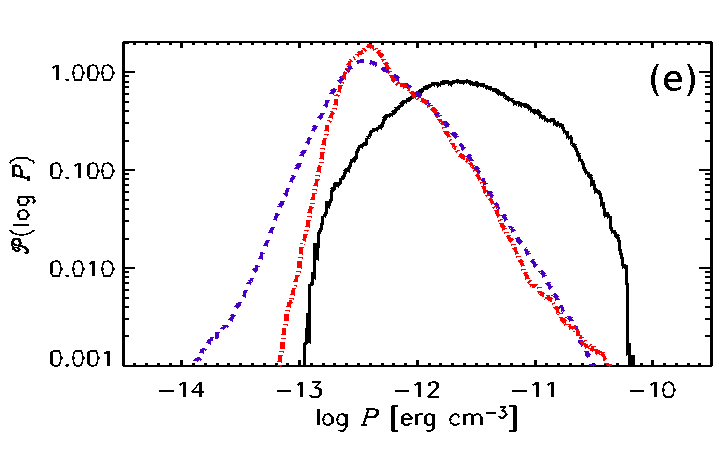}
\caption{{The} probability distribution{s} of
\textbf{(a)}~density,
\textbf{(b)}~{{random velocity, ${u}_0$}} 
\textbf{(c)}~Mach number 
{{of random motions ${u}_0$}}
(defined with respect to the local speed of sound),
\textbf{(d)}~thermal pressure, and
\textbf{(e)}~total pressure,
for {{each 
phase}} of {Model~\Op}, using 
{{11 snapshots spanning $t=634$ to $644\Myr$}}
{{and presented for each phase:}} cold $T<500\K$ (black, solid line), 
warm $500 \leq T < 5\times10^5\K$ (blue, dashed), and hot $T\geq5\times10^5\K$
 (red, dash-dotted).
\label{fig:pdf3ph}
\label{fig:apdf3ph}}
\end{figure}
%-----------------------------------------------------------------------------

In Fig.~\ref{fig:pdf3ph}, we show the probability distributions
of gas number density, random velocity, Mach number, thermal and total pressures
within each phase in
Model~{\Op}. 
{{The overlap in the gas density distributions (Fig.~\ref{fig:pdf3ph}a) 
is small {(at the probability densities of order $\mathcal{P}=0.1$)}.
The ratios of the probability densities near
 the maximum for each phase (mode) are about 100; the modal densities,
$n\approx10^{-3},\,10^{-1}$ and $10\cmcube$, thus typify the hot,
warm and cold gas respectively.}}

The velocity probability distributions in Fig~\ref{fig:pdf3ph}b reveal a clear
connection between the magnitude of the 
{{random}} 
 velocity of gas and its
temperature: the r.m.s.\ velocity in each phase scales with its speed of sound.
This is confirmed by the Mach number distributions in Fig.~\ref{fig:pdf3ph}c: 
both warm and hot phases are transonic with respect to their sound speeds. The 
cold gas is mostly supersonic,
having speeds typically under $10\kms$. 
{{The double peak in 
    the probability density for the cold gas velocity (Fig~\ref{fig:pdf3ph}b)
{(and the corresponding extension of the Mach number distribution
to $\mathcal{M}\ga1$)} is a robust 
feature, not dependent on the temperature boundary {of the cold  gas}. 
This plausibly includes ballistic gas {motion} in
{the shells of} SN remnants, as well as bulk 
{motions of cold clouds}}} at subsonic or transonic speed with respect to the 
{{ambient}} warm gas.

Probability densities of thermal pressure, shown in Fig.~\ref{fig:pdf3ph}d, are
notable for the relatively narrow spread: one order of magnitude, 
compared to a spread of six orders of magnitude in gas density. Moreover, the 
three phases have overlapping {{thermal pressure}} distributions, suggesting that the 
system is in a statistical thermal pressure balance. However, thermal pressure 
is not the only part of the total pressure in the gas, 
{{which here includes the turbulent pressure
{\replyb{{$\sfrac{1}{3}\rho|\vect{u}-\average{\vect{u}}_\ell|^2$}, where 
$\average{\vect{u}}_\ell$, defined in Eq.~\eqref{eq:Bxgauss},  is the mean
{fluctuation}
velocity.}}}} As shown in 
Fig.~\ref{fig:energetics}, total kinetic energy within the computational 
domain, associated with random flows, is about a third of the
thermal pressure. Correspondingly, the total pressure distributions in 
Fig.~\ref{fig:pdf3ph}e peaks at about $4\times10^{-13}\dyn\cm^{-2}$
{\replyb{(or $\!\erg\cm^{-3}$)}}, 
for both the warm and hot gas. The cold gas appears somewhat overpressured, with the 
modal pressure at $2\times10^{-12}\dyn\cm^{-2}$, and
with some regions under pressure as high as $10^{-11}\dyn\cm^{-2}$. 
It becomes apparent (see the discussion of Fig~\ref{fig:pall4fits}, below) that this is due to {{both compression by transonic random flows and}}
the vertical pressure gradient. All the cold gas occupies the higher 
pressure mid-plane, while the warm and hot gas distributions mainly include
lower pressure regions away from the disc. 

Cold, dense clouds are formed through radiative cooling facilitated by compression, 
which has more importance than in the other, hotter phases.
The compression is, however, truncated at the grid scale of 4\,pc, preventing
the emergence of higher densities in excess of about $10^2\cmcube$.

%-----------------------------------------------------------------------------
  \begin{figure}
  \centering
  \includegraphics[width=0.7\columnwidth,clip=true,trim=0 0 0 9mm]{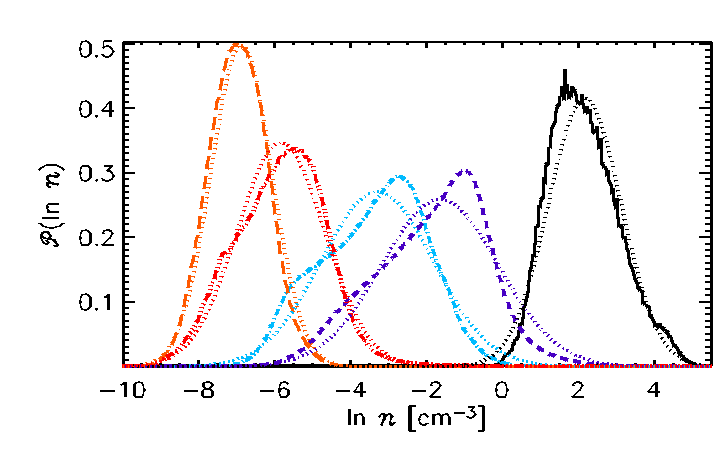} 
\caption{
  {{The probability distributions of gas desnity in model~{\Op}
for the cold (black, solid), warm and hot gas.
The warm/hot gas have been divided into regions $|z|\le200\p$ 
(purple/red, dashed/dash-triple-dotted) and $|z|>200\p$ (light blue/orange,
dash-dotted/long-dashed).
The best-fit lognormal distributions are shown dotted in matching colour.}}
\label{fig:all4fits}}
\end{figure}
%-----------------------------------------------------------------------------

%-----------------------------------------------------------------------------
\begin{table}
\caption{{\label{lognormal}
Parameters of the lognormal fits to the distribution of gas number density 
$n$ in various phases, where $\mu_{n}$ and $s_{n}$ are defined in 
Eq.~(\ref{ln}).}}
\centering
\begin{tabular}{ccc}
\hline
Phase & $\mu_n$ [$\ln\cmcube$] & $s_n$ [$\ln\cmcube$]\\
\hline
cold                  & \phantom{$-$}$2.02$ &  $0.92$ \\
warm ($|z|\leq 200\p$)& $-1.64$             &  $1.47$ \\
warm ($|z|> 200\p$)   & $-3.29$             &  $1.47$ \\
warm (total)          & $-3.03$             &  $1.47$ \\
hot ($|z|\leq 200\p$) & $-5.78$             &  $1.20$ \\
hot ($|z|> 200\p$)    & $-6.96$             &  $0.77$ \\
\hline
\end{tabular}
\end{table}
%-----------------------------------------------------------------------------

The probability distributions of gas density in Fig.~\ref{fig:pdf3ph}a can be 
reasonably approximated by the lognormal distributions, of the form
\begin{equation}\label{ln}
\mathcal{P}(n)=\Lambda(\mu_n,s_n)\equiv\frac{1}{n s_{n}\sqrt{2\upi}}
        \exp\left({\displaystyle-\frac{(\ln n-\mu_{n})^2}{2 s_{n}^2}}\right).
\end{equation}
The quality of the fits 
is illustrated in Fig.~\ref{fig:all4fits}, 
using 500 data bins in the range $10^{-4.8}<n<10^{2.5}\cm^{-3}$;
the best-fit parameters are given in Table~\ref{lognormal}.
Note that, in making these fits, we have subdivided the hot 
{{and warm}}  
gas into that near the mid-plane ($|z|\le200\p$) and that at greater heights
($|z|>200\p$);
the former is {{located in the SN active region, 
strongly shocked with 
a broad range of density and pressure fluctuations}},
whereas the latter is predominantly the more diffuse {{and homogeneous}}   
gas in the halo. 
{{As can be seen in Fig.~\ref{fig:all4fits}, the shape of the 
probability distribution of the warm gas (rather than the position of its 
maximum) does not var much with $|z|$.
Table~\ref{lognormal}, 
thus shows the parameters for the warm in the whole volume.}}
The lognormal fits satisfy the Kolmogorov-Smirnov test 
at or above the 95\% level of significance. 
{{For the hot gas fit the KS test fails for the total volume.
So only the fits for the hot gas split by height are included in 
Table~\ref{lognormal}}}.

%-----------------------------------------------------------------------------
\begin{figure}
\centering
\includegraphics[width=0.7\columnwidth,clip=true,trim=0 0 0 9mm]{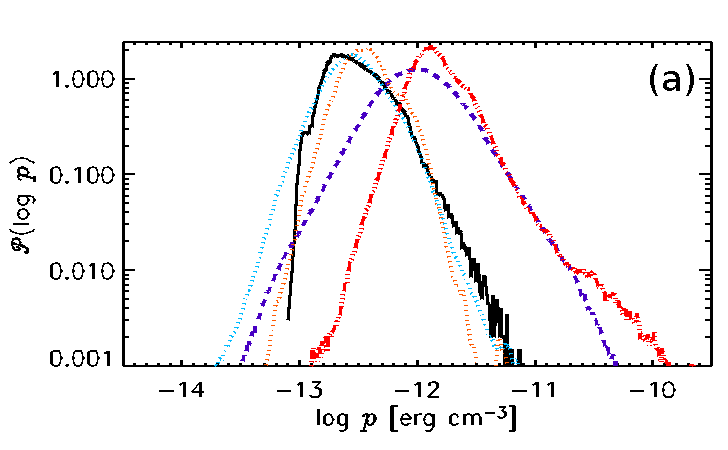}
\includegraphics[width=0.7\columnwidth,clip=true,trim=0 0 0 9mm]{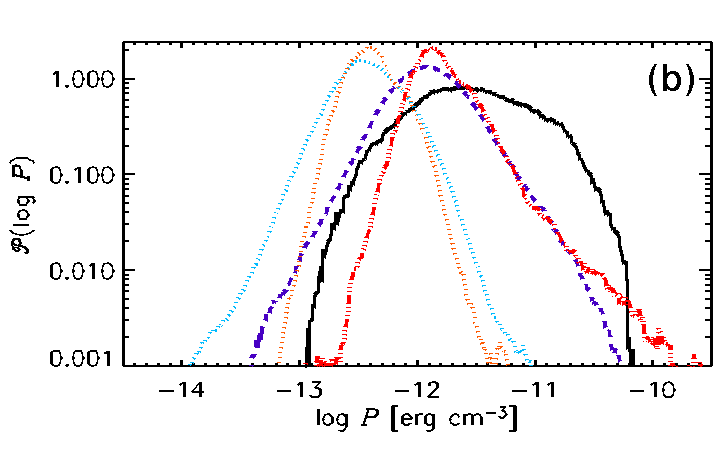}
\caption{
Probability distributions for \textbf{(a)}~thermal pressure $p$ and
\textbf{(b)}~total pressure $P$ in Model~{\Op}, 
{{for different gas phases:
cold (black, solid);
warm (blue, dashed) and
hot (red, dash-triple-dotted) at $|z|\le200\p$;
 warm (light blue, dotted) and hot (orange, dotted) at $|z|>200\p$.}}
\label{fig:pall4fits}}
\end{figure}
%-----------------------------------------------------------------------------

The probability densities of thermal and total pressure, displayed in Fig.~\ref{fig:pall4fits}, show 
that 
{{although the thermal pressure of the cold gas}} near the mid-plane
{{is lower than in the other phases}} the total pressures 
{{are 
much closer to balance.
The broad probability distribution of the cold gas density is consistent with 
multiple compressions in shocks. 
The hot and warm gas pressure distributions are also approximately lognormal.}}
{{The gas at $|z|>200\p$ (dotted lines) appears to be in both thermal and total
pressure balance.}}

In summary, we conclude that the system is close to the state of 
statistical pressure equilibrium: the total pressure has similar values and similar probability distributions in each phase. \citet{Joung09}
also conclude from their simulations that the gas is in both thermal and total
pressure balance. This could be expected, since the only significant deviation
from the statistical dynamic equilibrium of the system is the vertical outflow 
of the hot gas and entrained warm clouds (see Section~\ref{GO}).

%-------------------------------------------------------------------------
\section{The filling factor and fractional volume}\label{FF}
%----------------------------------------------------------------------

%-----------------------------------------------------------------------------
\subsection{{Filling factors: basic ideas}}\label{BI}
%-----------------------------------------------------------------------------
{{The \emph{fractional volume} of the ISM occupied by the phase $i$ is given by
\begin{equation}\label{fv}
  f_{V,i}=\frac{V_i}{V}
\end{equation}
where $V_i$ is the volume occupied by gas in the temperature range defining phase~$i$ and $V$ is the total volume. How the gas is distributed \emph{within} a particular phase is described by the \emph{phase filling factor}
\begin{equation}\label{phin}
\phi_{i}=\frac{\mean{n_i}^2}{\mean{n_i^2}} \; ,
\end{equation}
where the over bar denotes a \emph{phase average\/}, i.e., an average only taken over the volume occupied by the phase $i$. $\phi_{i}$ describes whether the gas density of a phase is homogeneous ($\phi_{i}=1$) or clumpy ($\phi_{i}<1$). Both of these quantities are clearly important parameters of the ISM, allowing one to characterise, as a function of position, both the relative distribution of the phases and their internal structure. As discussed below, the phase filling factor is also directly related to the idea of an ensemble average, an important concept in the theory of random functions and so $\phi_{i}$ provides a useful connection between turbulence theory and the astrophysics of the ISM. Both $f_{V,i}$ and $\phi_{i}$ are easy to calculate in a simulated ISM by simply counting mesh-points.}}

{{In the real ISM, however, neither $f_{V}$ nor $\phi_{i}$ can be directly measured. Instead the \emph{volume filling factor} can be derived \citep[][]{Re77, Kulkarni88, Re91},
\begin{equation}\label{ftilde}
\tphi_{i}=\frac{\average{n_i}^2}{\average{n_i^2}},
\end{equation}
for a given phase $i$, where the angular brackets denote a \emph{volume average\/}, i.e., taken over the total volume. \footnote{{{As with the density filling factors introduced here, filling factors of temperature and other variables can be defined similarly to Eqs.~(\ref{phin}) and (\ref{ftilde}), for example $\phi_{T,i}={\mean{T_i}^2}/{\mean{T_i^2}}$, etc.}}}

Most work in this area to-date has concentrated on the diffuse ionised gas (or warm ionised medium) since the emission measure of the free electrons $\mathrm{EM}\propto n_e^2$ and the dispersion measure of pulsars $\mathrm{DM}\propto n_e$, allowing $\tphi$ to be estimated along many lines-of-sight \citep[e.g.][]{Re77, Kulkarni88, Re91, BMM06, Hill08, Gaensler08}. It is useful to generalise the tools derived to interpret the properties of a \emph{single} ISM phase for the case of the \emph{multiphase} ISM, as this can help to avoid potential pitfalls when combining data from different sources with similar-sounding names (filling factor, filling fraction, fractional volume, etc.) but subtly different meanings. In particular, only under the very specific conditions explained below, do the volume filling factors $\tphi_{i}$ of the different phases of the ISM sum to unity.}} 

In terms of the volume $V_i$ occupied by phase $i$,
\begin{equation}\label{phase_averaging}
\mean{n_{i}}=\frac{1}{V_{i}} \int_{V_{i}} n_{i} \, dV ,
\end{equation}
whilst
\begin{equation}\label{volume_averaging}
\average{n_{i}}=\frac{1}{V} \int_{V} n_{i} \, dV =
\frac{1}{V} \int_{V_i} n_{i} \, dV\,,
\end{equation}
the final equality holding because $n_i=0$ outside the volume $V_{i}$ by definition.
Since the two types of averages differ
only in the volume over which they are averaged,
they are related by the fractional volume:
\begin{equation}
\average{n_i} = \frac{V_i}{V} \mean{n_i} = f_{V,i}\mean{n_i},
\end{equation}
and
\begin{equation}
\average{n_i^2} = \frac{V_i}{V} \mean{n_i^2} = f_{V,i}\mean{n_i^2}.
\end{equation}
Consequently, the \textit{volume filling factor\/} $\tphi_{n,i}$ and the 
\textit{phase filling factor\/} $\phi_{n,i}$ are similarly related:
\begin{equation}\label{FV}
\tphi_{i}=\frac{\average{n_i}^2}{\average{n_i^2}}
        =f_{V,i}\frac{\mean{n_i}^2}{\mean{n_i^2}}=f_{V,i}\phi_{i}.
\end{equation}

{{Thus the parameters of most interest, $f_{V,i}$ and $\phi_{n,i}$, characterizing the fractional volume and the degree of homogeneity of a phase respectively, are related to the observable quantity $\tphi_{n,i}$ by Eq.~(\ref{FV}). This relation is only straightforward when the ISM phase can be assumed to be homogeneous or if one has additional statistical knowledge, such as the probability density function, of the phase. In the next sub-section we use two simple examples to illustrate how the ideas developed here can be applied to the real ISM; we then use them to develop a new interpretation of existing observational data and finally discuss how the properties of our simulated ISM compare to observations. But first a brief note about different methods of averaging is necessary.}}

\subsubsection{{{Averaging methods for {\replyb{observations}} and theory}}}

An 
important feature of the 
definition of the
volume filling factor given by Eq.~(\ref{phin}),
is that the averaging involved is inconsistent
with that used in theory of random functions.
In the latter, the
calculation of volume (or time) averages is usually complicated or impossible
and, instead, ensemble averages
(i.e. averages over the relevant probability distribution functions)
are used;
the ergodicity of the random functions is relied upon to ensure
that the two averages are identical to each other
\citep[Section~3.3 in][]{MY07,TL72}.
But the volume filling factors $\tphi_{i}$ are not compatible
with such a comparison,
as they are based on averaging over the total volume,
despite the fact that each phase occupies only a fraction of it.
In contrast, the phase averaging used to derive $\phi_{i}$ \textit{is} performed
only over the volume of each phase,
and so should correspond better to results
from the theory of random functions.

%----------------------------------------------------------------------
\subsection{{Filling factors: applications}}
%----------------------------------------------------------------------

\subsubsection{{{Assumption of homogeneous phases}}}

{{The simplest way to interpret an observation of the volume filling factor $\tphi_{i}$ is to assume that each ISM phase has a constant density.}}
Consider 
Eqs.~(\ref{fv}), (\ref{phin}) and (\ref{ftilde})
for an idealised two-phase system, where each phase is homogeneous.
(These arguments can easily be generalised to an arbitrary number
of homogeneous phases.)
For example a set of discrete clouds of one phase,
of constant density and temperature,
embedded within the other phase, with different (but also constant)
density and temperature.
Let one phase have (constant) gas number density $N_1$ and occupy volume $V_1$,
and the other $N_2$ and $V_2$, respectively.
The total volume of the system is $V=V_1+V_2$.

The volume-averaged density of each phase,
as required for Eq.~(\ref{ftilde}),  is given by
\begin{equation}\label{aven}
\average{n_i}=\frac{N_i V_i}{V} = f_{V,i} N_i.
\end{equation}
where $i=1,2$.
Similarly, the volume average of the squared density is
\begin{equation}\label{avensq}
\average{n_i^2}=\frac{N_i^2V_i}{V} = f_{V,i} N_i^2.
\end{equation}
The fractional volume of each phase can then be written as
\begin{equation}\label{FVhomogen}
f_{V,i} = \frac{\average{n_i}^2}{\average{n_i^2}}
=\frac{\average{n_i}}{N_i}= \tphi_{i} ,
\end{equation}
with $f_{V,1}+f_{V,2}=1$,
and $\tphi_{1}+\tphi_{2}=1$.
The volume-averaged quantities satisfy
$\average{n}=\average{n_1}+\average{n_2}=f_{V,1}N_1+f_{V,2}N_2$ and
$\average{n^2}=\average{n_1^2}+\average{n_2^2}=f_{V,1}N_1^2+f_{V,2}N_2^2$,
with the density variance
$\sigma^2\equiv \average{n^2}-\average{n}^2 =f_{V,1} f_{V,2} (N_1-N_2)^2$.

In contrast, the phase-averaged density of each phase,
used to calculate the phase filling factor Eq.~(\ref{phin}), is simply
$\mean{n_{i}}=N_{i}$,
and the phase average of the squared density is
$\mean{n_{i}^{2}}=N_{i}^{2}$,
so that the phase filling factor is
$\phi_{i}=1$, as must be the case for a homogeneous phase. 

Thus for homogeneous phases,
the volume filling factor and the fractional volume
of each phase are identical to each other,
$\tphi_{i}=f_{V,i}$,
and both sum to unity when considering all phases;
in contrast,
the phase-averaged filling factor is unity for each phase,
$\phi_{i}=1$.
If a given phase occupies the whole volume
(i.e., we have a single-phase medium),
then all three quantities are simply unity:
$\phi_{i}=\tphi_{i}=f_{V,i}=1$.

{{Whilst an assumption of homogeneous phases may be justified for some ISM phases, perhaps in specific regions of the galactic disc, in the case of the simulated ISM discussed in this paper such an assumption would lead to significant underestimates of $f_{V,i}$ for all phases, by a factor of 2 for the cold and hot gas and by an order of magnitude for the warm gas.}}

\subsubsection{{{Assumption of lognormal phases\label{lnp}}}}

{{For the more realistic case of an inhomogeneous ISM, where each phase consists of gas with a range of densities, the interpretation of $\tphi_{i}$ requires additional knowledge about the statistical properties of a phase.}} 

{{For electrons in the diffuse ionised gas \citet{Re77} derived the correction factor $\sigma_c^2/n_c^2$, where $n_c$ is the average density of electron clouds and $\sigma^2_c$ the density variance within clouds, to allow for clumpiness in the electron distribution when calculating the fraction of the total path length occupied by the clouds. More generally, the probability distribution function of the gas in a phase allows $\phi_{i}$ to be calculated directly, as we now illustrate for the case of the lognormal PDFs identified in Section~\ref{TMPS}.}}  

For a lognormal distribution $\mathcal{P}(n_i) \sim \Lambda(\mu_{i},s_{i})$,
Eq.~(\ref{ln}),
the mean and mean-square densities are given
by the following phase (`ensemble') averages:
\begin{equation}\label{on}
  \mean{n_{i}}={\rm e}^{\mu_{i}+s_{i}^2/2},
~~
\sigma_i^2=\mean{(n_i-\mean{n_i})^2}=\mean{n_i}^2\left({\rm e}^{s_{i}^2}-1\right),
\end{equation}
where $\sigma_i^2$ is the density variance around the mean $\mean{n_i}$, so that
\begin{equation}\label{ophi}
\phi_{n,i}=\frac{\mean{n_i}^2}{\mean{n_i^2}}=
\frac{\mean{n_i}^2}{\sigma_i^2+\mean{n_i}^2}
=\exp(-s_{i}^2).
\end{equation}
So the phase filling factor 
$\phi_{n,i}=1$ \emph{only\/} for a homogeneous density
distribution, $\sigma_i=0$ (or equivalently, $s_{i}=0$).
This makes it clear that this filling factor,
defined in terms of the phase average,
is quite distinct from the fractional volume, $f_{V,i}$,
but rather quantifies the degree of homogeneity
of the gas distribution \emph{within\/} a given phase.
Both describe distinct characteristics of the multi-phase ISM,
and, if properly interpreted,
can yield rich information about the structure of the ISM.

{{In the case of the simulated ISM, using the lognormal description of the phases given in Table~\ref{lognormal} gives a reasonable agreement between the actual and estimated $f_{V,i}$ and $\phi_{i}$ for all phases, with the biggest discrepancy being an underestimate of $f_{V, \mathrm{warm}}\approx0.4$ against a true value of $f_{V, \mathrm{warm}}\approx0.6$. }}

%-----------------------------------------------------------------
\subsubsection{Application to observations\label{CWO}}
%-----------------------------------------------------------------

Observations can be used to estimate the volume-averaged filling factor $\tphi_{i}$, defined in Eq.~(\ref{ftilde}), for a given ISM phase. On its own, this quantity is of limited value in understanding how the phases of the ISM are distributed: of more use are the fractional volume occupied by the phase $f_{V,i}$, defined in Eq.~(\ref{fv}), and its degree of homogeneity which is quantified by $\phi_{i}$, defined by Eq.~(\ref{phin}). Knowing $\tphi_{i}$ and $\phi_{i}$, $f_{V,i}$ 
follows via Eq.~(\ref{FV}):
\begin{equation}\label{appl}
f_{V,i}=\frac{\tphi_{n,i}}{\phi_{n,i}}.
\end{equation}
This formula is exact, 
but its applicability in practise is limited if $\phi_{i}$ is unknown.
{{However $\phi_{i}$ can be deduced from the probability distribution of $n_i$: for example if the 
the density probability distribution of the
phase can be approximated by the
lognormal, as is expected for a turbulent compressible gas \citep{VS01, Elm04}, 
then $\phi_{i}$ can be estimated from Eq.~\eqref{ophi}.}} 

{\replyb{
To illustrate how these quantities may be related, let us consider some
observations reported for the diffuse ionised gas (the general approach suggested can be applied to any observable or computed
quantity).}} 
\citet{BMM06} and \citet{BM08} estimated $\tphi_{\rm DIG}$
for the diffuse ionised gas (DIG) in the Milky Way
using dispersion measures of pulsars and emission measure maps.
In particular, \citet{BMM06} obtain $\tphi_{\rm DIG}\simeq0.24$
towards $|z|=1\kpc$,
and \citet{BM08} find the smaller value $\tphi_{\rm DIG}\simeq0.08$
for a selection of pulsars that are closer to the Sun
than the
sample of \citet{BMM06}. 
On the other hand, \citet{BF08,BF12} used {\replyb{the same}} data for 
pulsars with known distances to derive PDFs of the distribution of 
{\replyb{average DIG cloud}} densities which are well described by a 
lognormal distribution; the fitted lognormals have 
{\replyb{$s_{\rm DIG}\simeq0.32$}} %at $|b|>5\deg$
\citep[Table~1 in ][]{BF12}.
Using Eqs.~(\ref{ophi}) and (\ref{appl}),
this implies that the fractional volume of DIG
with allowance for its inhomogeneity is about
\[
f_{V,\rm DIG}\simeq 0.1\mbox{--}{\replyb{0.3}}.
\]
{{In other words the combination of $\tphi_{\rm DIG}$ and $s_{\rm DIG}$ from these results imply that the DIG is approximately homogeneous. This value of $f_{V, \rm DIG}$ is in good agreement with the earlier estimates of \citet{Re77} and \citet{Re91} who obtained $f_{V, \rm DIG}\ge 0.1\mbox{--}0.2$ and close to that of \citet{Hill08} who obtained $f_{V, \rm DIG}\approx 0.25$ for a vertically stratified ISM, by comparing observed emission and dispersion measures to simulations of isothermal MHD turbulence.}} 

%\citet{BF08} also fitted lognormal distributions to the volume densities of the warm \HI\ 
%along lines of sight to 140 stars (although no filling factors could be calculated for this gas): 
%they found $s_\mathrm{HI}\simeq 0.3$, again suggesting that this phase of the ISM is 
%{\replyb{nearly}} homogeneous.
%Corrections for the inhomogeneity in the fractional volume only become significant 
%when $s_i\ga0.5$; by a factor of 1.3 for $s_i=0.5$ and a factor of 2 for $s_i=0.8$.

{{Volume density PDFs derived from observations are still rare. 
However, PDFs of the column density (and similar observables such as emission
measure and dispersion measure) are more easily derived.
The applicability of the method outlined in this Section, of deriving the
fractional volume occupied by different ISM phases from the (observable) volume 
filling factor and the PDF of the density distribution, 
would improve as the relation between the statistical parameters of volume and
column density distributions becomes better understood.}}

%-----------------------------------------------------------------
\subsubsection{{Simulation results}}\label{TAB5}
%-----------------------------------------------------------------------------

% -----------------------------------------------------------------------------
\begin{figure}
\centering
\includegraphics[width=0.85\columnwidth]{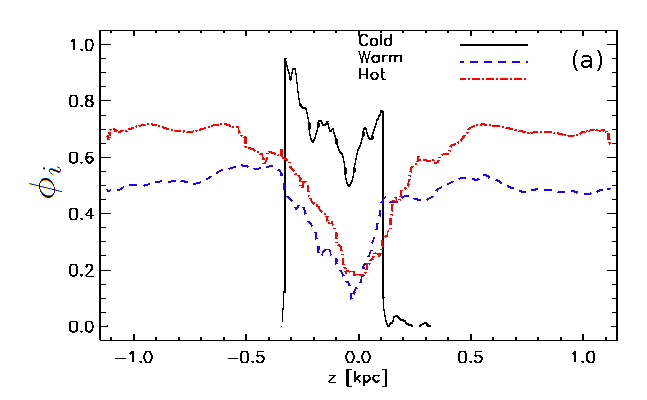}
\includegraphics[width=0.85\columnwidth]{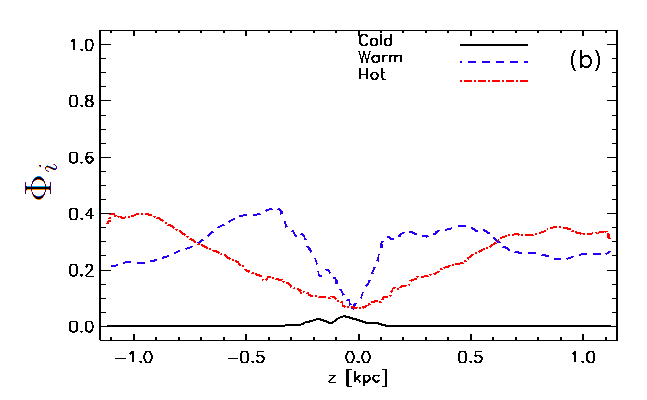}
\includegraphics[width=0.85\columnwidth]{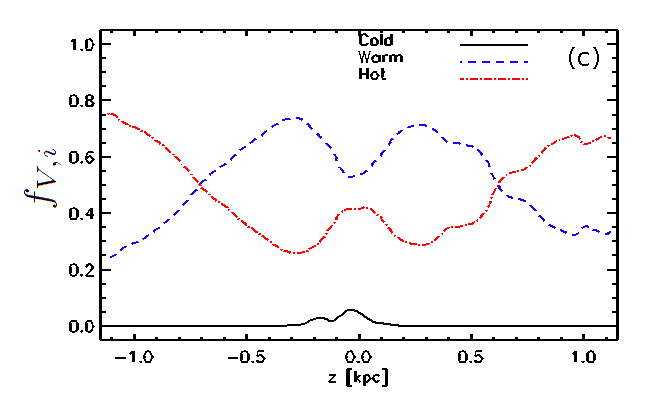}
\caption{
  Vertical profiles of \textbf{(a)}~the phase-averaged density filling factors 
  {{$\phi_i=\mean{n_i}^2/\mean{n_i^2}$}} of the 
gas phases identified in the text: cold
(black, solid line, $T<500\K$);
warm (blue, dashed, $5\times10^2\leq T<5\times10^5\K$) and
hot (red, dash-dotted, $T\geq5\times10^5\K$); and
\textbf{(b)}~the volume-averaged density filling factors 
{{{$\tphi_i=\average{n_i}^2/\average{n_i^2}$}}},
and \textbf{(c)}~the fractional volumes $f_{V,i}$ 
with the same line style for each phase. The various filling
factors are defined and discussed in Section~\ref{FF}.
These results are from 21 snapshots in the interval $636\leq t\leq646\Myr$ for Model~\Op.
\label{fig:3fill}}
\end{figure}
%-----------------------------------------------------------------------------

The filling factors and fractional volumes from 
Equations~\eqref{fv}, \eqref{ftilde} and \eqref{phin} 
{{have been computed for}} 
the phases identified in Section~\ref{TMPS} for the reference
{\replyb{model~\Op\ and}}
presented in Fig.~\ref{fig:3fill}.
Volumes are considered as discrete horizontal slices. 
{\replyb{To isolate the $z$-dependence we averaged over slices of single cell 
thickness ($4\p$-thick).}}

  {{The hot gas (Fig.~\ref{fig:3fill}c) accounts for about 70\% of the volume at
$|z|\simeq1\kpc$ and about 40\% near the mid-plane.
The local maximum of the fractional volume of the hot 
gas at $|z|\la200\p$ is due to the highest concentration of SN remnants there,
filled with 
the very 
hot gas. Regarding its contribution to integrated gas parameters, it should perhaps be considered as a separate phase. 

At $|z|\lesssim 0.7\kpc$ the warm gas accounts for over 50\% of the volume. 
The cold gas occupies a negligible volume, even in the mid-plane where it is concentrated. 
It is, however, quite homogeneous at low $|z|$ compared to the warm and hot
phases, which only become relatively homogeneous at $|z|\gtrsim 0.3\kpc$ 
(Fig.~\ref{fig:3fill}a).}} 
  
%--------------------------------------------------------------------------
\section{The correlation scale of the random flows}\label{CORR}
%-----------------------------------------------------------------------------

%-----------------------------------------------------------------------------
\begin{figure}
\centering
\includegraphics[height=0.625\columnwidth,width=0.90\columnwidth]{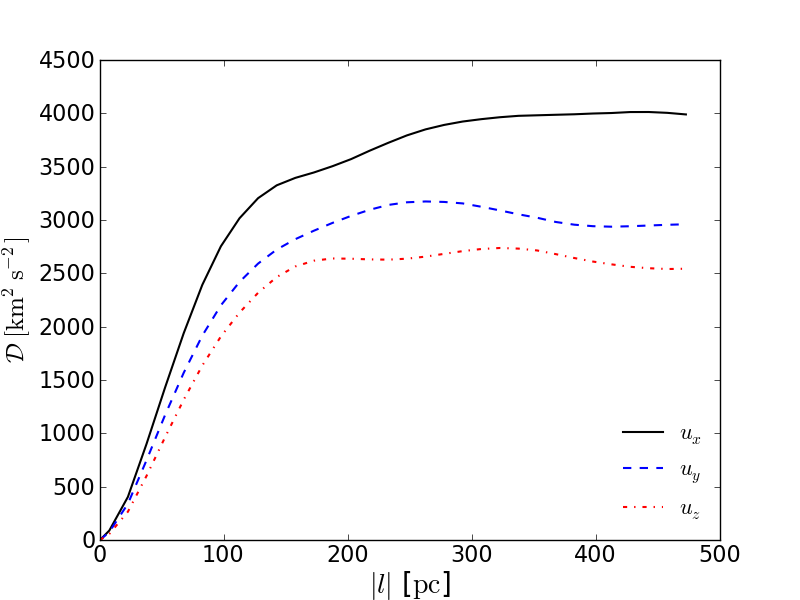}
\caption{The second-order structure functions  
calculated using Eq.~(\ref{eq:CORR:sf}),
for the layer $-10<z<10$\,pc, of the velocity components $u_x$ (black, solid line),
$u_y$ (blue, dashed) and $u_z$ (red, dash-dot). The offset $l$ is confined
to the $(x,y)$-plane only.
\label{fig:CORR:sf}}
\end{figure}
%------------------------------------------------------------------------

We have estimated the correlation length of the random velocity $\vect{u}$ at a
single time step of the model~{\Op}, by calculating the second-order structure
functions
$\str(l)$ of the velocity components $u_x$, $u_y$ and $u_z$,
where
\begin{equation}
\str(l) = \langle \left[u(\vect{x}+\vect{l})-u(\vect{x})\right]^2\rangle,
\label{eq:CORR:sf}
\end{equation}
with $\vect{x}$ the position in the $(x,y)$-plane and $\vect{l}$ a horizontal offset.
{{We did not include offsets in the $z$-direction 
and aggregated the squared differences by $|\vect{l}|$ only.
Since the flow is expected to be statistically homogeneous horizontally, 
while the correlation length is expected to 
vary with $z$.}}
A future paper will analyse in more detail the three-dimensional properties
of the random flows, including its anisotropy and dependence on
height. We measured
$\str(l)$
for {{five}} different heights,
$z=0, 100, -100$, $200\p$ and $800\p$, averaging over six adjacent slices in
the $(x,y)$-plane at each position, corresponding to a layer thickness of 20\,pc.
The averaging took advantage of the periodic boundaries in $x$ and $y$;
for simplicity we chose a simulation snapshot at a time for which
the offset in the $y$-boundary, due to the shearing boundary condition, was zero.
The structure function for the
mid-plane ($-10<z<10\p$) is shown in Fig.~\ref{fig:CORR:sf}.

The correlation scale can be estimated from the form of the structure function
since velocities are uncorrelated if $l$ exceeds the correlation length $l_0$, so that
$\str$ becomes independent of $l$, ${\mathcal D}(l)\approx 2u_\mathrm{rms}^2$ for $l\gg l_0$.
Precisely which value of
$\str(l)$
should be chosen to estimate $l_0$ in a finite domain is not always clear;
for example, the structure function of $u_y$ in Fig.~\ref{fig:CORR:sf}
allows one to make a case for either
the value at which $\str(l)$ is maximum or the value at the greatest
$l$. 
Alternatively, and more conveniently, 
one can estimate $l_0$ via the autocorrelation function 
$\corr(l)$,  related  to $\str(l)$ by
\begin{equation}
\corr(l) = 1-\frac{\str(l)}{2u^2_\mathrm{rms}}.
\label{eq:CORR:ac}
\end{equation}
In terms of the autocorrelation function, the
correlation scale $l_0$ is defined as
\begin{equation}
l_0=\int_0^\infty \corr(l)\,\dd l,
\label{eq:CORR:l}
\end{equation}
and this provides a more robust method of deriving $l_0$ in a finite
domain. Of course, the domain must be large enough to make $\corr(l)$
negligible at scales of the order of the domain size; this is a nontrivial
requirement, since even an exponentially weak tail can make a finite
contribution to $l_0$. In our estimates we are, of course, limited 
to the range
of $\corr(l)$ within our computational domain, so that the upper limit in the
integral of Eq.~(\ref{eq:CORR:l}) is equal to $L_x=L_y$, 
the horizontal box size.

%-----------------------------------------------------------------------------
\begin{figure}
\centering
\includegraphics[height=0.475\columnwidth,width=0.75\columnwidth,clip=true,trim=0 0 0 9mm]{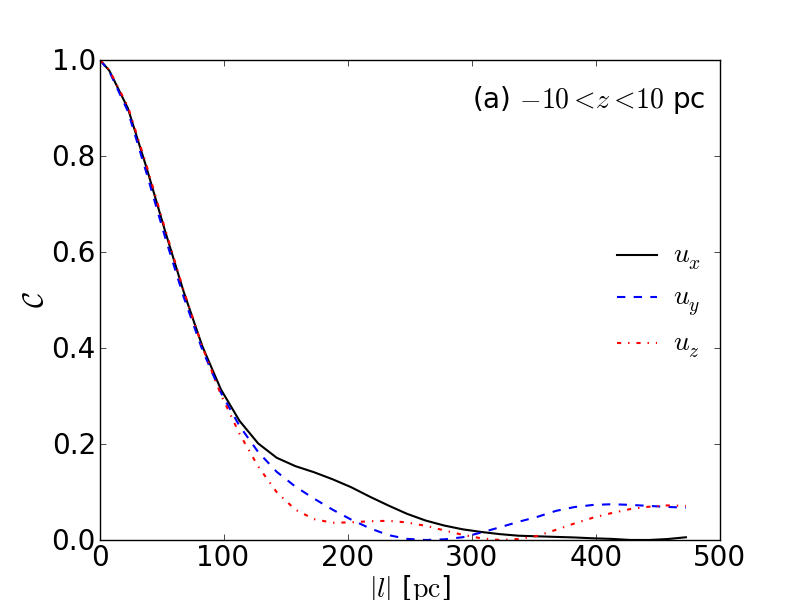}
\includegraphics[height=0.475\columnwidth,width=0.75\columnwidth,clip=true,trim=0 0 0 9mm]{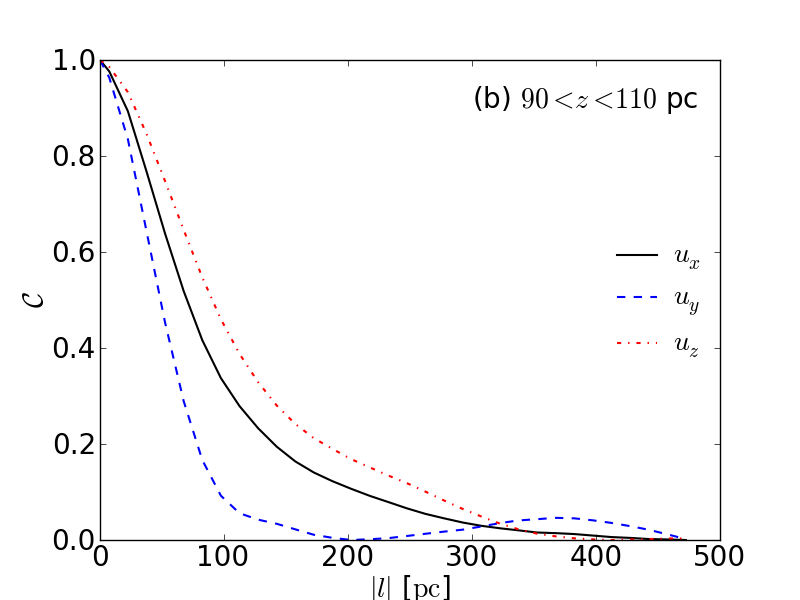}
\includegraphics[height=0.475\columnwidth,width=0.75\columnwidth,clip=true,trim=0 0 0 9mm]{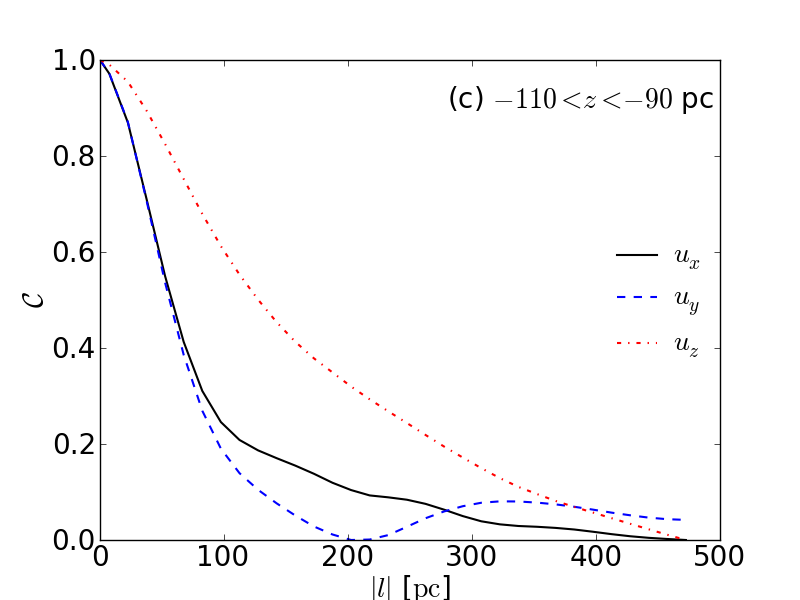}
\includegraphics[height=0.475\columnwidth,width=0.75\columnwidth,clip=true,trim=0 0 0 9mm]{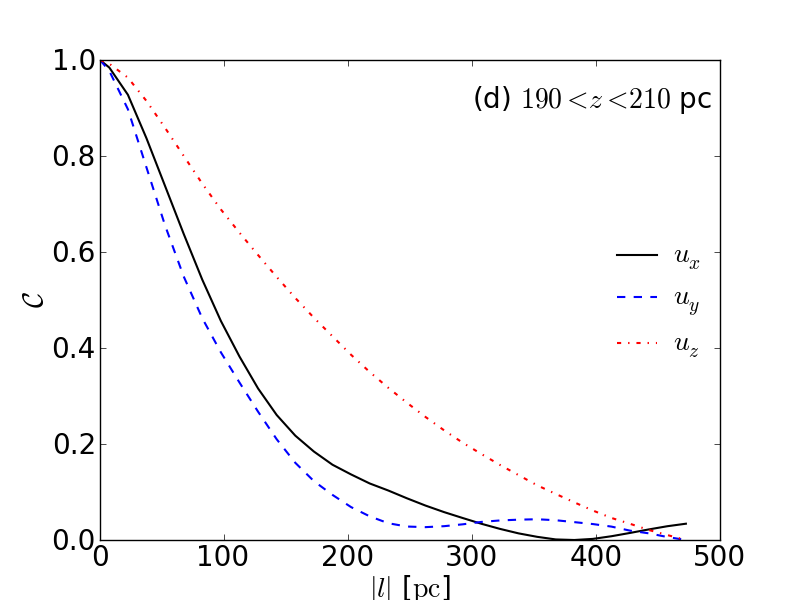}
\includegraphics[height=0.475\columnwidth,width=0.75\columnwidth,clip=true,trim=0 0 0 9mm]{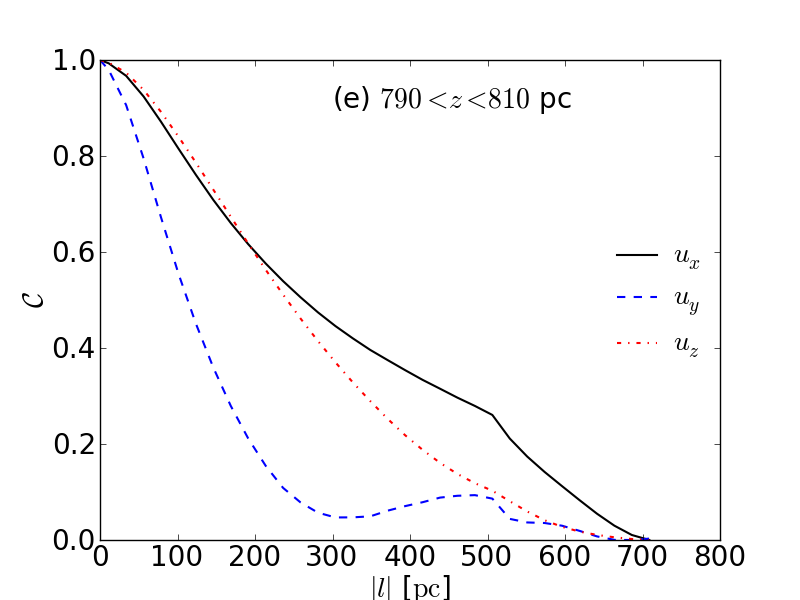}
\caption{Autocorrelation functions for the velocity components $u_x$ (black, solid line),
$u_y$ (blue, dashed) and $u_z$ (red, dash-dot) for $20\p$ thick layers centred on
four different heights, from top to bottom: $-10<z<10\p$, $90<z<110\p$, $-110<z<-90\p$, $190<z<210\p$ and $790<z<810\p$.
\label{fig:CORR:ac}}
\end{figure}
%-----------------------------------------------------------------------------

Figure~\ref{fig:CORR:ac} shows
$\corr(l)$ for {{five}} different heights in the disc, where
$u_\mathrm{rms}$ was taken to correspond to the absolute maximum of the structure
function, $u^2_\mathrm{rms}=\sfrac{1}{2}\max(\str)$, 
from Eq.~(\ref{eq:CORR:ac}) {{at each height}}.

The autocorrelation function of the vertical velocity varies with $z$
more strongly than, and differently from,
the autocorrelation functions of the horizontal velocity components;
it broadens as $|z|$ increases, meaning that the vertical velocity is
correlated over progressively greater horizontal distances. 
Already at $|z|\approx 200\p$, 
$u_z$ is coherent across a significant horizontal cross-section of the
domain, {{and at $|z|\approx 800\p$ so is $u_x$.}} An obvious explanation for this behaviour is the expansion of the
hot gas streaming away from the mid-plane,
which thus occupies a progressively larger
part of the volume as it flows towards the halo.

%-----------------------------------------------------------------------------
\begin{table}
\centering
\caption{The correlation scale $l_0$ and rms velocity $u_\mathrm{rms}$ at 
various distances from the mid-plane.
\label{table:CORR:l}}
\begin{tabular}{cccccccc}
\hline
 & \multicolumn{3}{c}{$u_\mathrm{rms}$ [$\!\kms$]}& & \multicolumn{3}{c}{$l_0$ [$\!\p$]} \\
 \cline{2-4} \cline{6-8}
$z$             &$u_x$ & $u_y$ & $u_z$ &&  $u_x$ & $u_y$ & $u_z$ \\
\hline
$\phantom{-10}0$& $45$ & $40$  & $37$  && $99$   & $98$  & $94$  \\
$\phantom{-}100$& $36$ & $33$  & $43$  && $102$  & $69$  & $124$ \\
$-100$          & $39$ & $50$  & $46$  && $95$   & $87$  & $171$ \\
$\phantom{-}200$& $27$ & $20$  & $63$  && $119$  & $105$ & $186$ \\
$\phantom{-}800$& $51$ & $21$  & $107$ && $320$  & $158$ & $277$ \\
\hline
\end{tabular}
\end{table}
%-----------------------------------------------------------------------------

Table~\ref{table:CORR:l} shows the rms velocities derived from the structure
functions for each component of the velocity at each height, and the
correlation lengths obtained from the autocorrelation functions.
Note that these are obtained without separation into phases.
The uncertainties in $u_\mathrm{rms}$ due to the choices of local maxima in
$\str(l)$ are less than $2\kms$. However, these can produce quite large systematic
uncertainties in $l_0$, as small changes in $u_\mathrm{rms}$ can lead to $\corr(l)$
becoming negative in some range of $l$ (i.e.\ a weak anti-correlation), and
this can significantly alter the value of the integral in Eq.~(\ref{eq:CORR:l}).
Such an anti-correlation at moderate values of $l$ is natural for
incompressible flows; the choice of $u_\mathrm{rms}$ {{and the estimate of $l_0$ are}} thus not straightforward.
Other choices of $u_\mathrm{rms}$ in Fig.~\ref{fig:CORR:sf} can lead
to a reduction in $l_0$ by as much as $30\p$.
Better statistics, derived
from data cubes for a number of different time-steps, will allow for a more
thorough exploration of the uncertainties, but we defer this analysis to a
later paper.

%-----------------------------------------------------------------------------
\begin{figure}
\centering
\includegraphics[width=0.85\columnwidth]{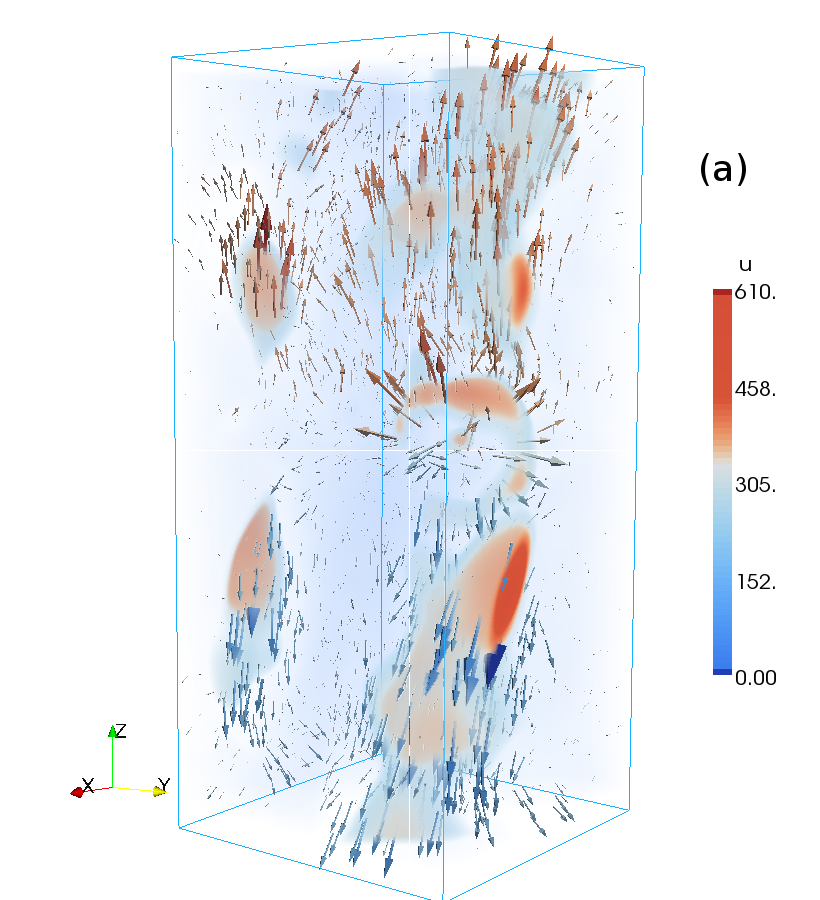}
\includegraphics[width=0.85\columnwidth]{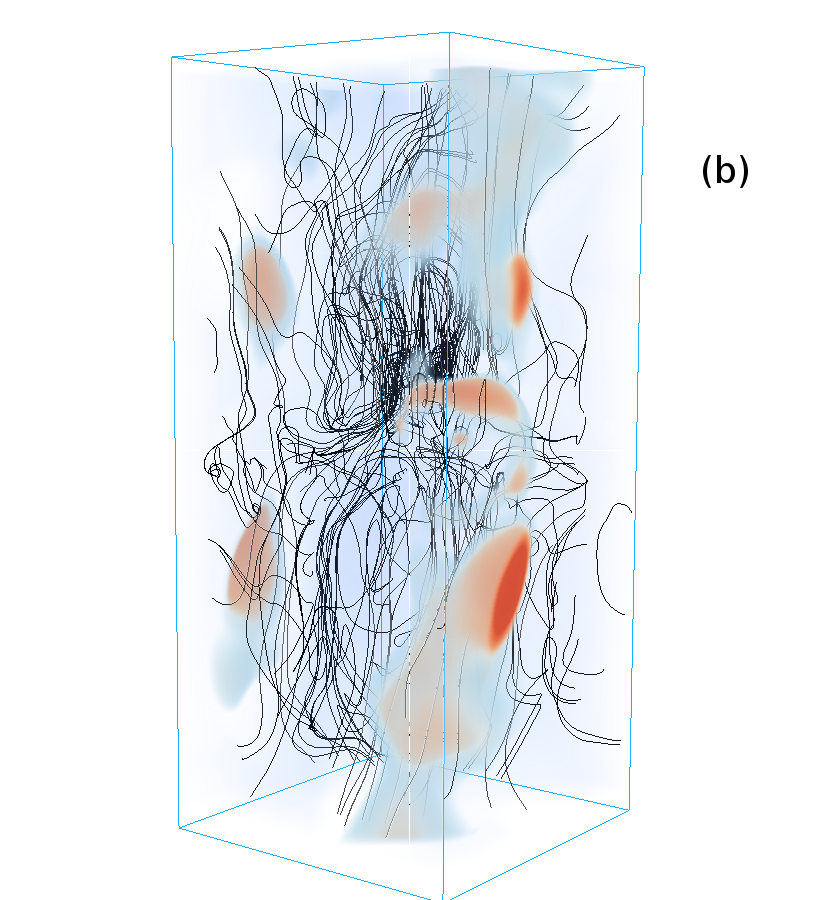}
\caption{
The perturbation velocity field $\vect{u}$ in  Model~{\Op} at $t=550\Myr$.  
{{The colour bar indicates the magnitude of the velocity field depicted 
in the volume shading,}} 
with rapidly moving regions highlighted with shades of red. 
{{The low velocity regions, shaded blue, have reduced opacity to 
assist visualisation.}} 
Arrow length of vectors \textbf{(a)} is proportional to the magnitude of 
$\vect{u}$, with red (blue) arrows
corresponding to $u_z>0$ ($u_z<0$) {{and independent of the colour bar.}}
Trajectories of fluid elements \textbf{(b)} are also shown, indicating the 
complexity of the flow and its pronounced vortical structure.
 \label{fig:vplot}}
 \end{figure}
%------------------------------------------------------------------------

The rms velocities given in Table~\ref{table:CORR:l} are compatible with the
global values of $u_\mathrm{rms}$ and $u_0$ for the reference {Model~{\Op}}
shown in Table~\ref{table:models}.
The increase in the
rms value of  $u_z$ with height, from about $40\kms$ at $z=0$ to
about $60\kms$ at $z=200\p$, reflects the systematic {net}
outflow with a speed increasing with $|z|$.
There is also an apparent marginal tendency for the rms values of
$u_x$ and $u_y$ to decrease with increasing distance from the mid-plane. 

The correlation scale of the random flow is very close to $100\p$ in the
mid-plane, and we have adopted this value for $l_0$ elsewhere in
the paper. This estimate is in good agreement with the hydrodynamic ISM
simulations of \citet{Joung06}, who found that most kinetic energy is contained
by fluctuations with a wavelength (i.e.\ $2l_0$ in our notation) of $190\p$. In
the MHD simulations of \citet{Korpi99}, $l_0$ for the warm gas was $30\p$ at
all heights, but that of the hot gas increased from $20\p$ in the mid-plane to
$60\p$ at $|z|=150\p$. \citet{AB07} found $l_0=73\p$ on average, 
with strong fluctuations in time.
As in \citet{Korpi99},
there is a weak tendency for $l_0$ of the horizontal velocity components to
increase with $|z|$ in our simulations, but
this tendency remains tentative, and must be examined more carefully 
to confirm its robustness.

% ----------------------------------------------------------------------------
\section{Gas flow to and from the mid-plane \label{GO}}
% ----------------------------------------------------------------------------

Figure~\ref{fig:vplot} illustrates the 3D structure of the perturbation velocity
field for the reference Model~\Op. Shades of red show the regions of high
speed, whereas regions moving at speeds below about $300\kms$ are transparent
to aid visualisation. Velocity vectors are shown in panel~(a) using
arrows, with size indicating the speed, and colour indicating the sign of the
$z$-component of the velocity ({{indicating preferential outflow 
from the mid-plane}}). Red patches are indicative of recent SN
explosions, and there is a strongly divergent flow close to the middle of the
$xz$-face. In addition, stream lines in panel~(b) display the presence of 
considerable small scale vortical flow near the mid-plane.

%-----------------------------------------------------------------------------
\begin{figure}
\centering
\includegraphics[width=0.9\columnwidth]{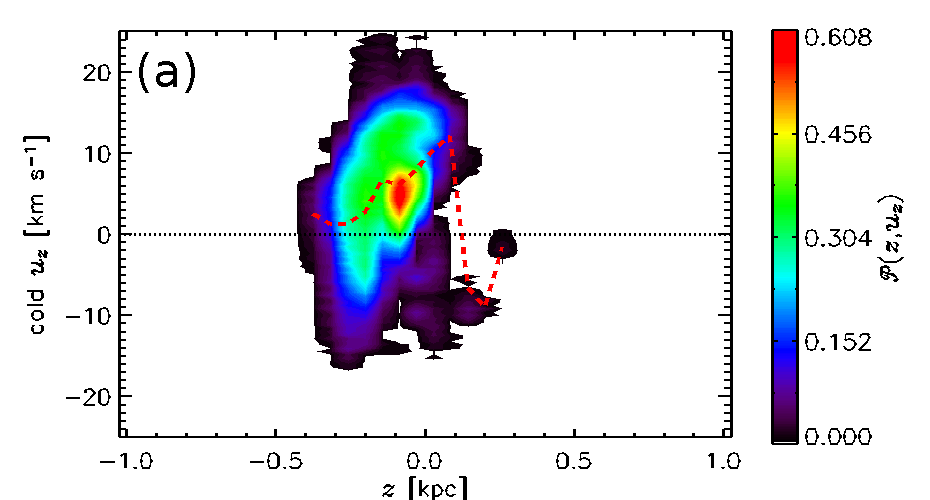}
\includegraphics[width=0.9\columnwidth]{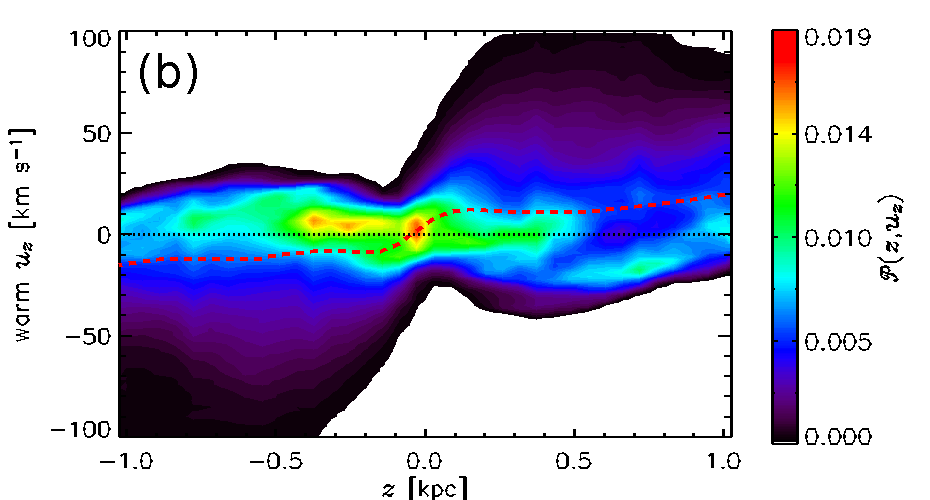}
\includegraphics[width=0.9\columnwidth]{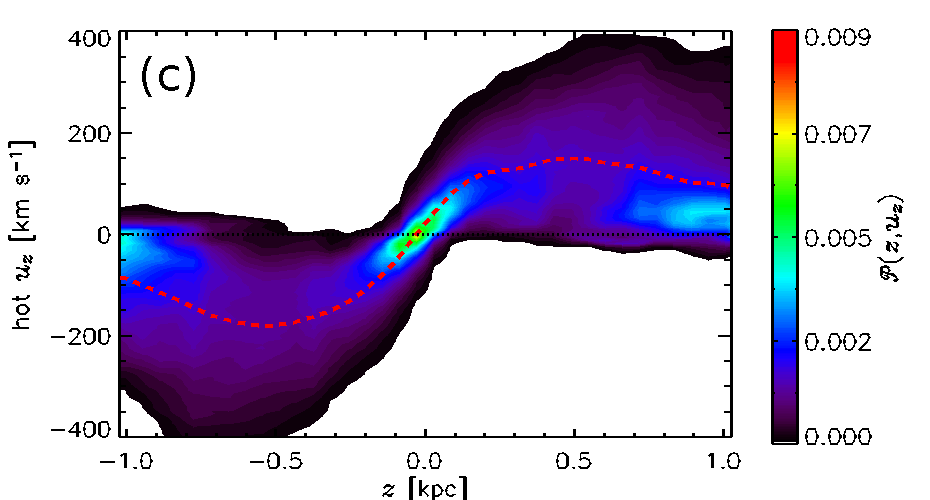}
\caption{
{{Contours of the probability density of 
the vertical velocity $u_z$ as a function of $z$ in Model~{\Op} from
11 snapshots at $t=$ 634--644\Myr.
The cold $(T<500\K)$, warm $(500\K\leq T<5\times10^5\K)$ and hot 
$(T\geq 5\times10^5\K)$ are shown in panels \textbf{(a)} to \textbf{(c)},
respectively.
The horizontal averages of the vertical velocity $u_z$ in each case are
shown in red, dashed in each panel as well as the mid-plane position (black, dotted).}}
\label{fig:uzm}\label{fig:uzscat}}
 \end{figure}
%------------------------------------------------------------------------

The mean vertical flow is dominated by the high velocity hot gas, so it is
instructive to consider the velocity structure of each phase separately.
Figure~\ref{fig:uzscat} shows {{the probability 
distributions $\mathcal{P}(z,u_z)$ as functions of $u_z$ in the $(z,u_z)$-plane
from 11
snapshots of Model~\Op,}} separately for the cold (a), warm (b) and hot gas (c).
{{The cold gas is mainly restricted to $|z|<300\p$
and its vertical velocity varies within $\pm20\kms$. 
As indicated by the red dashed curve in Panel (a),}}
on average, the cold gas moves towards the mid-plane, 
presumably after cooling at larger heights. 
{{The warm gas is involved in a weak net vertical outflow above $|z|=100\p$, 
of order $\pm10\kms$.
This might be an entrained flow within the hot gas.
However, due to its skewed distribution, the modal flow and thus mass transfer
is typically towards the mid-plane.
The hot gas has large net outflow speeds, accelerating to
about $100\kms$ within $|z|\pm200\p$, but with small amounts of inward
flowing gas at all heights.}}
The mean hot gas outflow speed increases at an approximately constant rate
to somewhat over $100\kms$ within $\pm100\p$ of the mid-plane,
and then decreases with further distance from the mid-plane,
at a rate that gradually decreases with height for $|z|\ga0.5\kpc$.
{{This is below the escape velocity in the gravitational potential 
adopted.}}
The structure of the velocity
field shall be investigated further elsewhere. 

%-----------------------------------------------------------------------------
\section{Sensitivity to model parameters\label{SMP}}
%-----------------------------------------------------------------------------

%-----------------------------------------------------------------------------
\begin{figure}
\centering
\includegraphics[width=0.9\columnwidth]{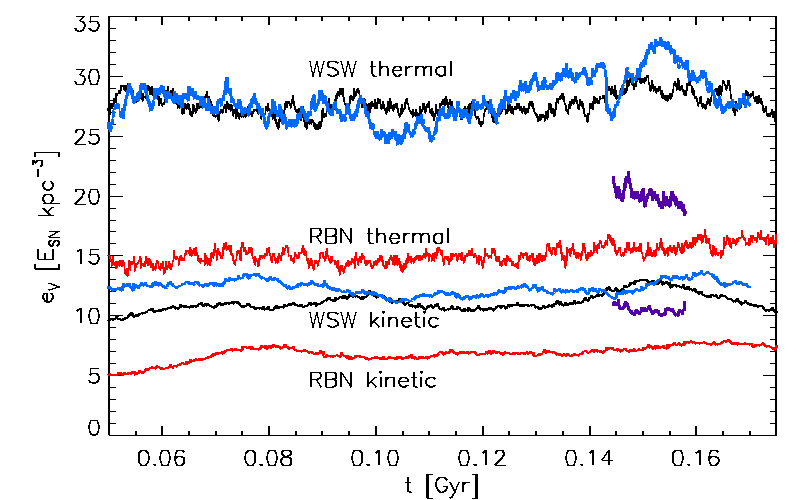}
\caption{Evolution of the volume-averaged thermal energy density (black:
model~\WSWa, blue: model~\Op, purple: model~\OpH, red: model~RBN) 
and kinetic energy density (as above;  lower lines)
in the statistical steady regime, 
normalised to the SN energy $E\SN \kpc^{-3}$.
Models~{\WSWa}~(black) and RBN (red) essentially differ 
only in the choice of the radiative cooling function.
\label{fig:energetics}}
\end{figure}
%-----------------------------------------------------------------------------

%-----------------------------------------------------------------------------
\begin{figure}
\centering
\hspace{-0.25cm}
\includegraphics[width=0.7\columnwidth,clip=true,trim=0 0 0 10mm]{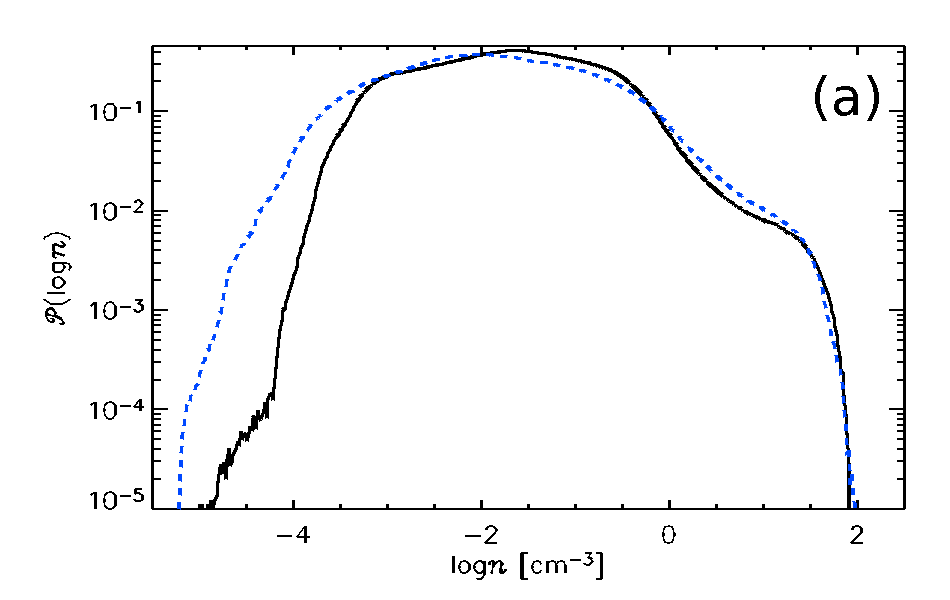}
\hspace{-0.25cm}
\includegraphics[width=0.332\columnwidth,clip=true,trim=0 0 0 10mm]{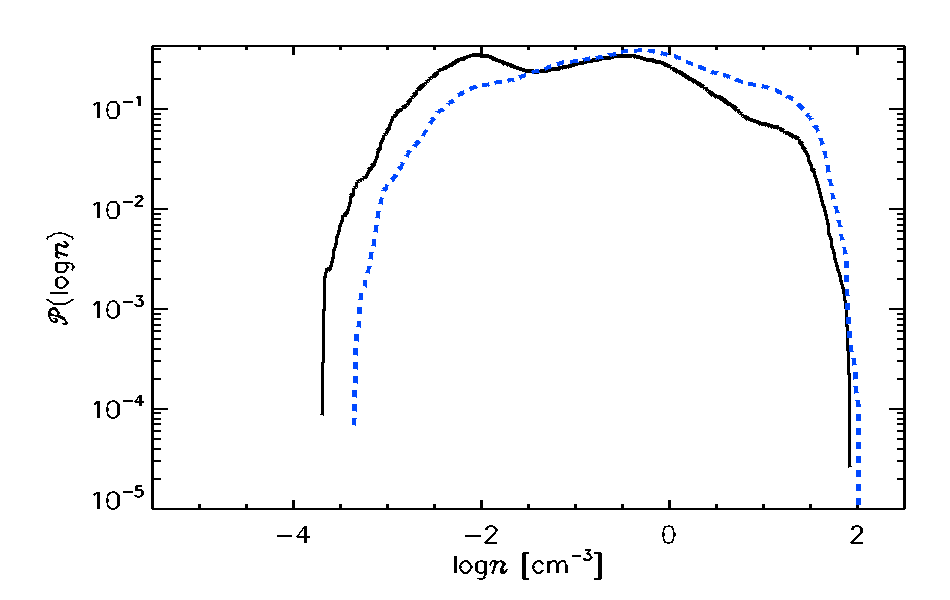}\\
\hspace{-0.25cm}
\includegraphics[width=0.7\columnwidth,clip=true,trim=0 0 0 10mm]{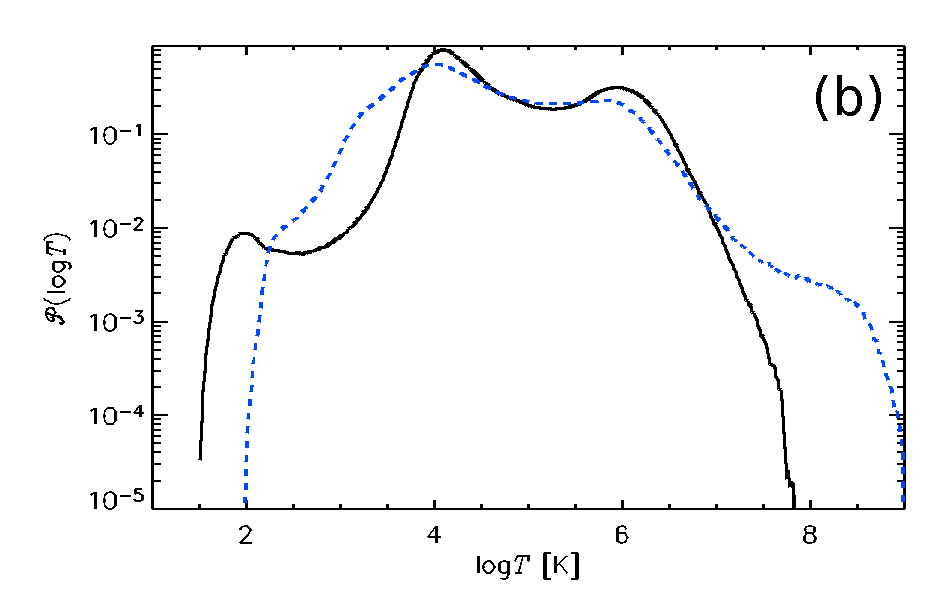}
\hspace{-0.25cm}
\includegraphics[width=0.332\columnwidth,clip=true,trim=0 0 0 10mm]{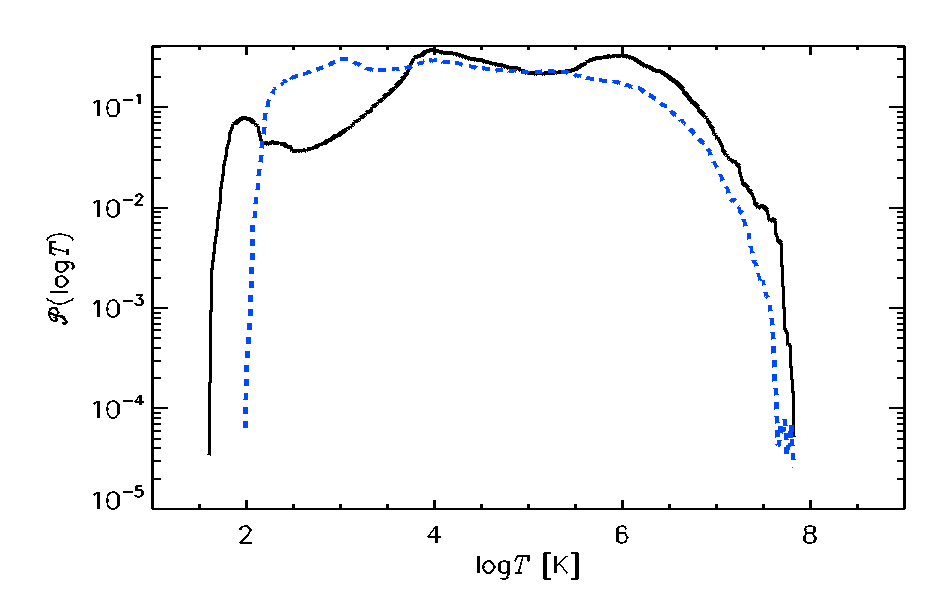}\\
\hspace{-0.25cm}
\includegraphics[width=0.7\columnwidth,clip=true,trim=0 0 0 10mm]{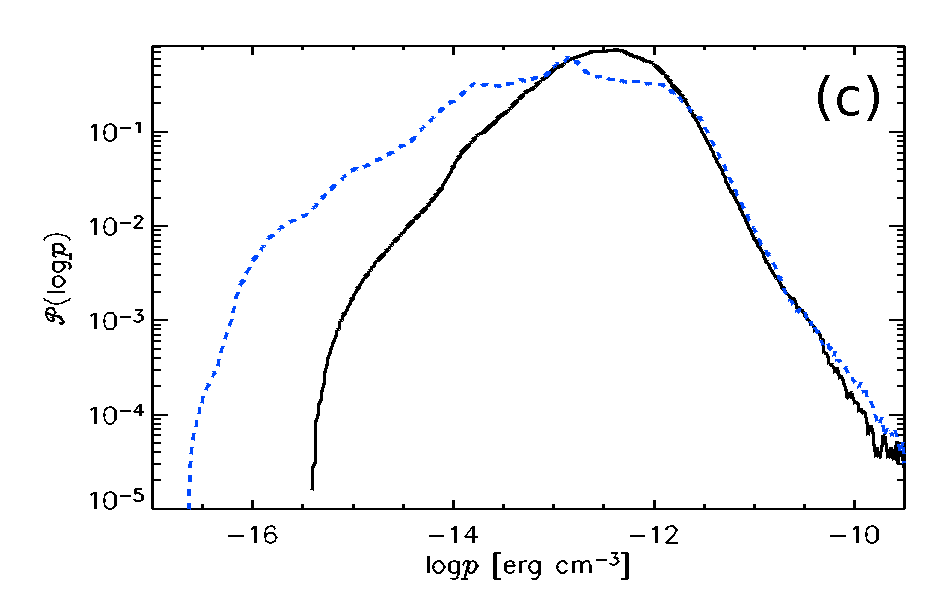}
\hspace{-0.25cm}
\includegraphics[width=0.332\columnwidth,clip=true,trim=0 0 0 10mm]{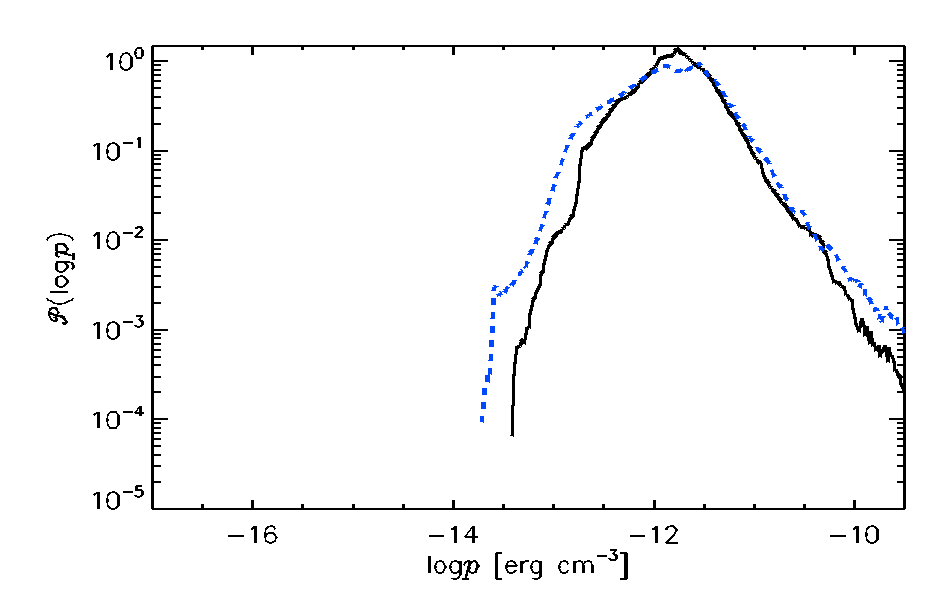}
\caption{
{\replyb{
Probability density distributions
in the whole computational domain}}{{, obtained
without separation into distinct phases,}} for \textbf{(a)}~gas density,
\textbf{(b)}~temperature and \textbf{(c)}~thermal pressure,
for Model~{RBN} (blue, dashed) and Model~{\WSWa} (black, solid),
in a statistical steady state,
each averaged over 21 snapshots spanning 20\,Myr 
(RBN: 266 to $286\Myr$, and {\WSWa}: 305 to $325\Myr$) {{and the total
simulation domain $|z|\le1.12\kpc$}}.
{\replyb{The smaller frames to the right display the same information but
near the midplane, $|z|<20\p$, only.}}
\label{fig:pdfs_cool}}
\end{figure}
%-----------------------------------------------------------------------------

%-----------------------------------------------------------------------------
\subsection{The cooling function}\label{COOL}
%-----------------------------------------------------------------------------

%-----------------------------------------------------------------------------
\begin{figure*}
\centering
\includegraphics[width=0.7\columnwidth,clip=true,trim=0 0 0 9mm]{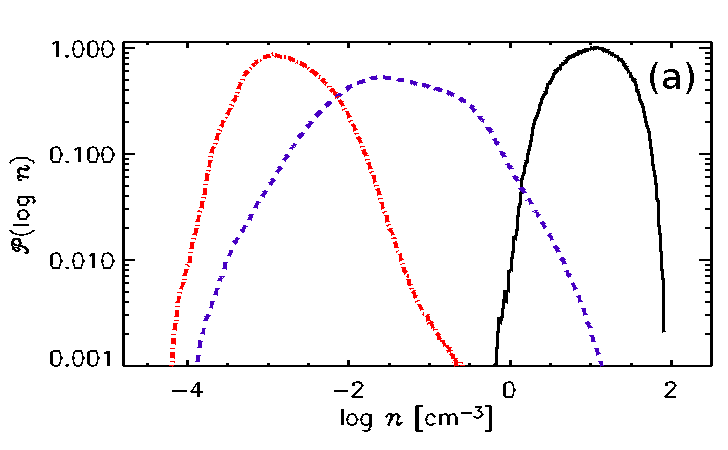}\hspace{1.5cm}
\includegraphics[width=0.7\columnwidth,clip=true,trim=0 0 0 9mm]{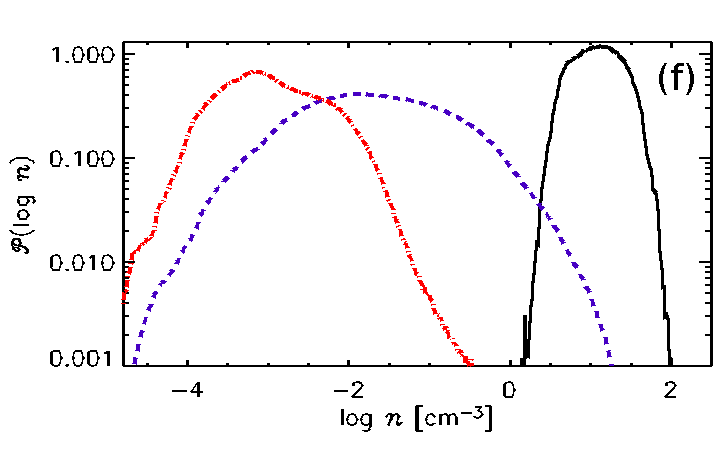}\\
\includegraphics[width=0.7\columnwidth,clip=true,trim=0 0 0 9mm]{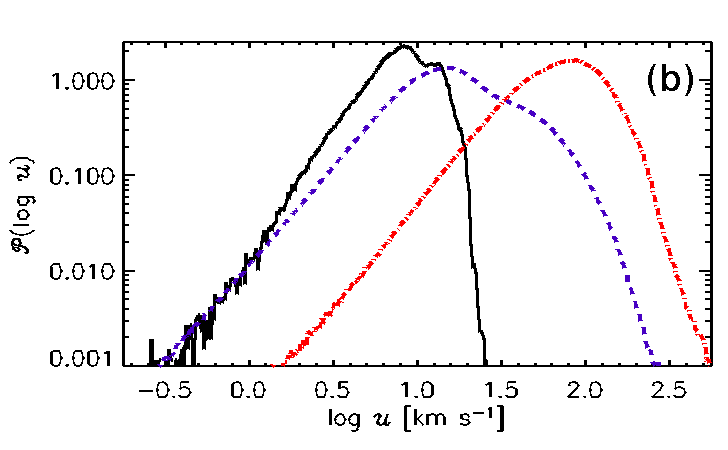}\hspace{1.5cm}
\includegraphics[width=0.7\columnwidth,clip=true,trim=0 0 0 9mm]{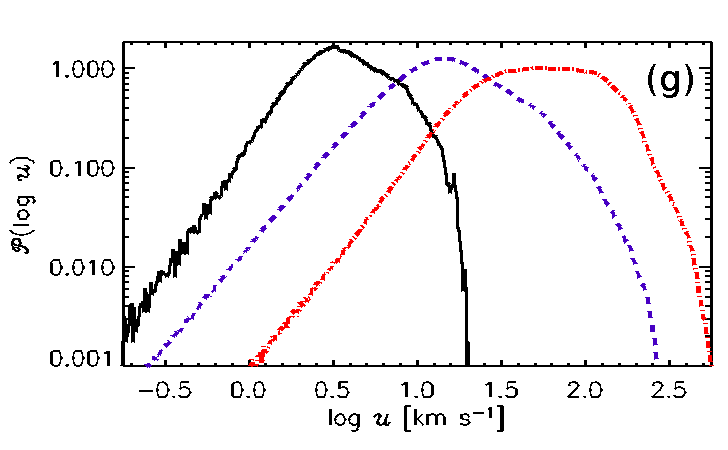}\\
\includegraphics[width=0.7\columnwidth,clip=true,trim=0 0 0 9mm]{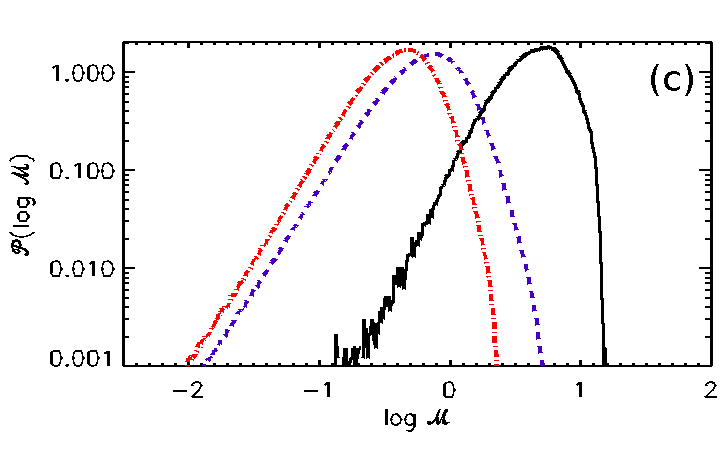}\hspace{1.5cm}
\includegraphics[width=0.7\columnwidth,clip=true,trim=0 0 0 9mm]{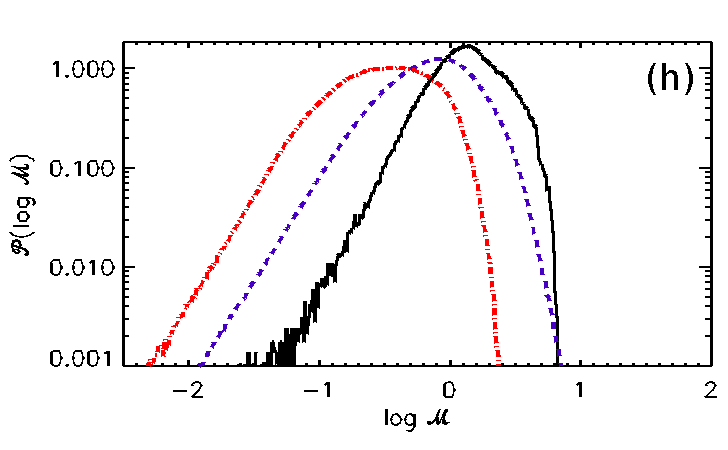}\\
\includegraphics[width=0.7\columnwidth,clip=true,trim=0 0 0 9mm]{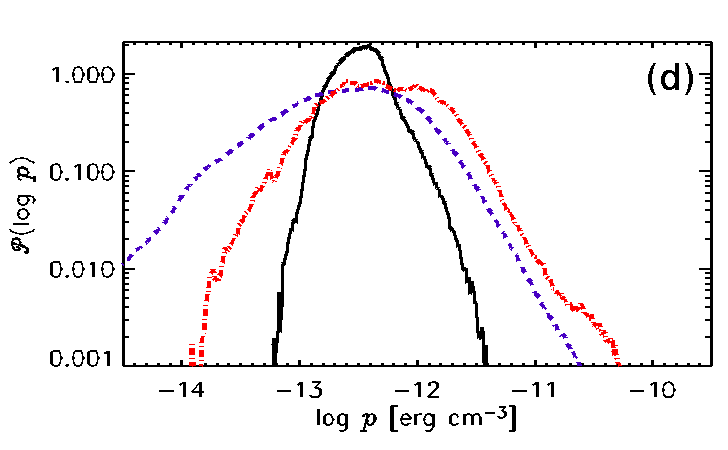}\hspace{1.5cm}
\includegraphics[width=0.7\columnwidth,clip=true,trim=0 0 0 9mm]{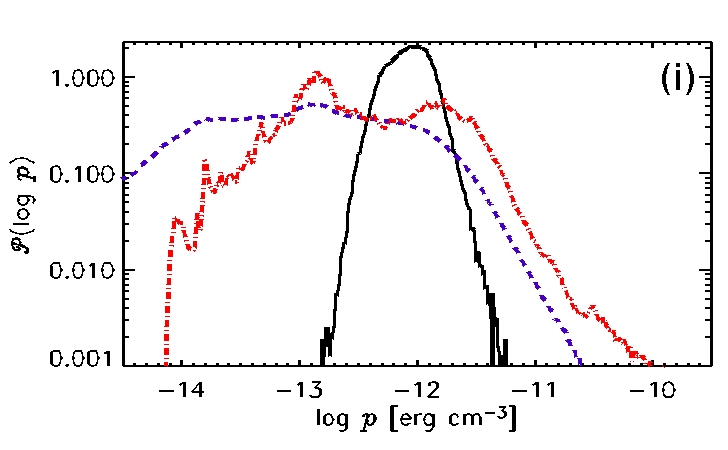}\\
\includegraphics[width=0.7\columnwidth,clip=true,trim=0 0 0 9mm]{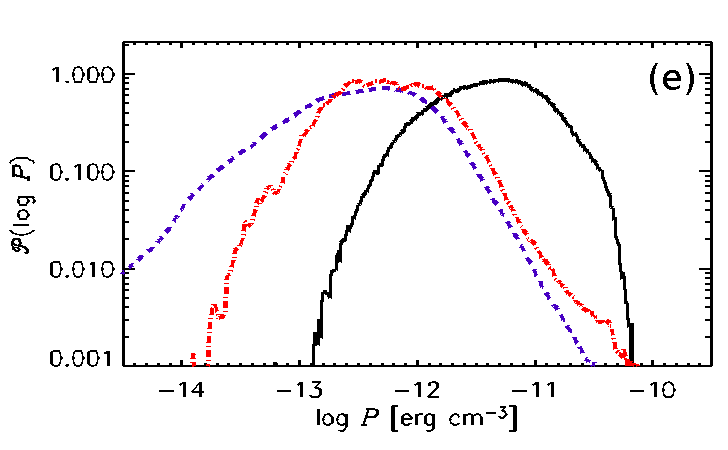}\hspace{1.5cm}
\includegraphics[width=0.7\columnwidth,clip=true,trim=0 0 0 9mm]{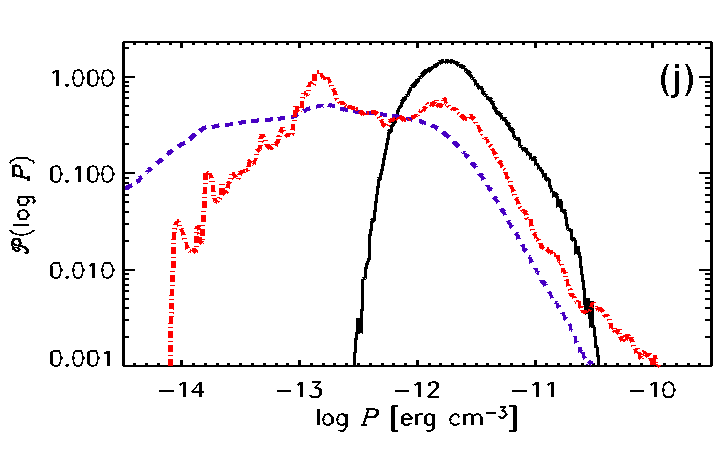}
\caption{Probability densities for various variables in individual phases, 
for Model~{\WSWa} (left-hand column of panels) and Model~{RBN} (right-hand column): 
\textbf{(a)} and \textbf{(f)} for gas density;
\textbf{(b)} and \textbf{(g)} for random velocity $u_0$;
\textbf{(c)} and \textbf{(h)} for the Mach number of the random velocity defined with respect to the local sound speed;
\textbf{(d)} and \textbf{(i)} for thermal pressure; and 
\textbf{(e)} and \textbf{(j)} for the total pressure.
The cold phase spans $T<500\K$ (black, solid), 
the warm gas has $500 < T < 5\times 10^5\K$ (blue, dashed) and 
the hot gas is at $T\geq5\times10^5\K$ (red, dash-dotted). 
Eleven snapshots have been used for averaging, spanning
$t=200$--$300\Myr$ for Model~RBN and  $t=300$--$400\Myr$ for Model {\WSWa}.
\label{fig:wsw_pdf3ph}}
\end{figure*}
%-----------------------------------------------------------------------------

We consider two models, RBN and \WSWa, with parameters given in Table~\ref{table:models},
to assess the effects of the specific choice of the cooling function. 
Apart from different parameterizations of the radiative cooling, the two models
share identical parameters, except 
the value of $T_0$ was slightly higher in Model~{RBN},
because of the sensitivity of the initial conditions to the cooling function
(Section~\ref{ICons}).

The volume-averaged thermal and kinetic energy densities, the latter
{{excluding the imposed shear flow $\vect{U}$}}, are shown in
Fig.~\ref{fig:energetics} as functions of time. The averages for each are 
shown in Columns~(11) and (12), respectively
of Table~\ref{table:models}, using the appropriate steady state time 
intervals given in Column~(4).
Models reach a statistical steady-state, 
with mild fluctuations around a well defined mean value, very soon
(within 60\,Myr of the start of the simulations). The effect of the cooling
function is evident: both the thermal and kinetic energies in Model~RBN
are about 60\% of those in Model~\WSWa. This is understandable as Model~RBN has
a stronger cooling rate than Model~\WSWa, only dropping below the WSW rate 
in the range $T<10^3\K$ (see Fig.~\ref{fig:cool}).
Interestingly, both models are similar in that the thermal energy is about
$2.5$ times the kinetic energy.

These results are also remarkably consistent with results by 
\citet[][their Fig.~6]{Balsara04} and \citet[][Fig.\,3.1]{Gressel08b}.
\citet{Gressel08b} applies WSW cooling and has a model very similar to 
Model\,{\Op},
with half the resolution and $|z|\leq2\kpc$. 
He reports average energy densities of 24 and 10\,$E\SN\kpc^{-3}$ 
(thermal and kinetic, respectively) 
with SN rate  $=\dot{\sigma}\SN$, comparable to 30 and 13\,$E\SN\kpc^{-3}$ 
obtained here for Model\,{\Op}. 

\citet{Balsara04} simulate an unstratified
cubic region $200\p$ in size, driven at SN rates of 8, 12 and 40 times the
Galactic rate, with resolution more than double that of Model\,{\Op}. 
For SN rates $12\dot{\sigma}\SN$ and $8\dot{\sigma}\SN$, they obtain average
thermal energy densities of about 225 and 160$\,E\SN\kpc^{-3}$, 
and average kinetic energy densities of 95 and 60$\,E\SN\kpc^{-3}$, 
respectively 
(derived from their energy totals divided by the $[200\p]^3$ volume). 

To allow comparison with our models, where the SNe energy
injection rate is $1\dot{\sigma}\SN$, 
if we divide their energy densities by 12 and 8, respectively,
the energy densities would be 19 and 20\,$E\SN\kpc^{-3}$ (thermal),
and 8 and 7.5\,$E\SN\kpc^{-3}$ (kinetic). 
These {{are slightly lower than}} our results with RBN cooling 
(25 and 9\,$E\SN\kpc^{-3}$), but are 
below those with WSW
(30 and 13\,$E\SN\kpc^{-3}$ for WSWa, as given above).
\citet{Balsara04} used an alternative cooling function 
\citep{Raymond77}, so allowing for some additional uncertainty over the net
radiative energy losses, the results appear remarkably consistent. 

While cooling and resolution may marginally affect the magnitudes, it appears
that thermal energy density may consistently be expected to be about $2.5$ 
times the 
kinetic energy density, in these models. It also appears, by comparing the stratified and
unstratified models, that the ratio of thermal to kinetic energy is not
strongly dependent on height
{{over the range included in our model.}}

The two models are further compared in Fig.~\ref{fig:pdfs_cool}, where we show
probability distributions for the gas density, temperature and thermal
pressure.  With both cooling functions, the most probable gas number density is
around $3 \times 10^{-2}\cm^{-3}$; 
the most probable temperatures are 
also similar, at around $3 \times 10^{4}\K$.
With the RBN cooling function, the density
range extends to smaller densities than with {\WSWa};
{\replyb{and yet the temperature range for {\WSWa} extends 
{to}
lower 
{values}
than for RBN.  
It is evident that the isobarically unstable 
{part} 
of 
{the} WSW cooling
{function}}}
does significantly reduce the amount of gas {{at $T=$}} 313--6102\,K
({{the}} temperature range 
{{corresponding to the thermally unstable regime of the WSW cooling}}),
and increase the amount
{of gas} below 100\,K. 
However this is not associated with higher densities than 
{when using the} RBN
{cooling function}. 
This may indicate that multiple compressions, rather than thermal 
{instability,}
dominate the formation of dense clouds.

The most probable thermal pressure is lower in Model~RBN than in {\WSWa}, 
consistent with the lower thermal energy content of the former.

The probability distributions of various quantities,
shown in Fig.~\ref{fig:wsw_pdf3ph}, confirm the clear phase separation in terms
of gas density and perturbation velocity. Here we used the same borderline
temperatures for individual phases as for Model~{\Op} (Fig.~\ref{fig:pdf3ph}).
Despite minor differences between the corresponding panels 
in Figs.~\ref{fig:pdf3ph} and \ref{fig:wsw_pdf3ph},
the peaks in the gas density probability distributions are close to
$10^1, 3 \times 10^{-2}$ and $10^{-3}\cm^{-3}$ in all models. 
Given the extra cooling of hot gas and reduced cooling of
cold gas with the RBN cooling function, 
more of the gas resides in the warm phase in Model~RBN. 
The thermal pressure distribution in the hot gas reveals the two `types'
(see the end of section~\ref{TMPS}),
which are mostly found within $|z|\la200\p$ 
(high pressure hot gas within SN remnants) 
and outside this layer (diffuse, lower pressure hot gas). 
The probability distribution for the Mach number in the warm gas extends to higher values with the RBN cooling function, perhaps because more shocks reside in the more widespread warm gas, at the expense of the cold phase. 
It is useful to remember that, 
although each distribution is normalised to unit underlying area, 
the fractional volume of the warm gas is about a hundred times that of the 
cold phase.

The probability distributions of {{density and pressure in}} {\replyb{without
preliminary separation into phases, presented in}} Fig.~\ref{fig:pdfs_cool} 
do not show 
clear separations into phases 
\citep[cf. e.g.\,][]{Joung06,AB04},
such that division into three phases would arguably only be conventional,
if based on these alone. 
{\replyb{The probability distributions near the mid-plane,
$|z|<20\p$  
Fig.~\ref{fig:pdfs_cool}, exhibit a marginally better phase separation
for the gas density (smaller frames in Fig.~\ref{fig:pdfs_cool})
 \citep[see also][their Figs.~1; and 6, respectively]{Korpi99a,HJMBM12}.
}}
{{However our analysis in terms of phase-wise PDFs confirms that the 
trimodal structure evident
in the temperature distribution (Fig.~\ref{fig:pdfs_cool}b) has a complementary
structure in the gas density.}}

{{Stratification of the thermal structure is clarified in 
    Fig.~\ref{fig:zfill_RB_WSW}, where we introduce narrower temperature bands
    specified in Table~\ref{table:bands}. 
  The fractional volume of gas in each
  temperature range $i$ at a height $z$ is given by
  \begin{equation}\label{fva}
    f_{V,i}(z)=\frac{V_i(z)}{V(z)}=\frac{N_i(z)}{N(z)},
  \end{equation}
  similarly to Eq.~\eqref{fv},
  where $N_i(z)$ is the number of grid points in the temperature range
  $T_{i,\mathrm{min}}\leq T<T_{i,\mathrm{max}}$,
  with $T_{i,\mathrm{min}}$ and
  $T_{i,\mathrm{max}}$
  given in Table~\ref{table:bands}, and $N(z)$ is the total number of grid
  points at that height.}}

{{The fractional volumes in Column~(13) of Table~\ref{table:models} 
show that near the mid-plane cold gas forms in similar abundances, independent
of the cooling function. 
However, much less hot gas is achieved for Model~RBN.}}
Figure~\ref{fig:zfill_RB_WSW} {{also}} 
helps show how the thermal gas structure depends on the cooling
function. 
Model~{\WSWa}, panel {\bf {(b)}}, has significantly more very cold gas ($T<50\K$)
than RBN, panel {\bf{(a)}}, but slightly warmer cold gas ($T<500\K$) is more abundant in RBN. 
The warm and hot phases ($T>5\times10^3\K$) have roughly similar
distributions in both models, although Model~RBN has less of both phases.
Apart from relatively minor details, the effect of the form of the
cooling function thus appears to be straightforward and predictable: 
stronger cooling means more cold gas and
vice versa. What is less obvious, however, is that the very hot gas is more
abundant near $\pm1\kpc$ in Model~RBN than in {\WSWa}, indicating that the
typical densities must be much lower. 
This, together with the greater abundance of cooler
gas near the mid-plane, suggest that there is less stirring with RBN cooling.

Altogether, we conclude that the properties of the cold and warm phases 
are not strongly affected by the choice of the cooling function. 
The main effect is that the RBN cooling function produces
less hot gas with significantly lower pressures.
This can readily be understood,
as this function provides significantly stronger cooling at $T\ga10^3\K$.

%-----------------------------------------------------------------------------
\begin{figure}
\centering
\includegraphics[width=0.85\columnwidth]{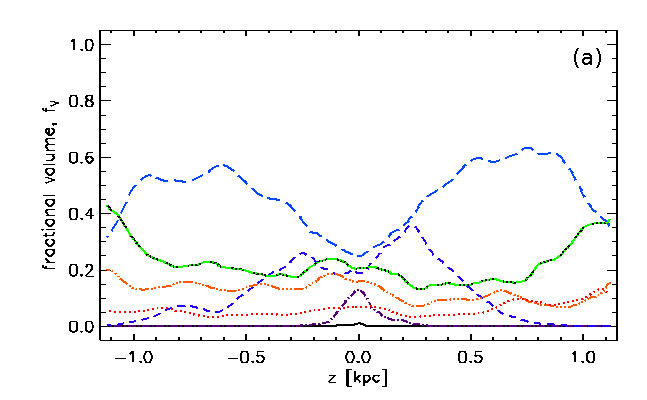}
\includegraphics[width=0.85\columnwidth]{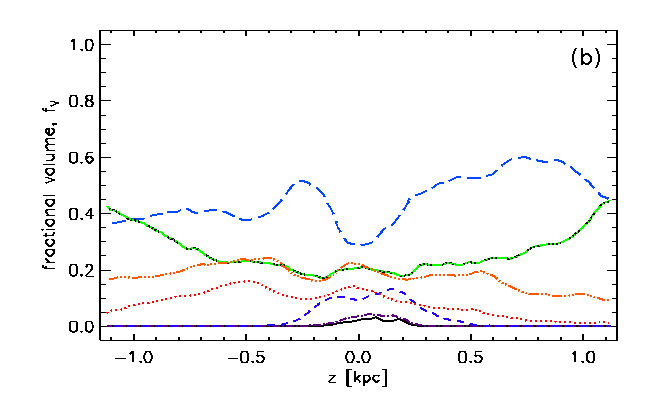}
\caption{Vertical profiles of the fractional volumes occupied by the
various temperature ranges, with the key shown in Table~\ref{table:bands}.
\textbf{(a)}~Model~{RBN}, using 21  snapshots spanning 266 to 286\,Myr.
\textbf{(b)}~Model~{\WSWa}, using 21 snapshots spanning 305 to 325\,Myr.
\label{fig:zfill_RB_WSW}}
\end{figure}
%-----------------------------------------------------------------------------

%-----------------------------------------------------------------------------
\begin{table}
\centering
\caption{Key to Figs~\ref{fig:zfill_RB_WSW}
 and \ref{fig:zfill_comp}, defining the gas temperature bands used there,
and the classification {{within}} three phases.}
\label{table:bands}
\begin{tabular}{ccl}
\hline
Temperature band   &Line style & Phase \\
\hline
  \phantom{$5\times10^1\K<$}$T<5\times10^1\K$  &\rule[0.08cm]{1.05cm}{0.75pt}  & cold \\
  $5\times10^1\K\leq T<5\times10^2\K$ &
    {\Large\color{violet}{$\cdot$-$\cdot$-$\cdot$-$\cdot$-$\cdot$}} & cold \\
  $5\times10^2\K\leq T<5\times10^3\K$ &
    {\Large\color{blue}{- - - - -}} & warm \\
  $5\times10^3\K\leq T<5\times10^4\K$ &
    {\Large\color{royalblue}{-- -- -- -}} & warm \\
  $5\times10^4\K\leq T<5\times10^5\K$ &
    {\Large\color{mygreen}{--}\color{black}{-}\hspace{-0.08cm} \color{mygreen}{$\cdot$}\hspace{-0.03cm}\color{black}{-}\hspace{-0.08cm} \color{mygreen}{$\cdot$}\hspace{-0.03cm}\color{black}{-}\hspace{-0.08cm}  \color{mygreen}{--}} & warm \\
  $5\times10^5\K\leq T<5\times10^6\K$ &
    {\Large\color{burntorange}{--$\cdots$--$\cdot$}} & hot \\
  \phantom{$5\times10^1\K<$}$T\geq5\times10^6\K$ &
    {\Large\color{red}{$\cdots\cdots$}} & hot \\
\hline
\end{tabular}
\end{table}
%-----------------------------------------------------------------------------

%-----------------------------------------------------------------------------
\subsection{The total gas mass}\label{MOL}
%-----------------------------------------------------------------------------

Models~RBN and {\WSWa} have about 17\% more mass of gas 
than the reference Model~{\Op}, where we have removed that part of
the gas mass which should be confined to molecular clouds 
unresolved in our simulations 
(as described in section~\ref{REF}).
The difference is apparent in comparing 
Fig.~\ref{fig:zfill_RB_WSW}b with {{Fig.~\ref{fig:zfill_comp}b (or
 Fig.~\ref{fig:zfill}a).}}
Higher gas mass causes the abundance of hot gas to reduce with height,
contrary to observations, 
and to the behaviour of Model~\Op.
Otherwise, the fractional volumes within $\pm200\p$ of the mid-plane
appear independent of the gas mass. 

%-----------------------------------------------------------------------------
\begin{figure}
\centering
\includegraphics[width=0.85\columnwidth]{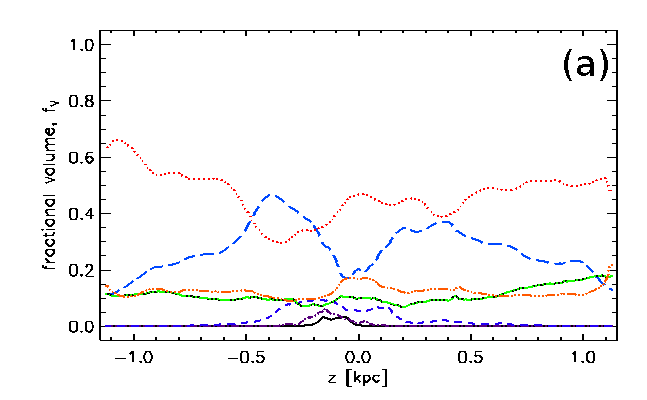}
\includegraphics[width=0.85\columnwidth]{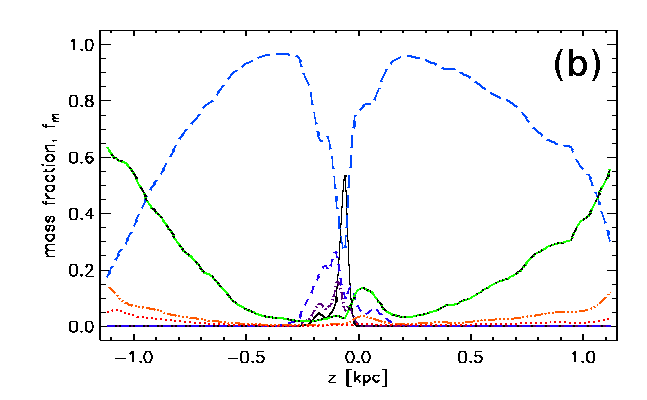}
\includegraphics[width=0.85\columnwidth]{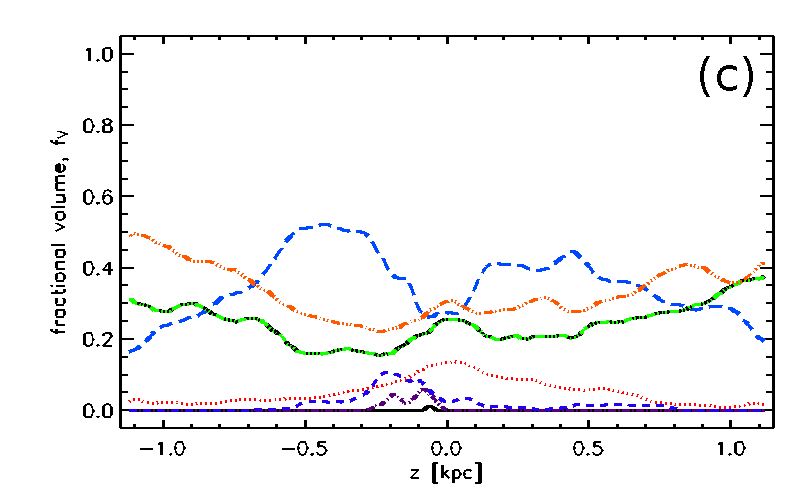}
\caption{
{{
  Vertical profiles of the fractional volumes (Eq.~\ref{fv}) for Model~{\OpH}
  {\textbf{(a)}}, which differs only in its doubled spatial resolution from the
  reference Model~{\Op} {\textbf{(c)}} and the fractional mass (Eq.~\ref{fm})
  from Model~{\Op} \textbf{(b)}. 
  These are calculated for the temperature ranges given, along with the figure
  legend, in Table~\ref{table:bands}.
  The former uses 10 snapshots and the latter 6 spanning 633 to 638\,Myr.}}
\label{fig:zrho}\label{fig:zfill}\label{fig:zfill_comp}}
\end{figure}
%-----------------------------------------------------------------------------

%-----------------------------------------------------------------------------
\subsection{Numerical resolution}\label{COMP}
%-----------------------------------------------------------------------------

%-----------------------------------------------------------------------------
\begin{figure}
\centering
\includegraphics[width=0.7\columnwidth,clip=true,trim=0 0 0 9mm]{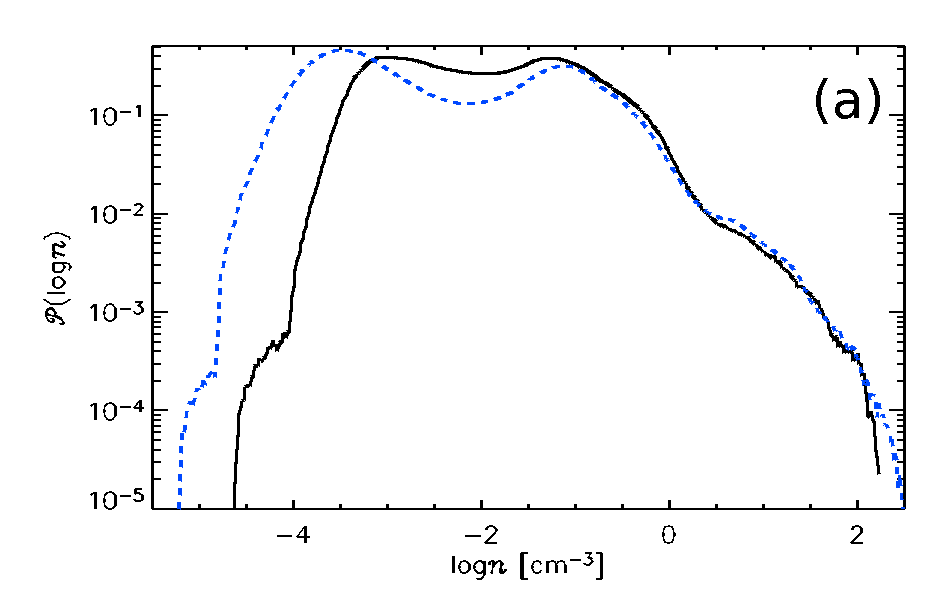}
\includegraphics[width=0.7\columnwidth,clip=true,trim=0 0 0 9mm]{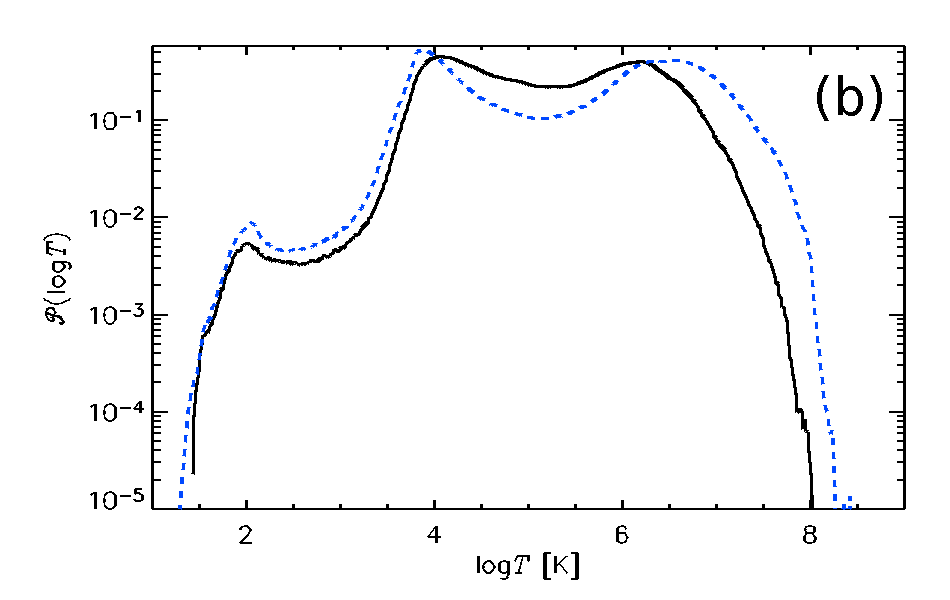}
\includegraphics[width=0.7\columnwidth,clip=true,trim=0 0 0 9mm]{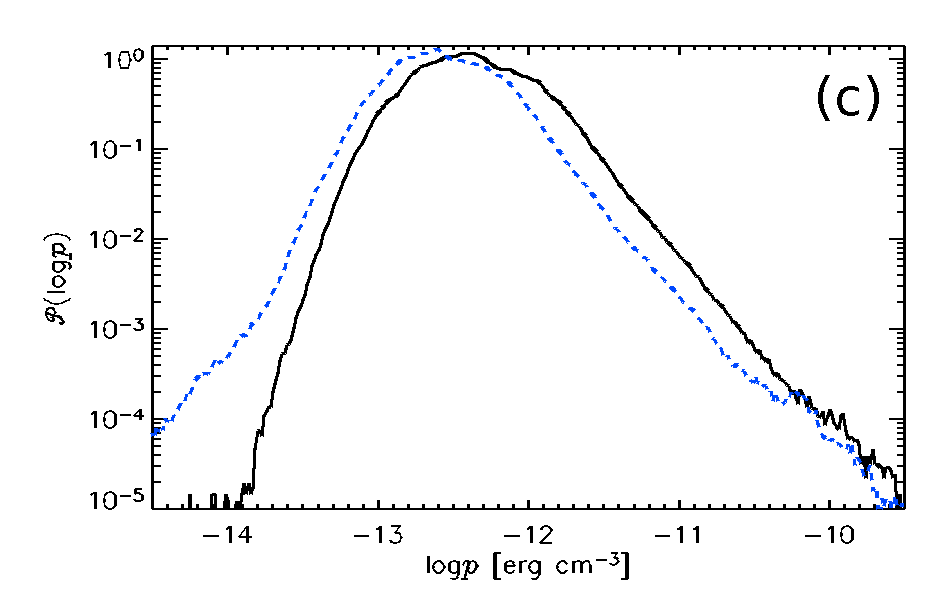}
\caption{
{{Volume weighted}} probability distributions of gas number density (a),
temperature (b) and thermal pressure (c) for  models {\Op} (black, solid) and
{\OpH} (blue, dashed) {{for the total numerical domain $|z|\le1.12\kpc$}}.
\label{fig:pdf2}}
 \end{figure}
%-----------------------------------------------------------------------------

Models~{\Op} and {\OpH} differ only in their resolution, using 2 and 4\,pc,
respectively. Model~{\OpH} is a continuation of the state of {\Op} after 600\,Myr of
evolution.

%-----------------------------------------------------------------------------
\begin{figure}
\centering
\includegraphics[width=0.85\columnwidth]{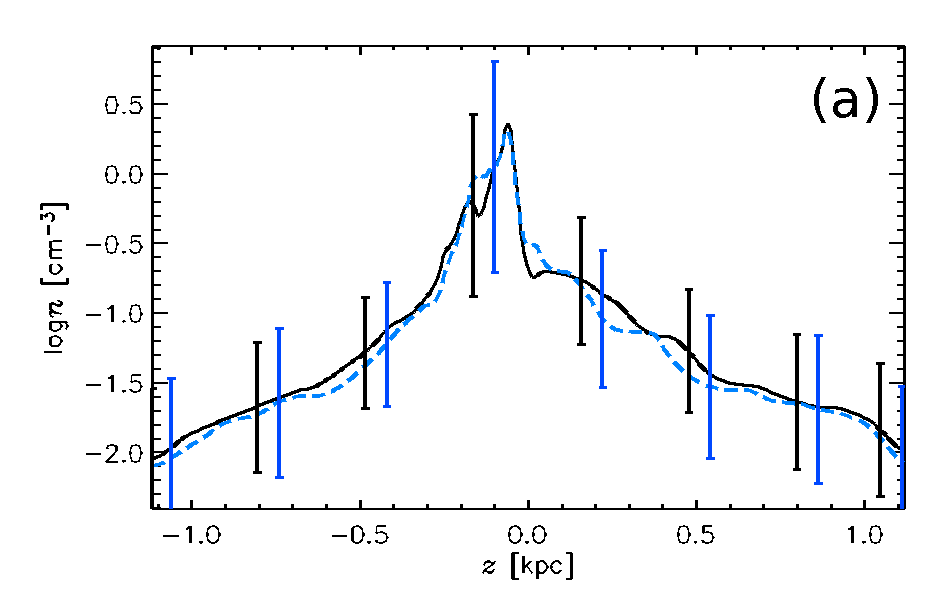}
\includegraphics[width=0.85\columnwidth]{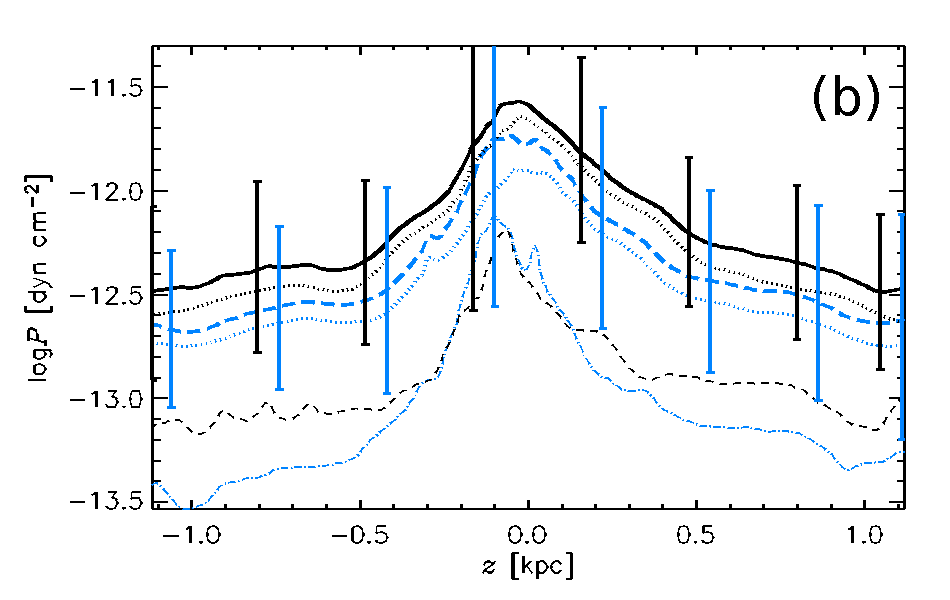}
\caption{
{{Horizontal averages of gas number density, $\mean{n}(z)$ 
{\textbf{(a)}}, and total pressure, $P(z)$ {\textbf{(b)}}, for Model~{\Op}
(solid, black), and Model~{\OpH} (dashed, blue). 
Each are time-averaged using 6 and 10 snapshots respectively, 
spanning 633 to 638\Myr.
The vertical lines indicate standard deviation within each horizontal slice.
The thermal $p(z)$ (dotted) and ram $p_0(z)$ (fine dashed) pressures are 
also plotted {\textbf{(b)}}.
}}
\label{fig:zrho_rho}\label{fig:zfill_comp_rho}}
\end{figure}
%-----------------------------------------------------------------------------

The most obvious effect of increased resolution is the increase in the
magnitude of the perturbed velocity{{and temperatures; 
    $\average{u_\rrms}=76\kms$ in
Model~{\Op} increasing to $103\kms$ in Model~{\OpH}}} 
(Table~\ref{table:results}, Column~9)
{{and $\average{c\sound}$ from 150 to $230\kms$ (Column~6).}}
{{Both $\average{u_\rrms}$ and the random velocity $\average{u_{0,\rrms}}$}} 
are increased by a similar factor of about {{1.3}}. 
However, the thermal energy $e_{\textrm{th}}$ is reduced by a factor of 0.6 
with the higher resolution, while kinetic energy $e_{\textrm{K}}$ remains about
the same. 
This suggests that in the higher-resolution model, the higher velocities and 
temperatures are associated with lower gas densities.

  The vertical distribution of the fractional volume in each temperature range 
  (defined in Table~\ref{table:bands}) 
  {{is shown in Fig.~\ref{fig:zfill_comp} for Model~{\OpH} (panel a) 
  for comparison with Model~\Op\, in (c).}} 
  {{The fractional mass (b) is calculated similarly to Eq.~\eqref{fva} 
  is
  \begin{equation}\label{fm}
  f_{M,i}(z)=\frac{M_i(z)}{M(z)},
  \end{equation}
  where $M_{i}(z)$ is the mass of gas within temperature range $i$ at a given
  $z$, and $M(z)$ is the total gas mass at that height.}}

  Note that the relative abundances of the various phases in these models
  might be affected by the unrealistically high thermal conductivity adopted.
  The coldest gas (black, solid), with $T<50\K$, is largely confined within
  about $200\p$ of the mid-plane.
  Its fractional volume (Fig.~\ref{fig:zfill}a,c) is small even at the 
  mid-plane, but it provides more than half of the gas mass at $z=0$
  (Fig.\ref{fig:zfill}b).
  Gas in the next temperature range, $50<T<500\K$ (purple, dash-dotted), is
  similarly distributed in $z$.
  Models~{\Op} and {\OpH} differ only in their resolution, using 2 and 4\,pc,
  respectively. 
  Model~{\OpH} is a continuation of the state of {\Op} after 600\,Myr of
  evolution.
  With higher resolution the volume fraction of the coldest gas is 
  significantly enhanced (Fig.~\ref{fig:zfill}c compared to a), but it is
  similarly distributed.

  Gas in the range $5\times10^2<T<5\times10^3\K$ (dark blue, dashed)
  has a similar profile to the cold gas for both the fractional mass and the
  fractional volume, and this is insensitive to the model resolution.
  This is identified with the warm phase, but exists in the thermally unstable
  temperature range.
  It accounts for about 10\% by volume and 20\% by mass of the gas near the 
  mid-plane, which is consistent with observational evidence.
  It is negligible away from the supernova active regions. 
  
  The two bands with $T>5\times10^5\K$ (red, dotted and orange, dash-3dotted)
  behave similarly to each other (Fig.~\ref{fig:zfill}a,c), occupying similar fractional volumes 
  for $|z|\la0.75\kpc$, and with $f_{V,i}$ increasing above this height 
  (more rapidly for the hotter gas). 
  In contrast the fractional masses (Fig.~\ref{fig:zfill}b) in these temperature bands are
  negligible for $|z|\la0.75\kpc$, and increase above this height (less 
  rapidly for the hotter gas).
  The temperature band $5\times10^4<T<5\times10^5\K$ (green/black, 
  dash-3dotted) is similarly distributed to the hotter gas (orange) in all
  profiles.
  It is however identified with the warm phase, indicating that this is mainly
  hot gas cooling, a transitional state, which accounts for a relatively small
  volume fraction of the warm gas and especially a small mass fraction.
  The dramatic effect of increased resolution (Fig.~\ref{fig:zfill}a compared to c) is the
  significant increase in the very hot gas (red, dotted), particularly 
  displacing the hotter gases (orange and green) but also to some degree the
  bulk warm gas (blue, dashed).
  This reflects the reduced cooling due to the better density contrasts 
  resolved, associating the hottest temperatures to the most diffuse gas.   
 
  The middle temperature range $5\times10^3<T<5\times10^4\K$ has a distinctive
  profile in both fractional volume and fractional mass, with minima near the
  mid-plane and maxima at about $|z|\simeq400\p$, being replaced as the 
  dominant component by hotter gas above this height.
  The fractional volume and vertical distribution of this gas is quite 
  insensitive to the resolution.
  The distribution of the warm gas ($5\times10^3\K\leq T<5 \times10^4\K$; 
  blue, long-dashed) does not change much with increased resolution.
  However, the higher-resolution model has more of the cold phase 
  ($T<500\K$; black, solid and dash-dotted) and, especially, of the very 
  hottest gas ($T\geq5\times10^6\K$; red, dotted), at the expense of the
  intermediate temperature ranges. 

This can also be seen in the gas density and temperature
probability distributions shown in Fig.~\ref{fig:pdf2}(a), (b): increased resolution
modestly increases the abundance of cold gas and significantly enhances
the amount of very hot gas. The minima in the distributions (at 
density $10^{-2}\,{\rm cm}^{-3}$,
and at temperatures $10^2$ and $3 \times 10^{5}$ K)
appear independent of resolution, 
suggesting that the phase separation is physical, rather than numerical. 
The distributions are most consistent in the thermally unstable 
range 313--6102\,K. Higher resolution also reduces the minimum further about
the  unstable range above $10^5\K$, as the
highest temperature gas has lower losses to thermal conduction.
The mean temperatures of the cold gas ($60$\,K)
warm gas ($10^4$\,K) 
and the mean warm gas density ($0.14\,{\rm cm}^{-3}$) 
also appear to be independent of the resolution. 
{{However the natural log mean $\mu_n$ is about -8 for the hot gas, 
both within and without $2\p$ of the mid-plane (with larger standard deviation for the gas near the mid-plane). This compares with values of -6.97 and -5.78 in 
 {our model with} 
$4\p$ resolution; i.e.\ a factor of about 1/3. 
This reflects the improved resolution of low density in the remnant interiors.}}  

The density and temperature probability distributions for {\Op} are similar to 
those obtained by \citet[][their Fig.~7]{Joung06}, who used a similar cooling
function, despite the difference in the numerical methods (adaptive mesh 
refinement down to $1.95\p$ in their case).
With slightly different implementation of the cooling and heating processes, 
again with adaptive mesh refinement down to $1.25\p$,
\citet[][their Fig.~3]{AB04} found significantly more cool, dense gas.
It is noteworthy that the maximum densities and lowest temperatures
obtained in our study with a non-adaptive grid are of the same order of
magnitude as those from AMR-models where the local resolution is up to 
{{three}} 
times higher.
{{At $4\p$ our mean minimum temperature is $34\K$, within the range 
 $15$--$80\K$ for 0.625--$2.5\p$ \citep[][their Fig.~9]{AB04}.  
 For mean maximum gas number, our $122\cmcube$ is within their range 
$288$--$79\cmcube$.
}}

The vertical density profiles obtained under the different numerical
resolutions are shown in Fig.~\ref{fig:zrho_rho}a.
Although the density distribution in Fig.~\ref{fig:pdf2}a reveals higher 
density contrasts with increased resolution, 
there is little difference in the $z$-profiles of the models. 
{{The mean gas number density at the mid-plane, $n(0)$ --- which with our
course grid resolution excludes the contribution from \HII --- is about 
$2.2\cmcube$: double the observation estimates summarised in \citet{F01}.
This might be expected in the absence of the magnetic and cosmic ray components
of the ISM pressure, to help support the gas against the gravitational force.}}

{{However the vertical pressure distributions are consistent with the
models of \citet[][their Figs.~1 and 2]{BC90}, which include the weight of the ISM up to $|z|=5\kpc$.
The total pressure $P(0)\simeq2.5~(2.0)\times10^{-12}\dyn\cm^{-2}$ for the
standard~(high) resolution model is slightly above their estimate of about 1.9 
for hot, turbulent gas.  
For the turbulent pressure alone we have 
$p_0(0)\simeq6.3~(7.9)\times10^{-13}\dyn\cm^{-2}$ falling to $1.0~(0.6)$ at 
$|z|=500\p$ and then remaining reasonably level. 
The pressures are generally slightly reduced with increased resolution, except
for $p_0$ near the mid-plane. 
Small scales are better resolved, so the turbulent structures are a stronger
component of the SN active region.
These pressures are consistent with \citet{BC90}, even though our model does
not explicitly include the pressure contributions from the ISM above $1\kpc$.

Comparing our thermal pressure distribution (Fig.~\ref{fig:pdf2}c) with 
\citet[][their Fig.~4a]{AB04} and \citet[][their Fig.~2]{Joung09},
the three models peak at
3.16, 1.3 and $4.1\times10^{-13}\dyn\cm^{-2}$, respectively. 
The latter models include $|z|=10\kpc$ and resolution up to 1.25\p. 
Our data summarise the volume within $z\pm1\kpc$, while the comparisons are 
within $10\kpc$ and $125\p$ respectively. 
}}

We conclude that the main effects of the increased resolution are confined to
the very hot interiors and to the thin shells of SN remnants;
the interiors become hotter and the SN shell shocks become thinner 
with increased resolution (see Appendix~\ref{EISNR}). 
Simultaneously, the higher density of the shocked gas
enhances cooling, producing more cold gas and reducing the total thermal
energy. Otherwise, the overall structure of the diffuse gas is little
affected: the probability distributions of thermal pressure are almost
indistinguishable, with our standard resolution fractionally higher pressure
(Fig.~\ref{fig:pdf2}c).

We are satisfied that the numerical resolution of the reference model,
$\Delta=4\p$, is sufficient to model the diffuse gas phases reliably. This
choice of the working numerical resolution is further informed by tests
involving the expansion of individual SN remnants (presented in
Appendix~\ref{EISNR}).

%-----------------------------------------------------------------------------
\section{Discussion and conclusions} \label{disc}\label{conc}

%-----------------------------------------------------------------------------
  \begin{figure*}
  \centering
  \includegraphics[width=0.3765\linewidth]{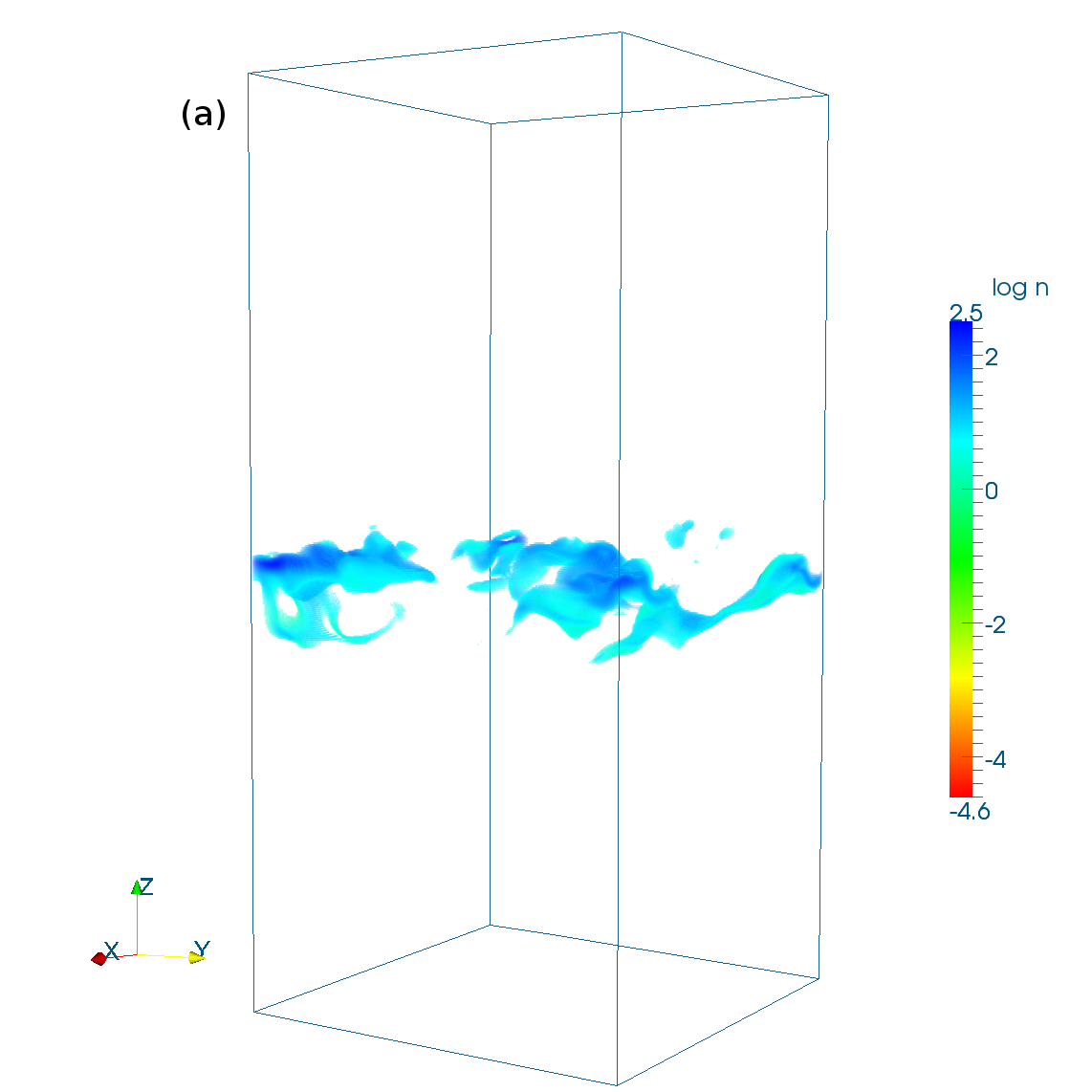}\hspace{-1.3cm}
  \includegraphics[width=0.3765\linewidth]{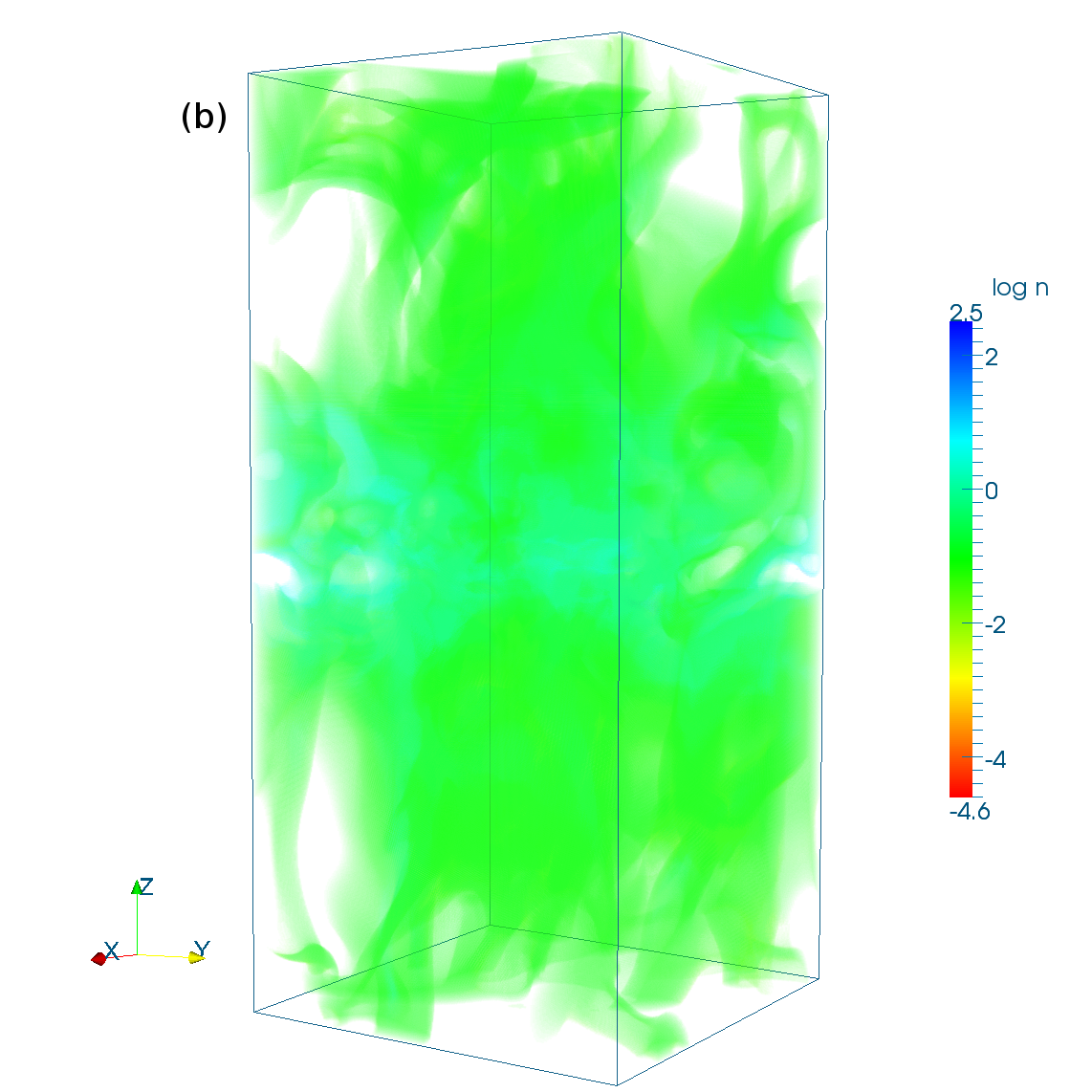}\hspace{-1.3cm}
  \includegraphics[width=0.3765\linewidth]{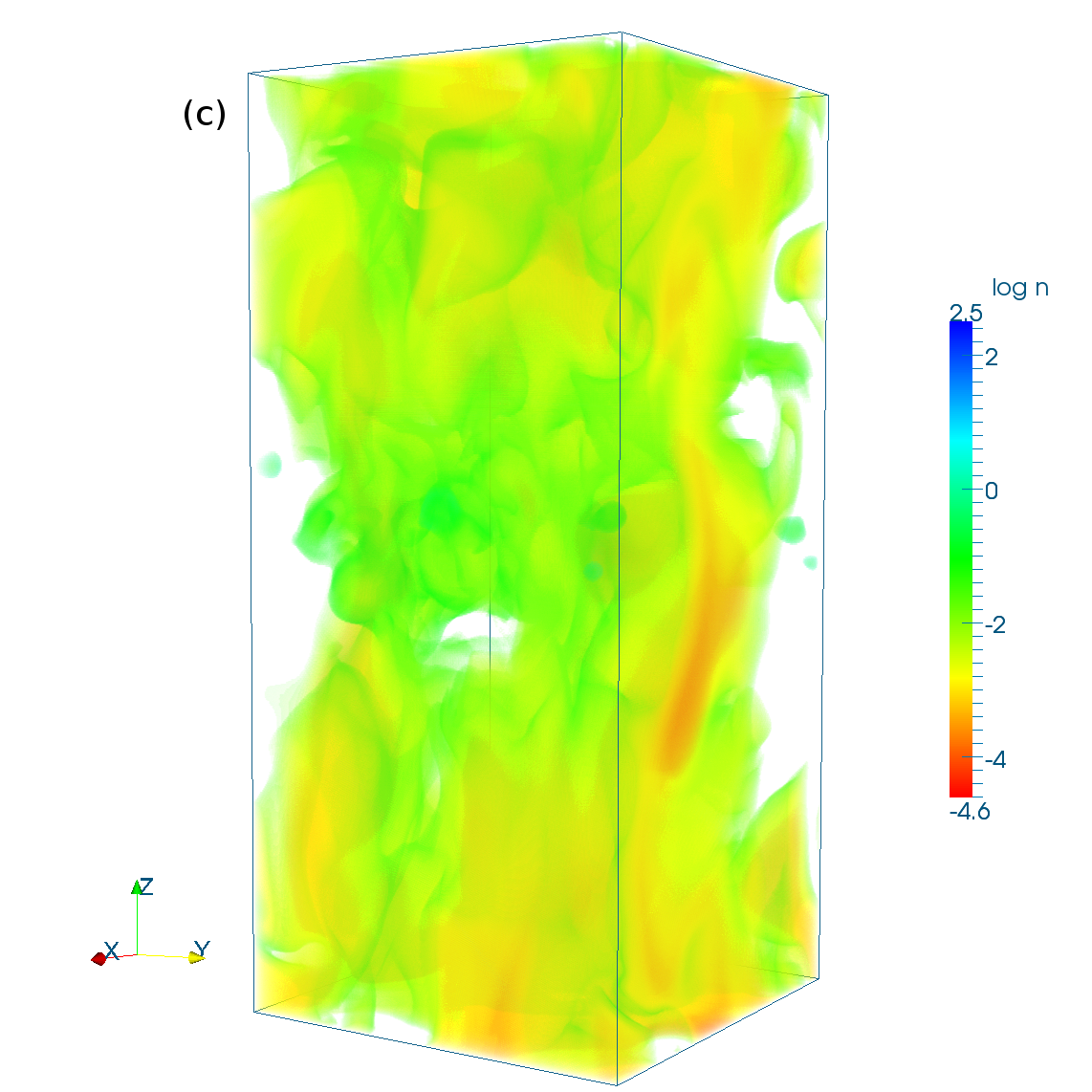}
  \caption{3D snapshots, from model~{\Op}, of gas number density in
  \textbf{(a)} the cold gas, \textbf{(b)} the warm  gas, and \textbf{(c)} the hot gas. In each plot regions
  that are clear (white space) contain gas belonging to another phase.
  The phases are separated at temperatures $500\K$ and 
  $5\times10^5\K$. The colour scale for log\,$n$ is common to all 
  three plots.
  \label{fig:rho3d}}
  \end{figure*}
%-----------------------------------------------------------------------------

The multi-phase gas structure obtained in our simulations appears to be
robust, with overall parameters relatively insensitive to the physical
(Section~\ref{TMPS}) and numerical (Section~\ref{COMP}) details, including the
parameterizations of the radiative cooling tested here (Section~\ref{COOL}). 
We have identified natural temperature boundaries of the major phases 
using the variation, with height above the mid-plane, 
of the fractional volume occupied by the gas 
in relatively narrow temperature ranges. This confirms that the system
can be satisfactorily described in terms of just three major phases with
temperature ranges $T<5\times10^2\K$, $5\times10^2\leq T<5\times10^5\K$ and
$5\times10^5\leq T<5\times10^6\K$. The most probable values of the variables we
have explored (gas density, thermal and total pressure, perturbed velocity and Mach
number) are practically independent of the cooling function chosen
(Fig.~\ref{fig:wsw_pdf3ph}). Moreover, this is true for the cold, warm and hot
phases separately.
A 3D rendering of a snapshot of the density distribution from the reference 
model~{\Op} is illustrated in Fig~\ref{fig:rho3d}, showing the typical 
location and density composition of each phase separately.

A conspicuous contribution to various diagnostics --- especially within $200\p$
of the mid-plane, where most of the SNe are localised --- comes from the very hot
gas within SN remnants. Regarding its contribution to integrated gas
parameters, it should perhaps be considered as a separate phase.

The fractional volume occupied by each phase is a convenient diagnostic and
an important physical parameter. We have clarified the relation between the
fractional volume and various probabilistic measures of a random distribution
of density (or of any other quantity), and established an exact relation
between the fractional volume and various density averages obtainable
observationally (in Section~\ref{FF}). This represents a significant improvement
upon the assumption of locally homogeneous gas, the only analytical tool 
used to date in determinations of the fractional volumes of the phases.

The correlation scale of the random flows is obtained in Section~\ref{CORR},
from the autocorrelation functions of the velocity components. 
Within $200\p$ of the mid-plane, the horizontal velocity components 
have a consistent correlation scale of about $100\p$.
In contrast, the scale of the vertical velocity 
(which has a systematic part due to the galactic outflow of hot gas) 
grows from about $100\p$ at the mid-plane 
to nearly $200\p$ at $z=200\p$,
and {may do so} further at larger heights
{\citep[cf.][]{Korpi99}}. This is due to the increase of the fractional
volume of the hot gas with distance from the mid-plane. At $|z|\simeq1\kpc$
most of the volume is occupied by the hot gas. As the
interstellar gas flows out of the galactic disc into the halo, it must expand,
and the scale of the expanding regions {may be expected to become} comparable to 
$1\kpc$ at $|z|\simeq1\kpc$. 
{It would be helpful to obtain estimates of the horizontal correlation 
of the flow above $\pm1\kpc$, so that modelling of the galactic fountain might
be adequately formulated.}

We find clear indication of cold gas falling back towards the mid-plane at
speeds of a few km/s, hot gas involved in vigorous outflow away from the
mid-plane, and some warm gas entrained in this outflow (Section~\ref{GO}). The
outflow speed of the hot gas increases up to $100\kms$ within $100\p$ of the
mid-plane and then slowly decreases.
In contrast, the mean vertical velocity of the warm gas increases linearly
with $|z|$, up to $20\kms$ {towards} 
the upper boundaries of our domain.

Given that probability densities for gas temperature and number density,
calculated for individual phases, are clearly separated, the probability
densities for both thermal and total pressure (the sum of thermal and turbulent)
are not segregated at all.
Despite its complex thermal and dynamical structure, 
the gas is in statistical pressure equilibrium. 
Since the SN-driven ISM is random in nature, both total and thermal pressure
fluctuate strongly in both space and time (albeit with significantly smaller 
relative fluctuations than the gas density, temperature and perturbation 
velocity), so the pressure balance is also
statistical in nature. 
These might appear to be obvious statements, since a statistical steady state
(i.e., not involving systematic expansion or compression) must have such a
pressure balance.
{{Deviations from \textit{thermal\/} pressure balance and observations
of significant regions of gas within the classically forbidden thermally
unstable range (300 -- 6000$\K$), which is also evident in our probability 
distributions, may lead to conclusions of an ISM comprising a
broad thermodynamic continuum in pressure disequilibrium 
\citep[][discussion on \emph{The controversy}]{VS12}.}}
The only systematic deviations from pressure balance are associated with the 
systematic outflow of the hot gas (leading to lower pressures), and with the 
compression of the cold gas by shocks and other converging flows (leading to 
somewhat increased pressures).
Even this can be further reconciled if we allow for the global vertical
pressure gradient (cf. Fig.~\ref{fig:pall4fits}). 
It is evident that phases are locally in total pressure equilibrium. 

An important technical aspect of simulations of this kind is the minimum
numerical resolution $\Delta$ required to capture the basic physics of the
multi-phase ISM. We have shown that $\Delta=4\p$ is sufficient with the
numerical methods employed here (Section~\ref{COMP}). 
In addition to comparing results obtained for $\Delta=4\p$ and $2\p$ 
with our own code, 
we have satisfied ourselves that our results are consistent 
with those obtained by other authors
using adaptive mesh refinement with maximum resolutions of $2\p$ and $1.25\p$.

As with all other simulations of the SN-driven ISM, we employ a host of
numerical tools (such as shock-capturing diffusivity) to handle the
extremely wide dynamical range ($10^2\la T\la10^8\K$ and $10^{-4}\la
n\la10^2\cm^{-3}$ in terms of gas temperature and number density in our model)
and widespread shocks characteristic of the multi-phase ISM driven by SNe.
Their detailed description can be found in Section~\ref{NS}. We have carefully
tested our numerical methods by reproducing, quite accurately, the
Sedov--Taylor and snowplough analytical solutions for individual SN remnants
(Appendix~\ref{EISNR}).

The major elements of the ISM missing from the models presented here are
magnetic fields and cosmic rays. 
{{Analysis of the structure of the velocity field and its interaction 
with the magnetic field, effects of rotation, shear and SN rates}} will be the 
subject of a future paper.

%----------------------------------------------------------------------------
\section*{Acknowledgements}
Part of this work was carried out under the programme HPC-EUROPA2 of
the European Community (Project No.~228398) -- Research Infrastructure Action
of the FP7. We gratefully acknowledge the resources and support of the CSC-IT
Center for Science Ltd., Finland, where the major part of the code adjustment
and all of the final simulations were carried out. The
contribution of Jyrki Hokkanen (CSC) to enhancing the graphical representation
of the results is gratefully acknowledged. We used the UK MHD Computer Cluster
in St~Andrews, Scotland, for code development and testing. MJM is grateful to
the Academy of Finland for support under Projects 218159 and 141017. 
The work of AF, GRS and AS has been supported by the Leverhulme
Trust's Research Grant RPG-097 and the STFC grant F003080. 
FAG has been supported by the EPSRC DTA grant to the Newcastle University.   
{{We thank {\replyb{Elly Berkhuijsen for insightful comments on filling factors and}} the anonymous referee for their conscientious and constructive
contribution.}}

%----------------------------------------------------------------------------
\appendix

%---------------------------------------------------------------------------
\section{Notation}\label{notation}

%-----------------------------------------------------------------------------
\begin{table}
\centering
\caption{\label{table:notation}Most important variables used in the text.}
\begin{tabular}{cp{0.8\columnwidth}}
\hline
Symbol				&Meaning                                                       \\
\hline
$c\sound$			&Adiabatic speed of sound                                      \\
$c_p$               &Heat capacity at constant pressure $[\kpc^2\Gyr^{-3}\K^{-1}]$ \\
$\mathcal{C}$   	&Velocity autocorrelation function, Eq.~\eqref{eq:CORR:ac}     \\ 
$D/Dt$				&Advective derivative, Eq.~\eqref{eq:advection}                \\
$\mathcal{D}$	    &Velocity structure function, Eq.~\eqref{eq:CORR:sf}           \\ 
$e_\mathrm{th}$     &Energy density, subscript thermal: `th', kinetic: `kin'       \\
$E\SN$				&Total energy injected into the ISM by a single SN             \\
$f_{M,i}$			&Fractional mass of gas in phase $i$, Eq.~\eqref{fm}           \\
$f_{V,i}$			&Fractional volume occupied by the phase $i$, Eq.~\eqref{fv}   \\
$g_z$				&Vertical acceleration due to the Galactic gravity, Eq.~\eqref{eq:grav}\\
$h_\mathrm{I}$	    &Scale height of the Type I SN distribution, Section~\ref{MSN} \\
$h_\mathrm{II}$	    &Scale height of the Type II SN distribution, Section~\ref{MSN}\\
$k_\mathrm{B}$      &Boltzmann's constant                                          \\
$K$					&Thermal conductivity ($=c_p\rho\chi$)                         \\
$l_0$               &Velocity correlation scale                                    \\
$m_\mathrm{p}$      &Proton mass                                                   \\
$\mathcal{M}$       &Mach number                                                   \\
$n$                 &Gas number density                                            \\
$\mean{n_i}$        &Gas density averaged within a given phase $i$, Eq.~\eqref{phase_averaging}\\
$\average{n_i}$     &Gas density averaged over volume $V$ of phase $i$, Eq.~\eqref{volume_averaging}\\
$p$                 &Thermal pressure                                              \\
$P$                 &Total pressure (thermal plus turbulent)                       \\
$\mathcal{P}$       &Probability density                                           \\
$r\SN$              &Characteristic radius of the SN energy injection site, Section~\ref{MSN}\\
rms                 &Root-mean square                                              \\
$s$                 &Specific entropy, $[\erg\g^{-1}\K^{-1}]$                      \\
$s_n$               &Parameter of the lognormal probability distribution, Eq.~\eqref{ln}\\
$S$                 &Velocity shear rate due to differential rotation              \\
SN                  &Supernova (also as a subscript)                               \\
$T$                 &Gas temperature                                               \\
$V$                 &Total volume of a region in Section~\ref{BI}                  \\
$V_i$               &Volume occupied by an ISM phase labelled $i$, Section~\ref{BI}\\
$\vect{u}$          &Velocity perturbation: deviation of the gas velocity from the background rotational flow\\	 
$\vect{u}_0$        &Random velocity, p.~\pageref{u0} \\	 
$\vect{U}$          &Large-scale shear flow (differential rotation), p.~\pageref{shear}\\	
$\mathbfss W$       &Rate of strain tensor, Eq.~\eqref{eq:str}                     \\
$\Gamma$            &Specific rate of photoelectric heating, $[\erg\g^{-1}\s^{-1}]$\\
$\Delta$            &Numerical mesh separation (resolution of a simulation)        \\
$\zeta_\nu$         &Shock-capturing viscosity                                     \\
$\zeta_\chi$        &Shock-capturing thermal diffusivity                           \\
$\Lambda$           &Radiative cooling rate, $[\erg \g^{-2}\s^{-1}\cm^{-3}]$       \\
$\mu$               &Molecular weight                                              \\	
$\mu_n$             &Parameter of the lognormal probability distribution, Eq.~\eqref{ln}\\
$\nu$               &Kinematic viscosity                                           \\
$\nu_\mathrm{I}$    &Type I SN rate per unit surface area, Section~\ref{MSN}       \\
$\nu_\mathrm{II}$   &Type II SN rate per unit surface area, Section~\ref{MSN}      \\
$\vect{\Omega}$     &Angular velocity of the Galactic rotation                     \\
$\phi_i$            &Phase filling factor within the ISM phase $i$, Eq.~\eqref{phin}\\
$\Phi_i$            &Volume filling factor of the ISM phase $i$, Eq.~\eqref{ftilde}\\
$\dot\rho\SN$       &Rate of mass injection, per unit volume, by SNe, Section~\ref{MSN}\\
$\rho$              &Gas density                                                   \\
$\dot\sigma\SN$     &Rate of energy injection by SNe (per unit volume), as kinetic energy in Eq.~\eqref{eq:mom} and as thermal energy in Eq.~(\ref{eq:ent}), see Section~\ref{MSN}\\
$\sigma^2_i$        &Variance of ISM phase $i$, Section~\ref{FF}\\	
$\tau\cool$         &Radiative cooling time\\	
$\Phi$				&Gravitational potential\\	
$\chi$				&Thermal diffusivity\\	 	 
\hline
\end{tabular}
\end{table}
%---------------------------------------------------------------------------

{{Table~\ref{table:notation} contains most of the symbols used in the text and their explanation, arranged alphabetically.
}}

%----------------------------------------------------------------------------
\section{Evolution of an individual supernova remnant}\label{EISNR}
The thermal and kinetic energy supplied by SNe drives, directly or indirectly, all
the processes discussed in this paper. It is therefore crucial that the model
captures correctly the energy conversion in the SN remnants and its
transformation into the thermal and kinetic energies of the interstellar gas. As discussed in Section~\ref{MSN},
the size of the region where the SN energy is injected corresponds to the
adiabatic (Sedov--Taylor) or the snowplough stage.
Given the multitude of artificial numerical effects required to model the extreme
conditions in the multi-phase ISM, it is important to verify that the basic
physical effects are not affected, while sufficient numerical control of strong
shocks, rapid radiative cooling, supersonic flows, etc., is properly ensured. Another important parameter to be chosen is the numerical resolution. 

Before starting the
simulations of the multi-phase ISM reported in this paper, we have carefully confirmed
that the model can reproduce, to sufficient accuracy, the known realistic analytical solutions for the late stages of SN remnant expansion, until merger with the ISM. The minimum numerical resolution required to achieve this in our model is $\Delta=4\p$. In this Appendix, we consider a single SN remnant, initialised as described in
Section~\ref{MSN}, that expands into a homogeneous environment. All the
numerical elements of the model are in place, but here we use periodic
boundary conditions in all dimensions. 

The parameters $\chi_1$ and $\nu_1$ are as applied in Model~{\Op} for 
$\Delta=4\p$, but reduced here proportionally for $\Delta=2$ and $1\p$. 
The constant $C\approx0.01$ used in Eq.~\eqref{coolxi} 
to suppress cooling around shocks is unchanged. 
This may allow excess cooling at higher resolution, 
evident in the slightly reduced radii in Fig.~\ref{fig:snplb}. 
For Model~{\OpH}, 
$\chi_1$ and $\nu_1$ were just as in Model~{\Op};
for future reference, they should be appropriately adjusted, as should $C$, 
to better optimise higher resolution performance.

%-----------------------------------------------------------------------------
\subsection{The adiabatic and snowplough stages}\label{SNPL}

%-----------------------------------------------------------------------------
\begin{figure*}
\centering
\includegraphics[width=0.33\textwidth]{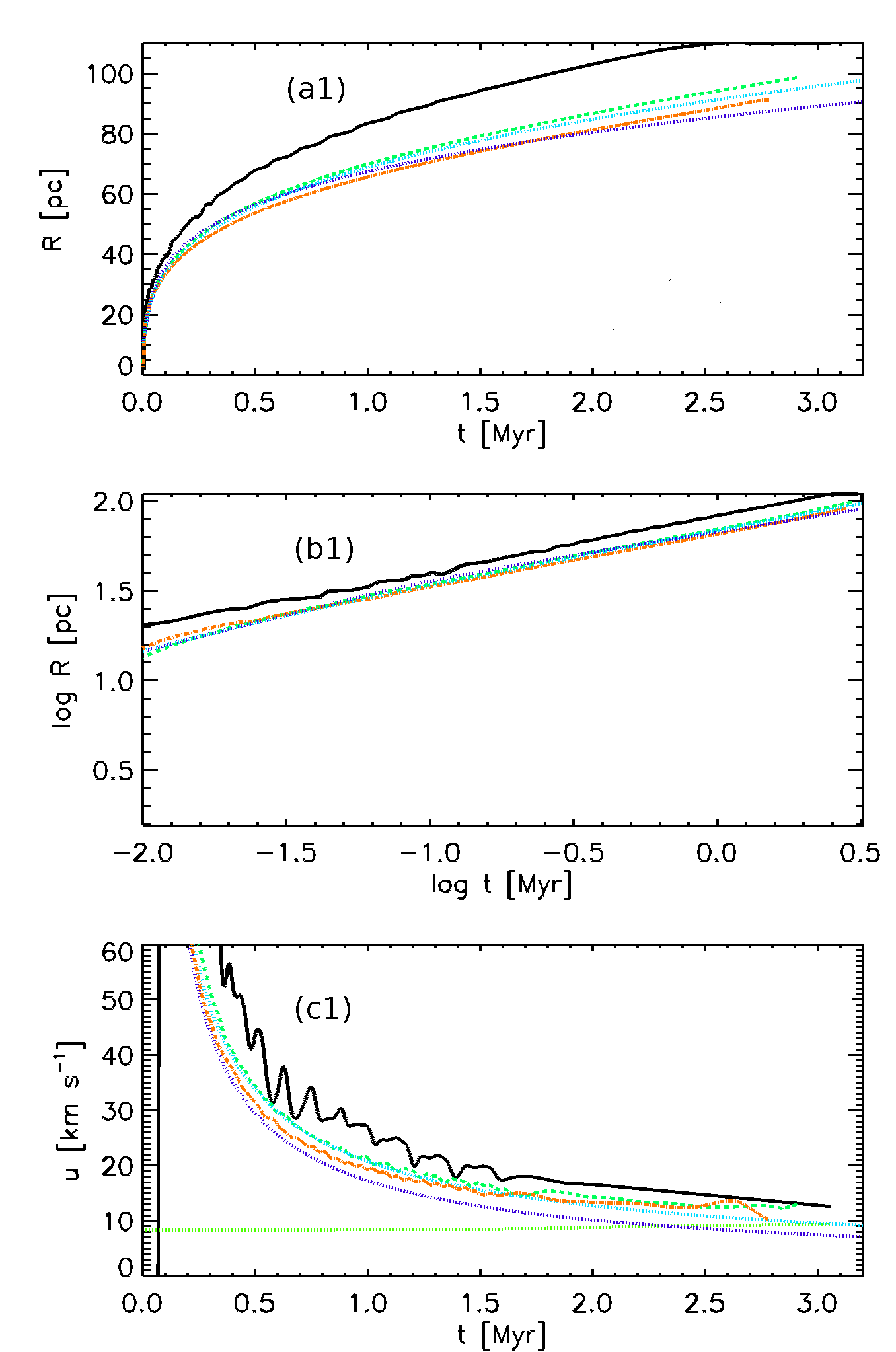}
\hfill
\includegraphics[width=0.33\textwidth]{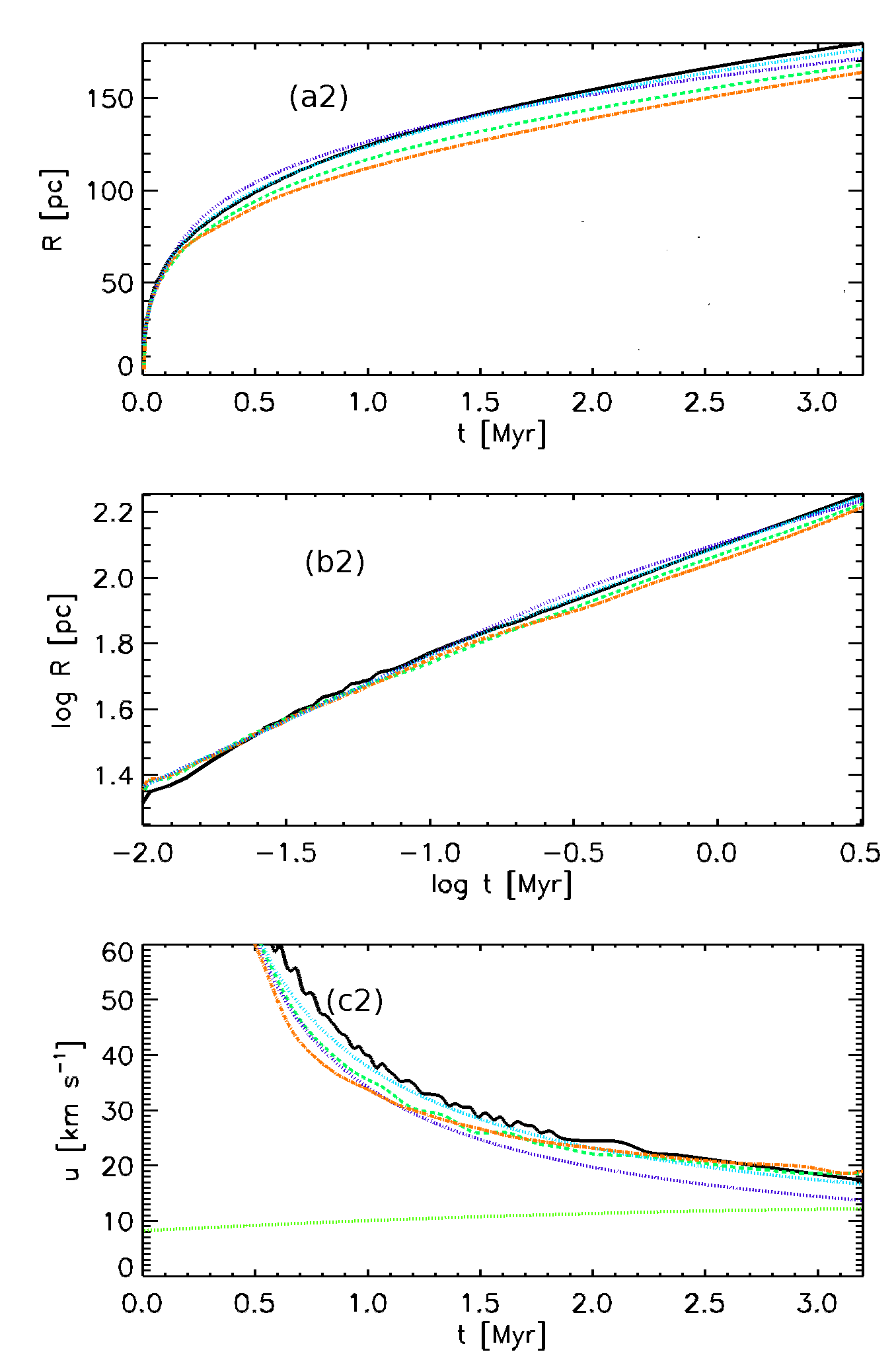}
\hfill
\includegraphics[width=0.33\textwidth]{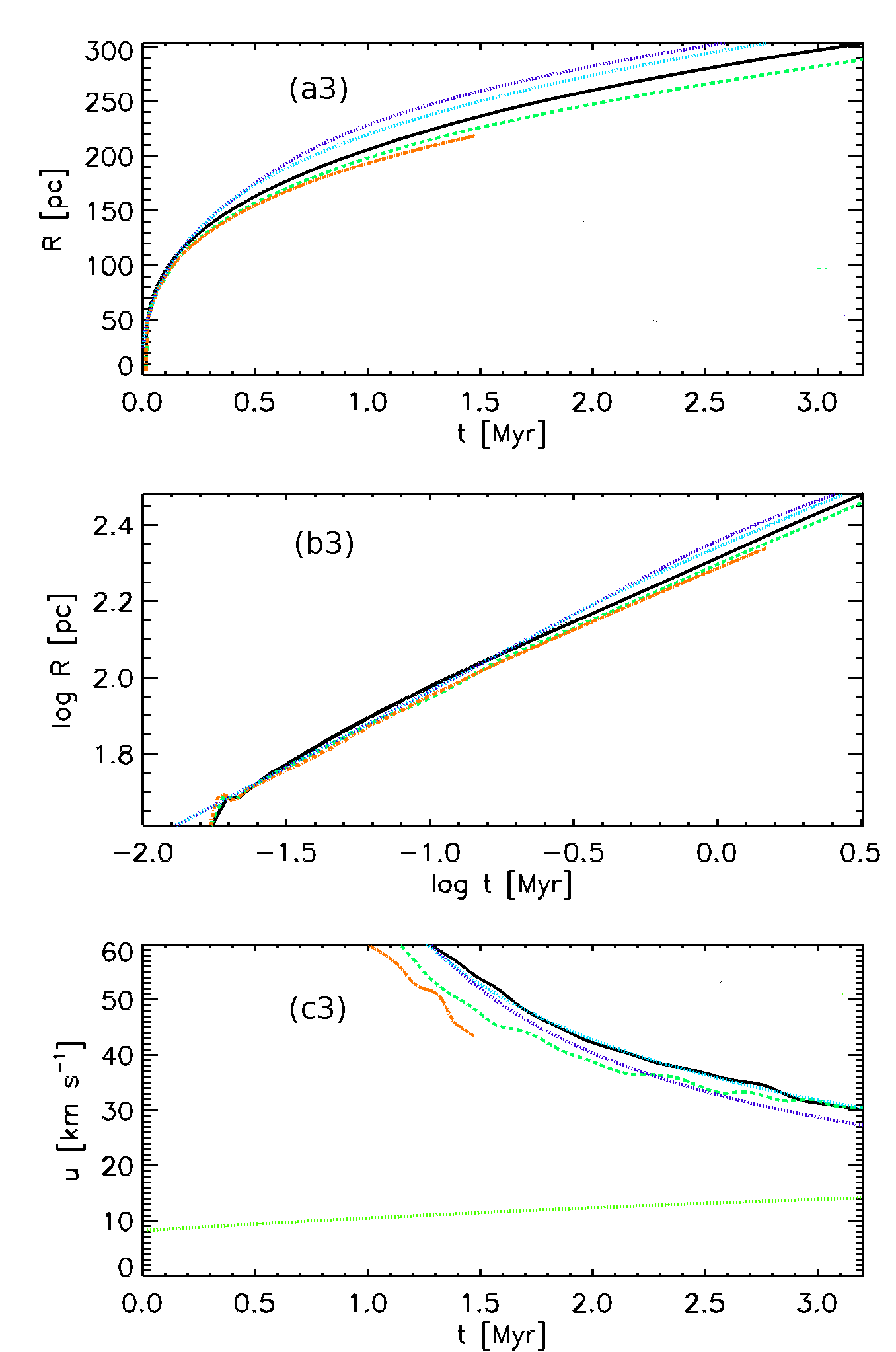}
\caption{The shell radius $R$ of an SN remnant versus time, shown in \textbf{(a)}~linear and 
\textbf{(b)}~logarithmic scales;
\textbf{(c)}~the corresponding expansion speed $\dot{R}$. Frame
columns 1--3 are for different ambient gas densities,
$\rho_0\times10^{24}\g\cmcube=1.0,0.1,0.01$ from left to right. Numerical
results obtained under three numerical resolutions are shown: $\Delta=4\p$
(black, solid), 2\,pc (green, dashed) and 1\,pc (orange, dash-dotted). Dotted
lines are for the standard snowplough solution (\ref{eq:snpl}) (dark blue) and
its modification by \citet{Cioffi98} (light blue). The horizontal line in
Panels~(\textbf{c1})--(\textbf{c3}) shows the sound speed in the ambient ISM.
\label{fig:snplb}}
\end{figure*}
%-----------------------------------------------------------------------------

The Sedov--Taylor solution,
\begin{equation}
  \label{eq:sed}
  R= \left(\kappa\frac{E\SN}{\rho_{0}}\right)^{{1}/{5}}t^{{2}/{5}},
\end{equation}
is accurately reproduced with our code at the resolution $\Delta=4\p$ or higher.
Here $R$ is the remnant radius, $E\SN$ the explosion energy,
$\rho_{0}$ the ambient gas density, and $\kappa\approx2.026$ for $\gamma=5/3$ \citep{Ostriker88}.

Modelling even a single remnant becomes
more challenging when radiative cooling becomes important. Here we compare
numerical results with two analytic solutions for an SN remnant expanding into
a perfect, homogeneous, monatomic gas at rest.
The standard momentum-conserving snowplough solution for a radiative SN remnant has the form
\begin{equation}
  \label{eq:snpl}
  R=R_{0}\left[1+4\frac{\dot{R_{0}}}{R_{0}}(t-t_{0})\right]^{{1}/{4}} ,
\end{equation}
where $R_{0}$ is the radius of the SN remnant at the time $t_{0}$ of the transition from the adiabatic stage, and
$\dot{R_{0}}$ is the shell expansion speed at $t_{0}$. 
The transition time is determined by \citet{Woltjer72} 
as that when half of the SN energy is lost to radiation; this happens when
\begin{equation}
  \dot{R_{0}}=230\kms \left(\frac{n_{0}}{1\cmcube}\right)^{{2}/{17}}\left(\frac{E\SN}{10^{51}\erg}\right)^
{{1}/{17}};
\end{equation}
the transitional expansion speed thus depends very weakly on parameters.

\citet{Cioffi98} obtained numerical and analytical solutions for an expanding SN remnant with special attention to the transition
from the Sedov--Taylor stage to the radiative stage. These authors adjusted an analytical solution for the pressure-driven snowplough stage to fit their numerical results to an accuracy of within 2\% and 5\% in terms of $R$ and $\dot R$, respectively.  (Their numerical resolution was $0.1\p$ in the interstellar gas and $0.01\p$ within ejecta.)  They thus obtained
\begin{equation}
  \label{eq:pds}
  R=R_{\rm{p}} \left(\frac{4}{3}\frac{t}{t_{\rm{p}}}-
  \frac{1}{3}\right)^{3/10},
\end{equation}
where the subscript ${\rm{p}}$ denotes the radius and time for the transition
to the pressure driven stage. The estimated time of this transition is 
\[ 
t_{\rm{p}}\simeq13\Myr\left(\frac{E\SN}{10^{51}\erg}\right)^{3/14}
   \left(\frac{n_0}{1 \cmcube}
          \right)^{-4/7}. 
\]
For ambient densities of $\rho_0=(0.01,0.1,1)\times10^{-24}\g\cmcube$, 
this yields transition times 
$t_\mathrm{p}\approx(25,6.6,1.8)\times10^4\yr$ and shell radii
$R_{\rm{p}}
\approx(130,48,18)\p$, respectively, with speed $\dot{R}_\mathrm{p}=(213,296,412)\kms$

This continues into the momentum driven stage with
\begin{equation}
  \label{eq:mcs}
  \left(\frac{R}{R_{\rm{p}}}\right)^{4}=
  \frac{3.63~\left(t-t_{\rm{m}}\right)}{t_{\rm{p}}}
  \left[1.29-\left(\frac{t_{\rm{p}}}{t_{\rm{m}}}\right)^{0.17}\right] +
  \left( \frac{R_{\rm{m}}}{R_{\rm{p}}}\right)^{4},
\end{equation}
where subscript ${\rm{m}}$ denotes the radius and time for this second transition,
\[
t_{\rm{m}}\simeq 61\, t_\mathrm{p}
               \left(\frac{\dot{R}_{\rm{ej}}}{10^3\kms}\right)^3
               \left(\frac{E\SN}{10^{51}\erg}\right)^{-3/14}
               \left(\frac{n_0}{1\cmcube}\right)^{-3/7},
\]
where $\dot{R}_{\rm{ej}}\simeq5000\kms$ is the initial velocity of the
$4M\sun$ ejecta.
For each $\rho_0=(0.01,0.1,1.0)\times10^{-24}\g\cmcube$, the transitions occur at 
$t_{\rm{m}}=(168,16.8,1.68)\Myr$, and $R_{\rm{m}}=(1014,281,78)\p$, respectively.
The shell momentum in the latter solution tends to a constant, and the solution 
thus converges with the momentum-conserving snowplough (\ref{eq:snpl});
but, depending on the ambient density, the expansion may become subsonic and the
remnant merge with the ISM before Eq.~(\ref{eq:snpl}) becomes applicable. 

We compare our results with the momentum-conserving snowplough solution and those of \citeauthor{Cioffi98} in Fig.~\ref{fig:snplb}, testing our model
with numerical resolutions $\Delta=1$, 2 and 4\,pc
for the ambient gas densities $\rho_0=(0.01,0.1,1.0,2.0)\times10^{-24}\g\cmcube$. 
Shown in Fig.~\ref{fig:snplb} are a linear plot of the remnant radius $R$ versus time to check if its magnitude is accurately reproduced, a double logarithmic plot of $R(t)$ to confirm that the scaling is
right, and variation of the expansion speed with time to help assess more delicate properties of the solution. 
We are satisfied to obtain good agreement with the analytical results for all
the resolutions investigated when the ambient gas number density is below
$1\cmcube$. For $\Delta=4\p$, the remnant radius is accurate to within about 3\% for $\rho_0=10^{-25}\g\cmcube$ and underestimated by up to 6\% for 
$\rho_0=10^{-26}\g\cmcube$. At higher numerical resolutions, the remnant radius is underestimated by up to 7\% and 11\% for $\rho_0=10^{-25}\g\cmcube$ and $10^{-26}\g\cmcube$, respectively.
For $\rho_0=10^{-24}\g\cmcube$, excellent agreement is obtained for the higher 
resolutions, $\Delta=1$ and $2\p$; simulations with $\Delta=4\p$ overestimate 
the remnant radius by about 20--25\% in terms of $R$ and $\dot R$ at $t=2\Myr$. We emphasize that a typical SN explosion site in the models
described in the main part of the paper has an ambient density $n_0<1\cmcube$
so that $\Delta=2$, or $4\p$ 
produce a satisfactory fit to the results,
despite the much finer resolution of the simulations, of \citeauthor{Cioffi98}

The higher than expected expansion speeds into dense gas can be explained by 
the artificial suppression of the radiative cooling within and near to the shock
front as described by Eq.~(\ref{coolxi}). Our model reproduces the low
density explosions more accurately because the shell density is lower, 
and radiative cooling is therefore less important. 

%------------------------------------------------------------------------
\begin{figure*}
\centering
\includegraphics[height=0.225\textwidth,width=.99\textwidth]{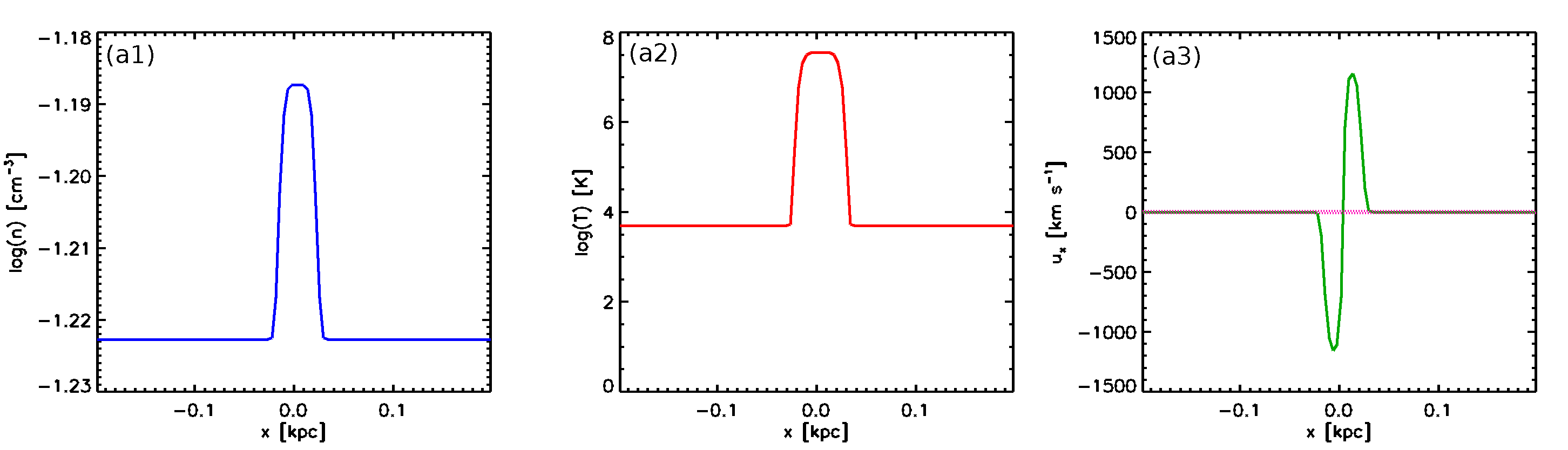}
\hfill
\includegraphics[height=0.225\textwidth,width=.99\textwidth]{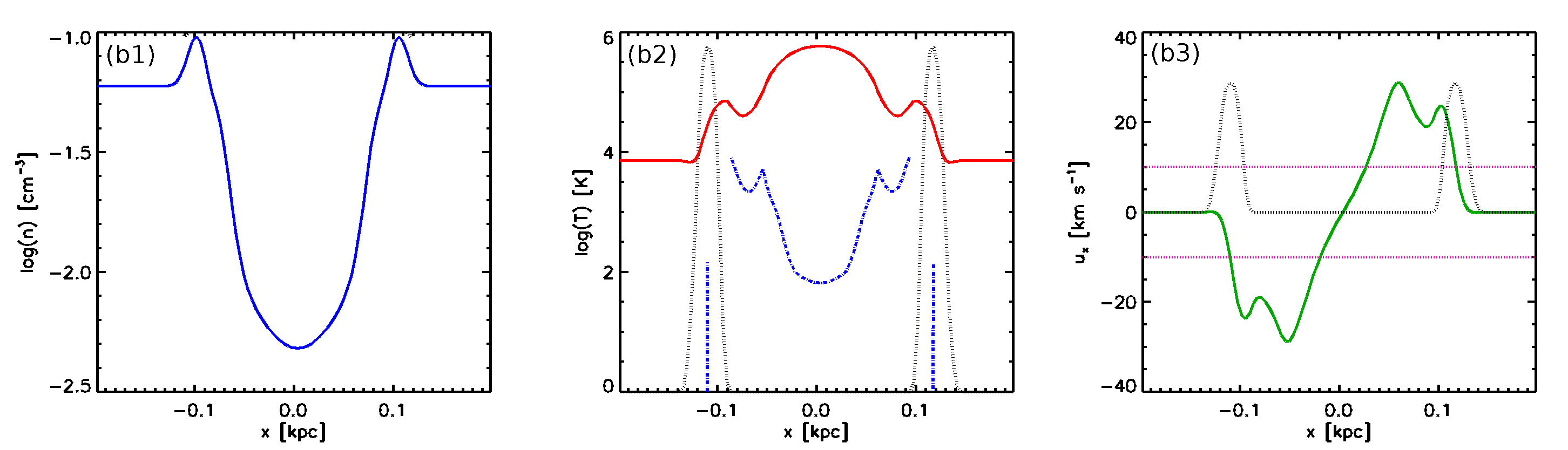}
\hfill
\includegraphics[height=0.225\textwidth,width=.99\textwidth]{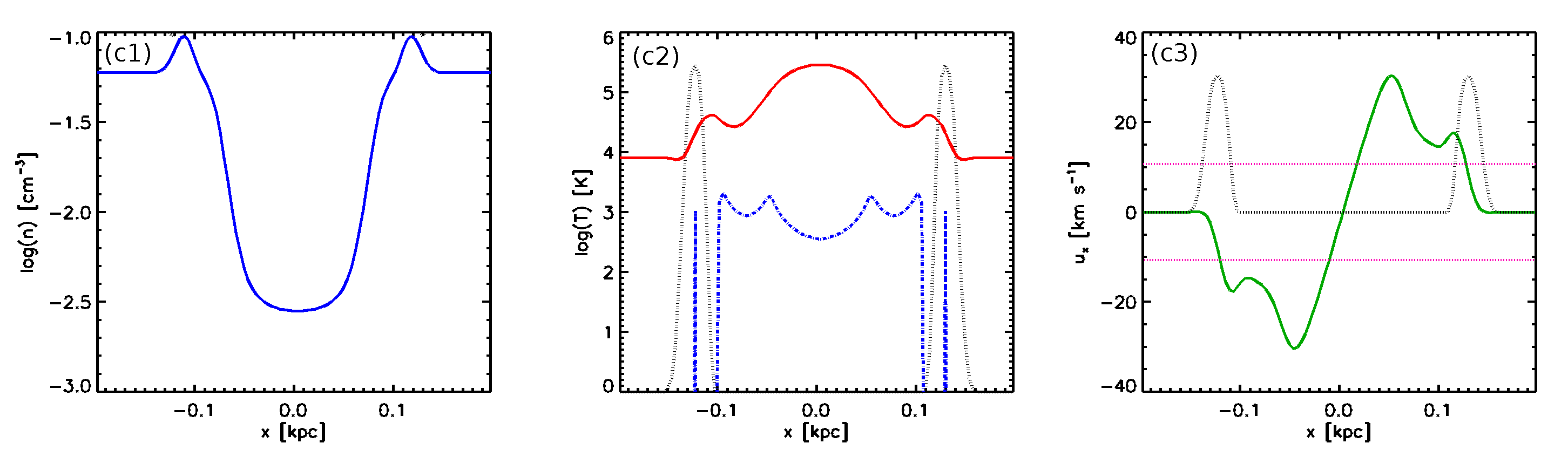}
\caption{
One-dimensional cuts through the origin of an SN remnant expanding into gas of
{{ambient density}} $\rho_0=10^{-25}\g\cmcube$, simulated with the 
numerical resolution $\Delta=4\p$. 
The variables shown are \textbf{(a1)--(c1)} gas number density {{(blue, solid)}}, 
\textbf{(a2)--(c2)} temperature {{(red, solid)}}, and
\textbf{(a3)--(c3)} velocity {{(green, solid)}}.
The shock viscosity profile of Eq.~(\ref{shockdiff}) (scaled to fit the frame, 
black, dotted) is shown in the temperature and velocity panels;
the net cooling (blue, dashed), log($-T^{-1}(\Gamma-\rho\Lambda)_+$), from 
Eq.~(\ref{eq:ent}) is included in the temperature panel;
and the ambient sound speed (pink, dotted) is also shown with the velocity. 
Panels in the top row \textbf{(a)} show the injection profiles used to 
initialise the remnant at $t=0$; the lower panel rows are for the later times 
\textbf{(b)}~$t=0.72\Myr$ and \textbf{(c)}~$t=1.02\Myr$.
\label{fig:1dtime}}
\end{figure*}
%-----------------------------------------------------------------------------

%-----------------------------------------------------------------------------
\subsection{The structure of the SN remnant}\label{RPROF}
Cuts through the simulated SN remnant are shown in Fig.~\ref{fig:1dtime} for gas density, temperature and velocity, obtained for resolution $\Delta=4\p$ and with ambient density $\rho_0=10^{-25}\g\cm^{-3}$. In the temperature
and velocity panels, we also include the profile of the shock viscosity from Eq.~(\ref{shockdiff}) (black dotted line), scaled to fit each plot. The
temperature panels also show where net cooling is applied to the remnant,
$T^{-1}(\Gamma-\rho\Lambda)<0$ 
from Eq.\eqref{eq:ent} (blue dashed line), 
while the velocity panels also show the ambient sound
speed (pink dashed lines).
The top panel depicts the initial distributions, at $t=0$, 
with which the mass of $4\Msol$ and $5\times10^{50}\erg$ each of thermal and kinetic energy are injected. 
The other panels are for $t=0.72$ and 1.02 after the start of the evolution, 
from top to bottom, respectively;  
the actual simulation continued to $t=1.32\Myr$,
when the remnant radius reached $130\p$.

The position of the peak of the density profile is used to determine the shell radius shown in Fig.~\ref{fig:snplb}.
The Rankine--Hugoniot jump conditions are not very well satisfied
with the numerical parameters used here. This is due to our numerical setup, 
essentially designed to control the shocks by spreading them sufficiently to be numerically resolvable in production runs that contain many interacting shocks and colliding SN shells. Better shock front profiles have been obtained with other choices of parameters and cooling control, and with better resolution.
The density and temperature contrasts across the shock fronts are reduced by the shock smoothing, which inhibits the peak density and enhances gas density
behind the shocks. In an isolated remnant, the peak gas number density does not exceed $10\cmcube$, but in the full ISM simulation we obtain densities in excess of $100\cmcube$, as a result of interacting remnants and highly supersonic flows.

The interior of the SN remnant, if more dense due to numerical smoothing 
about the shock profile, would cool unrealistically rapidly, 
so that the SN energy would be lost to radiation rather than agitate the ambient ISM. 
The centre panels in Fig.~\ref{fig:snplb} clarify how the cooling
suppression described in Eq.~(\ref{coolxi}) reduces the cooling rate in the relatively homogeneous interior of the remnant, while still allowing rapid cooling in the dense shell where the gradient of the shock viscosity is small. 
It is evident from the temperature cuts that the remnant still contains substantial amounts of hot gas  when its radius reaches 100\,pc, so it would be merging with the ISM in the full simulation.

The panels in the right column of Fig.~\ref{fig:snplb} demonstrate that the interior gas velocity can be more than twice the shell speed. Due to the high interior temperature, this flow is subsonic, while the remnant shell expands supersonically with respect to its ambient sound speed. The enhanced viscosity in the hotter interior (with viscosity proportional to the sound speed; see Section~\ref{NS}) inhibits numerical instabilities that could arise from the high velocities. In fact, accurate modelling of the SN interiors is not essential in the present context (where we are mainly interested in a realistic description the  multi-phase ISM), as long as the interaction of the remnant with the ambient gas is well described, in terms of the energy conversion and transfer to the ISM, the scales and energy of turbulence, and the properties of the hot gas.

%-----------------------------------------------------------------------------
\section{Boundary conditions and numerical control of advection and diffusion}\label{BCND}
%-----------------------------------------------------------------------------
\subsection{Top and bottom boundaries}
Unlike the horizontal boundaries of the computational domain, where
periodic or sliding-periodic boundary conditions are adequate 
(within the constraints of the shearing box approximation), 
the boundary conditions at the top and bottom of the domain are more demanding.
The vertical size of the galactic halo is of order of 10\kpc, and nontrivial physical processes occur even at that height, especially when galactic wind and cosmic ray escape are important. 
{As explained in Section~\ref{NS}, we do not attempt to model 
the full extent of the halo here.}
Therefore, it is important to formulate boundary conditions at the top and bottom of the domain 
that admit the flow of matter and energy, 
while minimising any associated artifacts that might affect the interior.
  
Stress-free, open vertical boundaries would seem to be the most appropriate, 
requiring that the horizontal stresses vanish,
while gas density, entropy and vertical velocity 
have constant first derivatives on the top and bottom boundaries. 
These are implemented numerically using `ghost' zones;
i.e., three outer grid planes that allow derivatives at the boundary to be calculated in the same way as at interior grid points.
The interior values of the variables are used to specify their ghost zone
values. When a sharp structure approaches the boundary, the strong gradients are therefore extrapolated into the ghost zones. This artificially enhances the prominence of such a structure, 
and may cause the code to crash.
Here we describe how we have modified these boundary conditions 
to ensure the numerical stability of our model.

To prevent artificial mass sources in the ghost zones, we impose a weak 
negative gradient of gas density in the ghost zones. Thus, the density values
are extrapolated to the ghost zones from the 
boundary point as
\[
\rho(x,y,\pm Z\pm k\Delta)=(1-\Delta/0.1\kpc)\rho(x,y,\pm Z\pm(k-1)\Delta)
\]
for all values of the horizontal coordinates
$x$ and $y$, where the boundary surfaces are at $z=\pm Z$, and the ghost zones are at $z=\pm Z\pm k\Delta$ with $k=1,2,3$. The upper (lower) sign is used at the top (bottom) boundary. This ensures that gas density gradually declines in the ghost zones. 

To prevent a similar artificial enhancement of temperature spikes in the ghost 
zones, gas temperature there is kept equal to its value at the boundary,
\[
 T(x,y,\pm Z\pm k\Delta)=T(x,y,\pm Z)\,,
\]
so that temperature is still free to fluctuate in response to the interior processes. 
This prescription is implemented in terms of entropy,
given the density variation described above.

Likewise, the vertical velocity in the ghost zones is kept equal to its 
boundary value if the latter is directed outwards,
\[
 u_z(x,y,\pm Z\pm k\Delta)=u_z(x,y,\pm Z)\,,
\qquad 
 u_z(x,y,\pm Z) \gtrless 0\, .
\]
However, when gas cools rapidly near the boundary, pressure can decrease and gas would flow inwards away from the boundary. 
To avoid suppressing inward flows, 
where $u_z(x,y,\pm Z)\lessgtr0$ we use the following: 
if $|u_z(x,y,\pm Z\mp \Delta)| < |u_z(x,y,\pm Z)|$, we set
\[
u_z(x,y,\pm Z\pm\Delta)=\tfrac{1}{2}\left[u_z(x,y,\pm Z)+u_z(x,y,\pm Z\mp\Delta)\right]\, ;
\]
otherwise, we set
\[
u_z(x,y,\pm Z\pm\Delta)=2u_z(x,y,\pm Z)-u_z(x,y,\pm Z\mp \Delta)\,.
\]
In both cases, in the two outer ghost zones ($k=2,3$), we set
\begin{align*}
u_z(x,y,\pm Z\pm k\Delta)=&2u_z(x,y,\pm Z\pm (k-1)\Delta)\\
                          -&u_z(x,y,\pm Z\pm (k-2)\Delta)\,,
\end{align*}
so that the inward velocity in the ghost zones is always smaller than its boundary value.
This permits gas flow across the boundary in both directions,
but ensures that the flow is dominated by the interior dynamics,
rather than by anything happening in the ghost zones.

The Pencil code is non-conservative, so that gas mass is not necessarily conserved; this can be a problem 
due to extreme density gradients developing with widespread strong shocks. 
Solving Eq.~(\ref{eq:mass}) for $\rho$, rather than $\ln\rho$, 
solves this problem for the snowplough test cases described in 
Appendix~\ref{SNPL}, with mass then being conserved within machine accuracy.
However for the full model, once the ISM becomes highly turbulent, 
there remains some numerical mass loss.
A comparison of mass loss through the vertical boundaries to the total mass 
loss in the volume
indicates that numerical dissipation accounts for 
$\ll1\%$ per $\Gyr$. 
The rate of physical loss, from the net vertical outflow, 
was of order $15\%$ per $\Gyr$. 

%--------------------------------------------------------------------------
\subsection{Time step control}
To achieve numerical stability with the explicit time stepping used, the
CFL conditions have to be amply satisfied. For example, for advection terms, the numerical time step should be selected such that
\[
  \Delta t < \kappa \frac{\Delta}{\max(c\sound, u, U)},
\]
where $c\sound$ is the speed of sound, $u=|\vect{u}|$ is the amplitude of the perturbed
velocity, i.e., the deviation from the imposed azimuthal shear flow $U$, and $\kappa$ is a dimensionless number, determined empirically, which often must be significantly smaller than unity.
Apart from the velocity field, other variables also affect the maximum time step, e.g., those associated with diffusion, cooling and heating, so that the following
inequalities also have to be satisfied:
\[
  \Delta t < \frac{\kappa_1\Delta^2}{\max(\nu, \gamma\chi, \eta)}\,,
\quad\quad  \Delta t < \frac{\kappa_2}{H_{\max}}\,,
\]
where $\kappa_1$ and $\kappa_2$ are further empirical constants and
\[
  H_{\max}=\max\left(\frac{2\nu|\mathbfss W|^2+\zeta_\nu(\nabla\cdot\vect{u})^2+\zeta_\chi(\nabla\cdot\vect{u})^2}{c_V T}\right)\,.
\]
We use $\kappa=\kappa_1=0.25$ and $\kappa_2=0.025$. The latter, more stringent
constraint has a surprisingly small impact on the
typical time step, but a large positive effect on the numerical accuracy.
Whilst the time step may occasionally decrease to below 0.1 or 0.01 years following
an SN explosion, the typical time step is more than 100 years. 

%-------------------------------------------------------------------------
\subsection{Minimum diffusivity}

Numerical stability also requires that the Reynolds and P\'eclet numbers
defined at the resolution length $\Delta$,
as well as the Field length, 
are sufficiently small.
These mesh P\'eclet and Reynolds numbers are defined as
\begin{equation}
\Pe\mesh=\frac{u \Delta}{\chi}\leq\frac{u_{\rm{max}} \Delta}{\chi}\,,
\quad
\Rey\mesh=\frac{u \Delta}{\nu}\leq\frac{u_{\rm{max}} \Delta}{\nu}\,, \label{mesh_re_pe}
\end{equation}
where $u_{\rm{max}}$ is the maximum perturbed velocity and $\Delta$
is the mesh length.  
For stability these must not exceed some value, typically between 1 and 10.
{{Note that the Reynolds and P\'eclet numbers characterizing the flow are 
25 times larger, since $\Delta=0.004$ is replaced by $l_0\simeq0.1$ 
as the relevant turbulent length scale in the non-mesh quantities.}} 

In numerical modelling of systems with weak diffusivity,
$\nu$ and $\chi$ are usually set constant, close to the smallest value 
consistent with the numerical stability requirements.
This level strongly depends on the maximum velocity, and hence is related to
the local sound speed, which can exceed $1500\kms$ in our model. 
To avoid unnecessarily strong diffusion and heat conduction 
in the cold and warm phases, we scale the corresponding diffusivity
with gas temperature, as $T^{1/2}$. 
As a result, the diffusive smoothing is strongest in the hot phase
(where it is most required).
This may cause reduced velocity and temperature inhomogeneities within
the hot gas, and may also reduce the temperature difference 
between the hot gas and the cooler phases.

The effect of thermal instability is controlled by the Field length,
\begin{align*}
\lambda_\mathrm{F}&\simeq\left(\frac{KT}{\rho^2\Lambda}\right)^{1/2}\\
									&\simeq2.4\p \left(\frac{T}{10^6\K}\right)^\frac{7}{4}\!\!
															 \left(\frac{n}{1\cmcube}\right)^{-1}\!\!
										           \left(\frac{\Lambda}{10^{-23}\erg\cm^3\s^{-1}}\right)^{-\frac{1}{2}}\!\!,
\end{align*}
where we have neglected any heating. 
To avoid unresolved density and temperature structures 
produced by thermal instability, we require that $\lambda_\mathrm{F}>\Delta$, 
and so the minimum value of the thermal conductivity $\chi$ follows as
\[
\chi_\mathrm{min}
      =\frac{1-\beta}{\gamma\tau\cool}
        \left( \frac{\Delta}{2 \upi}\right)^2,
\]
where $\tau\cool$ is the \textit{minimum\/} cooling time,
and $\beta$ is the relevant exponent from the cooling function 
(e.g.\ as in Table~\ref{table:coolSS} for WSW cooling).
 In the single remnant
simulations of Appendix~\ref{EISNR}, $\tau\cool \gtrsim 0.75\Myr$. 
In the full ISM simulations, minimum cooling times 
as low as $0.05\Myr$ were encountered.
$\chi_\mathrm{min}$ has maxima corresponding to $\beta=0.56,-0.2,-3,
\ldots$ for $T=313, 10^5, 2.88\times10^5\K,\ldots$. 
All of these, except for that at $T=313\K$, result in
$\chi_\mathrm{min}<4\times10^{-4}\kms\kpc$ at $c_s=c_1=1\kms$,
so are satisfied by default for any $\chi_1$ 
sufficiently high to satisfy the $\Pe\mesh\leq10$ requirement.
For $T=313\K$, at $c_s=c_1$ we have 
$\chi_\mathrm{min}=6.6\times10^{-4}\kms\kpc>\chi_1$. 
Thus if cooling times as short as 0.05\,Myr were to occur in the cold gas,
we would have $\lambda_\mathrm{F}<\Delta$,
and would be marginally under-resolved. 
Our analysis of the combined distribution of density and temperature,
however, indicates that cooling times this short occur exclusively 
in the warm gas.

With $\chi_1\approx4.1\times10^{-4}\kms\kpc$, 
as adopted in Section~\ref{NS}, 
then $\Pe\mesh\le10$ is near the limit of numerical stability. (We 
discuss our choice of thermal diffusivity further in Appendix~\ref{TI}.)
As a result, the code occasionally crashed (notably when hot gas was 
particularly abundant), and had to be restarted. 
When restarting, the position or timing of the next SN explosion was modified, 
so that the particularly troublesome SN that caused the problem was avoided.
In extreme cases, it was necessary to increase $\chi$ temporarily 
(for only a few hundred time steps), to reduce the value of $\Pe\mesh$
during the period most prone to instability,
before the model could be continued with the normal parameter values.

%----------------------------------------------------------------------------
\section{Thermal instability}\label{TI}
One of the two cooling functions employed in this paper, WSW, supports isobaric
thermal instability in the temperature range $313\leq T<6102\K$ where $\beta<1$.
(Otherwise, for the RBN cooling function or outside this temperature range for WSW cooling,
we have $\beta\geq1$ or $\Gamma\ll\rho\Lambda$, so the gas is either thermally 
stable or has no unstable equilibrium.)

Under realistic conditions of the ISM, thermal instability can produce very small, 
dense gas clouds which cannot be captured with the resolution $\Delta=4\p$ used here. 
Although the efficiency of thermal instability is questionable in the turbulent, magnetized 
ISM, where thermal pressure is just a part of the total pressure 
\citep[][ and references therein]{VS00,MLK04},
we prefer to suppress this instability in the model. However, we do that not by modifying the
cooling function, but rather by enhancing thermal diffusivity so as to avoid 
the growth of perturbations at wavelengths too short to be resolved by our grid.

Following \citet{Field65}, we introduce the characteristic wave numbers
\[
k_\rho=\frac{\mu(\gamma-1)\rho_0{\cal L}_\rho}{\mathcal{R}c\sound T_0},
\hfill 
k_T=\frac{\mu(\gamma-1){\cal L}_T}{\mathcal{R}c\sound},
\hfill 
k_K=\frac{\mathcal{R}c\sound\rho_0}{\mu(\gamma-1)K},
\]
where $\mathcal{R}$ is the gas constant, and
the derivatives 
$\mathcal{L}_T\equiv(\upartial\mathcal{L}/\upartial T)_\rho$ 
and
$\mathcal{L}_\rho\equiv(\upartial\mathcal{L}/\upartial\rho)_T$
are calculated for constant $\rho$ and $T$, respectively. 
The values of temperature and density in these equations,
$T_0$ and $\rho_0$, are those at thermal equilibrium,
$\mathcal{L}(T_0,\rho_0)=0$ with 
$\mathcal{L}=\rho\Lambda-\Gamma$.
Isothermal and isochoric perturbations have the characteristic wave numbers 
$k_\rho$ and $k_T$, respectively, 
whereas thermal conductivity $K$ is characterised by $k_K$.

The control parameter of the instability is 
$\varphi=k_\rho/k_K$. 

The instability is suppressed by heat conduction, with the largest unstable wave numbers
given by \citep{Field65}
\begin{align}
k_{\rm cc}  = & \left[k_K                             (k_\rho-k_T)\right]^{1/2}\,, \\
k_{\rm cw}  = & \left[-k_K\left(k_T+\frac{k_\rho}{\gamma-1}\right)\right]^{1/2}\,,
\end{align}
for the condensation and wave modes, respectively, whereas the most unstable wave
numbers are 
\begin{align}
k_{\rm mc}  = & \left[\frac{(1-\beta)^2}{\gamma^2}
               +\frac{\beta(1-\beta)}{\gamma}\right]^{1/4}
               (k_\rho k_{\rm cc})^{1/2}\,, \\
k_{\rm mw}  = & \left|\frac{\beta-1}{\gamma}\,
               k_\rho k_{\rm cw}\right|^{1/2}\,.
\end{align}
 
%------------------------------------------------------------
\begin{table}
\caption{\label{table:TI}The unstable wavelengths of thermal instability, according to 
\citet{Field65}, at thermally unstable equilibria $(T_0,\rho_0)$ with the WSW cooling function.}
\begin{tabular}{@{}lccccccc@{}}
\hline
$T_0$	&$\rho_0$			&$\varphi$ &$\lambda_\rho$	&$\lambda_\mathrm{cc}$	&$\lambda_\mathrm{mc}$	&$\lambda_\mathrm{cw}$	&$\lambda_\mathrm{mw}$\\
$[\!\K]$&[$10^{-24}\g/\!\cm^3$]&&[pc]						&[pc]										&[pc]										&[pc]										&[pc]\\
\hline
\phm313&4.97 &1.91              &\phm\phm2       &\phm5									&\phm\phm5							&\phm2										&\phm\phm4\\
4000	&1.20					&0.04				&101						&32											&\phm84		  						&14											&\phm74\\
6102	&0.94					&0.02				&192						&44											&136										&20											&120\\
\hline
\end{tabular}
\end{table}
%-----------------------------------------------------------------

Table~\ref{table:TI} contains the values of these quantities for the parameters
of the reference model~WSWa, where we present the wavelengths $\lambda=2\upi/k$ rather than the wave numbers $k$.
The unstable wavelengths of thermal instability are comfortably resolved at
$T_0=6102\K$ and $4000\K$, with the maximum unstable wavelengths  
$\lambda_{\rm cc}=44\p$ and $32\p$, respectively, being 
much larger than the grid spacing $\Delta=4\p$. The shortest unstable wavelength
of the condensation mode in our model, $\lambda_{\rm cc}=5\p$ at $T\approx313\K$
is marginally resolved at $\Delta=4\p$; gas at still lower temperatures is thermally stable.
Unstable sound waves with $\lambda_{\rm cw}=2\p$ at $T=4000\K$ are shorter than
the numerical resolution of the reference model. However, for these wave modes to be unstable, the
isentropic instability criterion must also be satisfied, which is
not the case for $\beta>0$, so these modes remain thermally stable.

Thus, we are confident that the parameters of our models (most importantly, the thermal diffusivity) have been chosen so as to avoid any uncontrolled development of thermal instability,
even when only the bulk thermal conductivity is accounted for.  
Since much of the cold gas, which is most unstable, has high Mach numbers, 
thermal instability is further suppressed by the shock capturing diffusivity
in the cold phase.

%-----------------------------------------------------------------------------
\bibliographystyle{mn2e}      
\bibliography{refs}
\label{lastpage}
\end{document}